\documentclass[epj]{svjour}
\usepackage{graphicx}
\usepackage{bm}% bold math
\usepackage{epsfig}
\usepackage{amssymb}
\usepackage{amsmath}
\usepackage{axodraw}
\usepackage{bbm}

\newcommand{\rmii}[1]{{\mbox{\tiny\rm{#1}}}}
\newcommand{\rmi}[1]{{\mbox{\scriptsize #1}}}
\newcommand{\Nc}{N_{c}}
\newcommand{\bmu}{\bar\mu}
\newcommand{\msbar}{{\overline{\mbox{\rm MS}}}}
\newcommand{\gammaE}{\gamma_\rmii{E}}
\newcommand{\be}{\begin{equation}}
\newcommand{\ee}{\end{equation}}
\newcommand{\ba}{\begin{eqnarray}}
\newcommand{\ea}{\end{eqnarray}}
\newcommand{\bi}{\begin{itemize}}
\newcommand{\ei}{\end{itemize}}
\newcommand{\tr}{{\rm Tr\,}}
\newcommand{\re}{\mathop{\rm Re}}

\newcommand{\nn}{\nonumber \\}

\newcommand{\half}{{\textstyle\frac{1}{2}}}

\newcommand{\quarter}{{\textstyle\frac{1}{4}}}

\newcommand{\Nt}{{N_t}}
\newcommand{\Ns}{{N_s}}

\newcommand{\rmO}{{\rm O}}

\newcommand{\<}{\langle}
\renewcommand{\>}{\rangle}
\newcommand{\eq}{Eq.~}

\newcommand{\fig}{Fig.~}
\newcommand{\tab}{Tab.~}
\newcommand{\la}{\label}

\newcommand{\txts}{\textstyle}

\newcommand{\im}{\mathop{\rm Im}}

\newcommand{\as}{a_{\sigma}}
\newcommand{\at}{a_{\tau}}

\newcommand{\ud}{\,\mathrm{d}}
\newcommand{\bj}{\boldsymbol{j}}
\newcommand{\by}{\boldsymbol{y}}
\newcommand{\bx}{\boldsymbol{x}}
\newcommand{\bp}{\boldsymbol{p}}
\newcommand{\bk}{\boldsymbol{k}}
\newcommand{\bq}{\boldsymbol{q}}

\begin{document}
\title{Transport Properties of the Quark-Gluon Plasma}
\subtitle{A Lattice QCD Perspective}
\author{Harvey~B.~Meyer\inst{1} 
% \author{Harvey~B.~Meyer\inst{1} \and Second author\inst{2}% etc
% \thanks is optional - remove next line if not needed
% \thanks{\emph{Present address:} Insert the address here if needed}%
}                     % Do not remove
%
% \offprints{}          % Insert a name or remove this line
%
\institute{    Institut f\"ur Kernphysik, 
Johannes Gutenberg Universit\"at Mainz, 
55099 Mainz, Germany}
% \institute{Insert the first address here \and the second here}
%
\date{\today}
% {Received: date / Revised version: date}
% The correct dates will be entered by Springer
% 

\abstract{ 
Transport properties of a thermal medium determine how its conserved
charge densities (for instance the electric charge, energy or
momentum) evolve as a function of time and eventually relax back to
their equilibrium values. Here the transport properties of the
quark-gluon plasma are reviewed from a theoretical perspective.  The
latter play a key role in the description of heavy-ion collisions, and
are an important ingredient in constraining particle production
processes in the early universe.  We place particular emphasis on
lattice QCD calculations of conserved current correlators. These
Euclidean correlators are related by an integral transform to spectral
functions, whose small-frequency form determines the transport
properties via Kubo formulae.  The universal hydrodynamic predictions
for the small-frequency pole structure of spectral functions are
summarized.  The viability of a quasiparticle description implies the
presence of additional characteristic features in the spectral
functions.  These features are in stark contrast with the functional
form that is found in strongly coupled plasmas via the gauge/gravity
duality.  A central goal is therefore to determine which of these
dynamical regimes the quark-gluon plasma is qualitatively closer to as
a function of temperature.  We review the analysis of lattice
correlators in relation to transport properties, and tentatively
estimate what computational effort is required to make decisive
progress in this field.
\PACS{
      {12.38.Gc}{}   \and
      {12.38.Mh}{}  \and 
      {25.75.-q}{}  
     } % end of PACS codes
} %end of abstract
\maketitle

\tableofcontents

\section{Introduction\label{sec:intro}}
% {\sc Completion end of December}

The study of a bulk state of matter such as water begins with its
equilibrium properties, such as the equation of state and the pair 
correlation function. However both to understand the nature of the state
more deeply and to interpret the outcome of experiments, which
necessarily involve time evolution at some level, it is essential to
study the dynamic properties of the medium.  A natural starting point
is to focus on the response of the medium to long-wavelength and
slow-frequency perturbations in energy density, momentum density and
conserved charges. This is the realm of hydrodynamics.  In addition
one will at first restrict oneself to small-amplitude perturbations,
and expand in them to first or perhaps second order. When studying a
medium in which quantum and relativistic effects play a dominant role,
the same approach proves useful. In this review we focus on the state
of matter obtained when heating up ordinary nuclear matter to
temperatures about 150'000 times the core temperature of the Sun.  If
the temperature is raised gradually, the system is thought to consist
at first of a gas of hadrons (nucleons, pions, kaons, \dots).  Around
a temperature $T=T_c\approx$100--200MeV the hadrons `melt', leading
the system to go over to a phase where individual color charges can no
longer be assigned to a unique hadron. At very high temperatures $T\gg
T_c$, the color charge is transported by weakly interacting
quasiparticles carrying either quark or gluon quantum numbers.  This
ultimate high-temperature behavior is a firm prediction of Quantum
Chromodynamics (QCD), the fundamental theory that quantitatively describes
the interactions of quarks and gluons, due to `asymptotic freedom'.
The latter property of the theory implies that the amplitude for
large-angle scattering between two energetic particles is small. The
phase above $T_c$ is called the quark-gluon plasma.  It equilibrium
properties have been studied extensively using Monte-Carlo
simulations~\cite{DeTar:2009ef}.

Two research fields provide the main motivation to study the dynamic
properties of the quark-gluon plasma: early-universe cosmology and
heavy-ion collisions.  In the latter context, a bulk state of hadronic
matter is formed for a time of the order of $5{\rm fm}/c$.  Therefore
dynamic properties of this hot phase are obviously
important. An \emph{a priori} open question is whether the relaxation
times of the medium are short enough for it to be treated as being
locally in equilibrium, thus enabling a hydrodynamic description of
the system's evolution. In the context of the early universe, the
strongly interacting particles were in the quark-gluon plasma phase up
to a time of the order of a few microseconds, at which point the
transition to the hadronic phase took place. While the universe's
expansion was slow enough to allow the quark-gluon plasma to
equilibrate, the production of weakly interacting particles depends on
certain dynamic structure functions, as we will illustrate below, with
potentially observable consequences.

Heavy ion collisions have been the experimental approach to study the
quark-gluon plasma since the mid 1970's. In the standard
interpretation of these reactions, a chunk is knocked out of each
nucleus, and these two fragments undergo strong interactions and form
a `fireball', out of which many particles are produced.  While the
`spectator' nucleons fly down the pipeline, the decay products of the
fireball are recorded in particle detectors.  The typically O($10^3$)
particles detected in the `central rapidity region' are produced with
a certain distribution in the azimuthal angle $\phi$ around the beam
axis. The distribution is parametrized by Fourier coefficients,
$dN/d\phi \propto [1+2\sum_n v_n \cos(2\phi)]$.

\begin{figure}
\vspace{0.25cm}
\centerline{\includegraphics[width=0.45\textwidth]{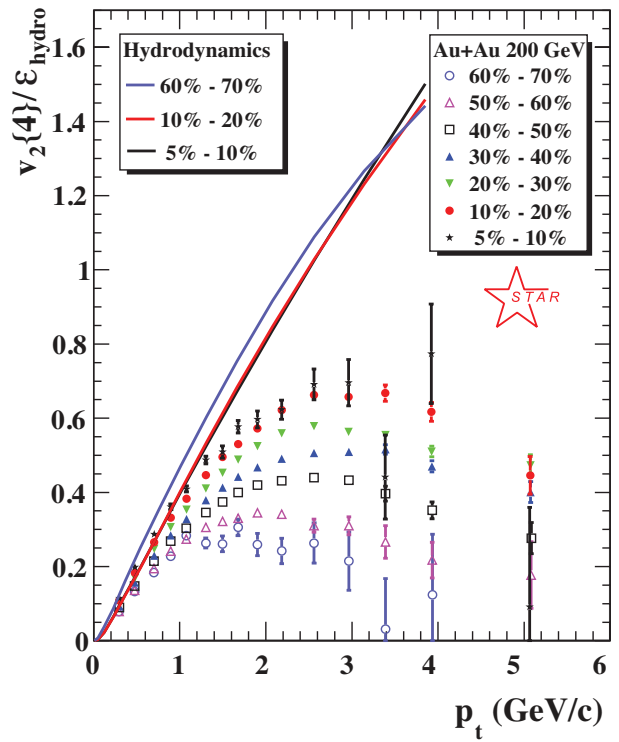}}
\vspace{-0.1cm}
\caption{The response of nuclear matter to an anisotropic initial condition 
in configuration space, expressed as the elliptic flow $v_2(p_t)$ per unit excentricity
$\epsilon_{\rm hydro} = \frac{\<y^2-x^2\>}{\<y^2+x^2\>}$, where averages are taken
with respect to the number of participants in the transverse plane.
The RHIC measurements (here by the STAR collaboration)
are compared to the ideal hydrodynamic predictions.
Figure from~\cite{BaiThesis}, presented in the review~\cite{Teaney:2009qa}.}
\label{fig:0905}
\end{figure}

Undoubtedly one of the most striking observations from the
Relativistic Heavy Ion Collider (RHIC) of Brookhaven National
Laboratory is the very large elliptic flow
$v_2$~\cite{Adcox:2004mh,Adams:2005dq,Back:2004je,Arsene:2004fa}.  The
largest nuclei collided are gold nuclei, at a center-of-mass energy
per nucleon of $\sqrt{s}/A=200$GeV.  Elliptical flow in peripheral
heavy-ion collisions is a response of the system to the initial,
anisotropic `almond-shaped' region where the interactions take place,
see \fig(\ref{fig:0905}).  The measured size of elliptical flow
suggests that the shear viscosity to entropy density ratio $\eta/s$ of
the formed system is smaller than ever observed in other
systems~\cite{Schafer:2009dj}.  In kinetic theory, the shear viscosity
is proportional to the mean free path of the quasiparticles; a short
mean free path corresponds to strongly interacting matter, which is
the reason why the hot quark matter produced at RHIC has been called a
strongly-coupled quark gluon plasma (sQGP).

\begin{figure}
\vspace{0.6cm}
\centerline{\includegraphics[width=0.45\textwidth]{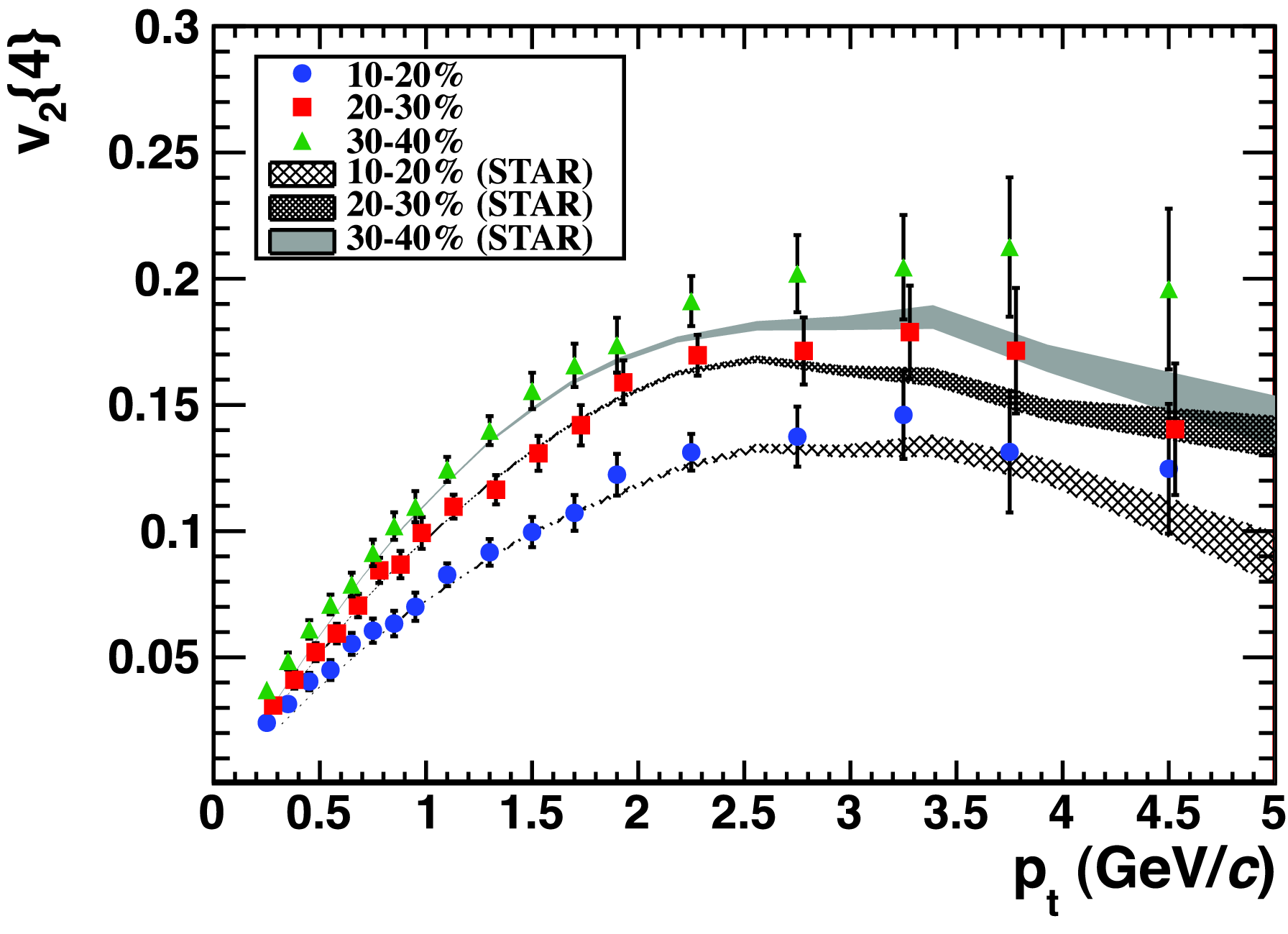}}
\caption{Comparison of the elliptic flow measured at RHIC (Au-Au collisions at $\sqrt{s}/A=200$GeV)
and at the LHC (Pb-Pb collisions at $\sqrt{s}/A=2.76$TeV)~\cite{Aamodt:2010pa}.}
\label{fig:ALICE}
\end{figure}

The Pb-Pb collisions at the Large Hadron Collider (LHC) at CERN have
very recently started to provide clues as to how sensitive the
remarkable properties of the quark-gluon plasma are with respect to a
change in temperature~\cite{Aamodt:2010pa}.  Inspite of an
order-of-magnitude increase in the center-of-mass energy, $\sqrt{s}/A
= 2.76$TeV, the charged-particle differential elliptic flow $v_2(p_t)$
is, within the uncertainties, identical to the flow observed in the
Au-Au collisions at RHIC~(see \fig\ref{fig:ALICE}).  Whatever the
outcome of the ongoing analyses, the phenomenology of these collisions
has generated a strong interest for transport properties in strongly
coupled quantum systems~\cite{Schafer:2009dj}.

\begin{figure}
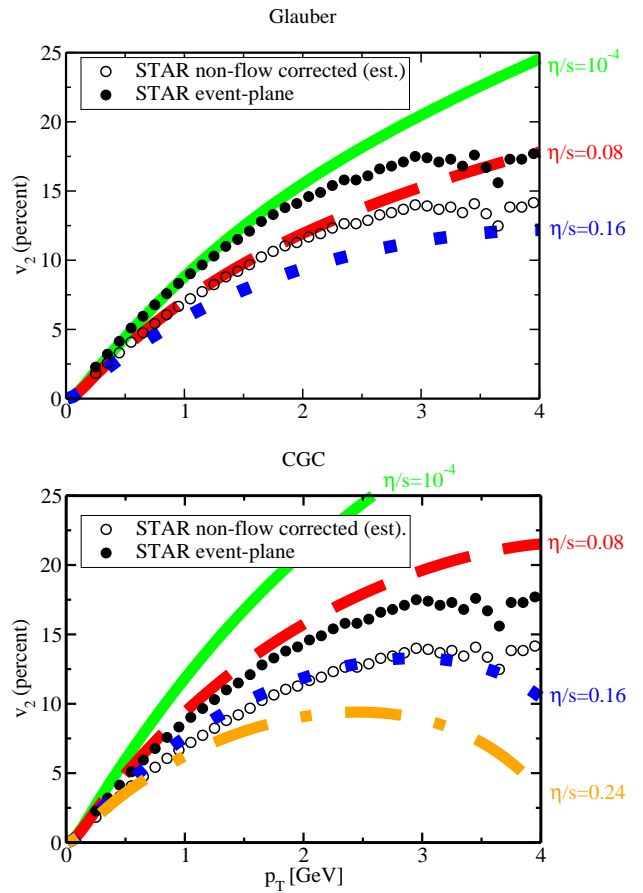

\centerline{\includegraphics[width=0.45\textwidth]{FIGS/v2minnew.eps}}
\centerline{\includegraphics[width=0.45\textwidth]{FIGS/v2minnew-paper.eps}}
\caption{The elliptic flow predicted by viscous hydrodynamics, for different 
values of the shear viscosity per unit entropy $\eta/s$ and for Glauber (top)
and color-glass condensate (bottom) initial conditions, compared to 
the measurements by the STAR collaboration at RHIC~\cite{Luzum:2008cw}.}
\label{fig:luzum}
\end{figure}

Bounds on the shear viscosity can be extracted phenomenologically by
comparing experimental elliptic-flow data to the dependence of $v_2$
on this parameter~\cite{Romatschke:2007mq,Dusling:2007gi,Song:2007fn},
as illustrated in \fig(\ref{fig:luzum}).  Viscous hydrodynamic
calculations are technically rather involved, in particular because a
naive implementation of the leading dissipative terms leads to
unstable solutions. However, several groups have come up with
numerical schemes (involving second-order terms) that cure this
problem, and the results are in good agreement (see the
review~\cite{Teaney:2009qa}.) Since hydrodynamics is a good
description at best for a part of the time evolution of the fireball,
there are many sources of systematic uncertainty in constraining the
values of $\eta/s$ that are consistent with the data.  The largest
seems to be the sensitivity to the initial conditions.  Nonetheless,
the confidence in the hydrodynamic description can be increased by
confronting the predictions for many correlation
observables~\cite{Teaney:2010vd} to the experimental data.

A conservative bound (see e.g.~\cite{Teaney:2009qa})
\be
\left(\frac{\eta}{s}\right)_{\rm pheno} \lesssim 0.40 
\la{eq:eta.s.bound}
\ee 
has been inferred from the RHIC data as characterizing the produced
medium.  This number can be compared to the 
perturbative prediction, which in leading-logarithmic approximation reads
\be
\left(\frac{\eta}{s}\right)_{\rm leading~log} = \frac{c}{g^4 \log(1/g)}\,,
\la{eq:etasPT}
\ee
$c$ being a constant (see~\cite{Arnold:2000dr} and Refs.~therein)
and $g$ the strong coupling constant. Numerically, $\frac{\eta}{s}\approx 2.0$ 
for $\alpha_s=0.15$ at full leading order~\cite{Arnold:2003zc}.
This value of the coupling is relevant to temperatures well above those reached in 
heavy-ion collisions, and it is uncertain how the numerical value extrapolates 
to lower temperatures.
Since the phenomenological bound is small, it has become customary 
to also compare it to the value of $\eta/s$ in the ${\cal N}=4$ super-Yang-Mills
(SYM) theory in the limit of infinite gauge coupling~\cite{Policastro:2001yc},
\be
\la{eq:etasSYM}
\left(\frac{\eta}{s}\right)_{{\cal N}=4,\lambda=\infty}
 = \frac{1}{4\pi}\,.
\ee
Due to the super-conformality of the ${\cal N}=4$ SYM theory, the
coupling constant does not evolve with the energy scale, instead it is
a truly constant parameter of the theory, unlike in QCD.  In fact the
result (\ref{eq:etasSYM}) is much more general: it holds for a large
class of field theories admitting a dual gravity
description~\cite{Kovtun:2004de}.

There are several other phenomenological indications that the medium
created at RHIC is strongly interacting, see for
instance~\cite{Muller:2008zzm}.  The elliptic flow of charm quarks
through the plasma is estimated experimentally via leptonic decays of
$D$ mesons~\cite{Abelev:2006db,Adare:2006nq}.
The charm quarks exhibit a significant amount of elliptic flow, from
which one infers that they kinetically thermalize on a time-scale
which is small compared to the naively expected one,
$M/T^2$~\cite{Moore:2004tg}. This aspect of relaxation in the
quark-gluon plasma is reviewed in section (\ref{sec:boltz}).

Another indication comes from the study of jets, i.e.~the passage of a
fast parton through the medium which then fragments successively into
the hadrons that are detected \cite{Adcox:2001jp,Adler:2002xw}.  It is
found that when triggering on a jet, the total transverse momentum is
balanced not by a second back-to-back jet, but is balanced instead by
a large number of particles of much lower momentum. This observation
leads to the statement that the system is very `opaque' to colored
probes. Quantitatively it is expressed by saying that the medium
admits a large jet quenching parameter $\hat q$.  While the latter
parameter is also a transport coefficient of the system, no method is
known at present to connect it to Euclidean correlation functions, and
we will therefore not discuss it further. We refer the reader
to~\cite{Wiedemann:2009sh} for an introduction to the subject.

% Motivation in cosmology.

Before we discuss the calculation of equilibrium and transport
properties of the quark-gluon plasma from first principles, we briefly
remark on the relevance of transport properties of the quark-gluon
plasma to early-universe cosmology. We recall that the precise
understanding of primordial nucleosynthesis has repeatedly allowed
cosmologists to put severe constraints on many new physics
scenarios~\cite{Iocco:2008va}.  Controlling quantitatively the
quark-gluon plasma era of the universe could potentially have
analogous benefits.

The QCD phase transition, being a crossover at zero net baryon
density~\cite{Aoki:2006we},
is not expected to lead to an observable signal in the gravitational
wave background. Only an effective large baryon chemical potential
could potentially turn it into a first order transition, where the
inhomogeneous nucleation processes would make an imprint on the
gravitational field (see e.g.~\cite{Durrer:2010xc} and Refs. therein).
The probability per unit volume per unit time to nucleate a
low-temperature phase bubble out of the high-temperature phase is
proportional to a linear combination of the shear and bulk viscosities
in the high-temperature phase~\cite{Csernai:1992tj}. In this way the
transport properties would have an impact on cosmological observables.

\begin{figure}
\vspace{0.6cm}
\centerline{\includegraphics[width=0.45\textwidth]{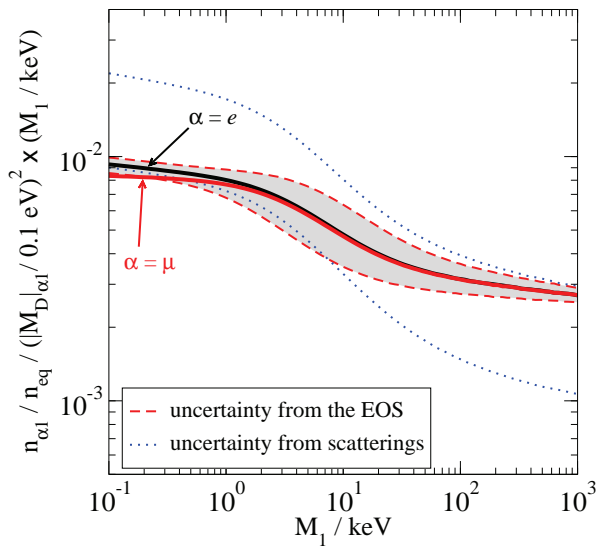}}
\vspace{-0.6cm}
\caption{Concentration of sterile neutrinos produced 
in the early universe~\cite{Asaka:2006nq}. 
The dominant uncertainty
comes from the limited knowledge of the spectral function.}
\label{fig:WMD}
\end{figure}

A sterile, right-handed neutrino of mass $m\approx 10$keV has been
proposed~\cite{Asaka:2005an} as a candidate for warm dark matter in
the context of the $\nu$MSM. In a given scenario the production
of these particles happens to peak around temperatures of
0.1GeV..1GeV.  Their distribution is then determined by the spectral
functions of active neutrinos in a thermal QCD bath. The latter in
turn can be expressed in terms of the thermal vector and axial-vector
spectral functions of QCD~\cite{Asaka:2006rw}.  Given the range of
temperatures in which these spectral functions need to be calculated,
non-perturbative methods are required for a precise and reliable
result, see \fig(\ref{fig:WMD}) reproduced
from~\cite{Asaka:2006nq}. The example is specific, but it is likely
that other particle physics models with cosmological implications can
be constrained if the transport properties of QCD are determined
reliably.

%% Brief summary of QGP static properties

\begin{figure}
\centerline{\includegraphics[width=0.5\textwidth]{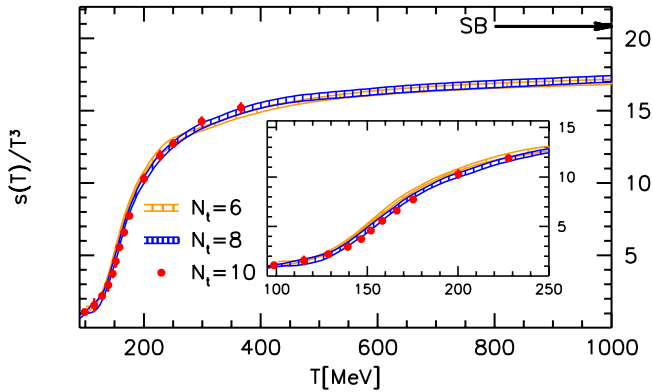}}
\vspace{-3.0cm}
\caption{The 2+1 flavor QCD entropy density in units of $T^3$ as a function 
of temperature~\cite{Borsanyi:2010cj}.}
\label{fig:eos_s}
\end{figure}

In order to understand the transport properties of the quark-gluon
plasma, some prior familiarity with its equilibrium properties is required. We now
summarize some of the essentials.  The most recent published lattice
QCD result for the entropy density as a function of temperature is
depicted in \fig(\ref{fig:eos_s})~(\cite{Borsanyi:2010cj}; see also \cite{Bazavov:2009zn}).  
The rapid rise between 150MeV and 250MeV in the number of degrees of freedom
exerting pressure is clearly visible. The ratio $s/T^3$ then saturates
and only very slowly tends towards the asymptotic `Stefan-Boltzmann'
value of
\be
\left(\frac{s}{T^3}\right)_{\rm SB} = {\txts\frac{2\pi^2}{45}} 
\left[ 2(N_c^2-1) + {\txts\frac{7}{8}}\cdot 4N_c N_{\rm f}\right].
\ee
One of the defining properties of a plasma is the existence of a
finite correlation length over which (chromo-)electric fields are
screened. Its inverse defines the Debye screening mass.  In a weakly
coupled plasma of quarks and gluons, the corresponding mass is
calculable in perturbation theory,
\be
m_{D}^2 = g^2 % \left[
\big({\txts\frac{N_c}{3}+\frac{N_{\rm f}}{6}}\big) T^2 .
%                        + {\txts\frac{1}{2\pi^2}\sum_f} \mu_f^2  \right].
\ee
More generally, a widely accepted definition of $m_D$
is that it is the smallest mass associated with a screening state
odd under Euclidean time reversal~\cite{Arnold:1995bh}, which
corresponds to the product of time reversal and charge conjugation in
Minkowski space. A unique feature of non-Abelian plasmas is that
magnetic fields are also screened (unlike in the Sun), and at high
temperature the corresponding lightest screening masses are
parametrically of order $g^2T$. The smallest screening masses of states with a
non-trivial flavor quantum number are close to $2\pi T$, which can be
understood in terms of the minimal frequency $\pi T$ associated 
with a fermion field in the imaginary-time formalism.  Finally,
as \eq(\ref{eq:etasPT}) illustrates, the mean path of a quark or gluon
quasiparticle before it changes its direction of flight by an O($1$)
angle is $(g^4T)^{-1}$~\cite{Arnold:1997gh}.  These length scales
associated with different physical phenomena are parametrically
different at sufficiently high temperatures, but there are indications
that at temperatures that can be reached in heavy-ion collisions,
these scales are overlapping and even inverted in some cases, see the
discussions in~\cite{Laine:2009dh,Cheng:2010fe}.

Without the RHIC and the LHC heavy-ion experiments, it would currently
be unclear which of the strong or weak coupling paradigms is more
appropriate to describe the quark-gluon plasma at temperatures of a
few $T_c$.  The answer must be expected to depend to some extent on
the physical quantity.  The convergence of perturbative calculations
at $T\approx 2-3T_c$ (see for instance~\cite{CaronHuot:2007gq}) is in
many cases poor, and these calculations are in tension with
phenomenological bounds such as \eq(\ref{eq:eta.s.bound}).  It is
therefore desirable to develop non-perturbative methods to compute
transport properties of the quark-gluon plasma starting from the
microscopic theory, QCD. This constitutes one of the central themes of
this review. For illustration, we describe the recipe to calculate the
shear viscosity in the next paragraph. The origin of this recipe will
be described in detail in section (\ref{sec:kubo}).

The shear viscosity $\eta$ parametrizes how efficiently the momentum
of a layer of fluid (assume the momentum to be in the plane defined by
that layer) diffuses in the direction orthogonal to the momentum.  The
diffusion constant is
\be
D_{\rm sh} = \frac{\eta}{e+p},
\ee
where $e+p$ is the enthalpy density. The time-evolution of transverse
momentum is related to the transport of shear stress, $T_{xy}$,
through the momentum conservation equation, $\partial_0
T^{0y}+\partial_x T^{xy}=0$.  A second key idea is that the 
response of the fluid to externally applied shear stress
is encoded in the equilibrium ensemble, 
namely in the way the thermal fluctuations of shear-stress
dissipate in real time.  This is the outcome of the linear response formalism,
reviewed in section (\ref{sec:kubo}). The time-correlation is encoded
in the commutator-defined correlator $\<[T_{xy}(t,\bx),T_{xy}(0)]\>$. Its
Fourier transform is given by the  spectral function
$\rho(\omega,\bk)$. One can then derive a `Kubo formula', which
gives the shear viscosity in terms 
of the low-frequency part of the spectral function,
\be
\eta = \pi \lim_{\omega\to0} \lim_{\bk\to0} \frac{\rho(\omega,\bk)}{\omega}.
\ee
Furthermore, the spectral function $\rho$ is formally 
related to the Euclidean correlator by analytic continuation, 
\be
\pi\rho(\omega) = \im \Big(\tilde G_E\big(\omega_n \to -i[\omega+i\epsilon]\big)  \Big),
\ee
where $\omega_n=2\pi Tn$ is one of the Matsubara frequencies.
Alternatively, in the mixed time+spatial momentum representation, 
the connection is given by an integral transform,
\be
G_E(t,\bk) = \int_0^\infty \ud \omega\, \rho(\omega,\bk) \,
\frac{\cosh\omega(\frac{\beta}{2}-t)}{\sinh\beta\omega/2}\,,
\la{eq:GErho}
\ee
where the kernel is the thermal propagator of a free scalar field.  The
computational recipe is thus formally straightforward: calculate the
Euclidean correlator $G_E$, determine the spectral function by
analytic continuation, and finally read off the shear viscosity from
the slope of $\rho$ at the origin.  In practice however there are very
significant difficulties in carrying out this program. The Euclidean
correlator is normally only obtained in some approximation, either
because it is expanded in powers of the coupling constant, or because it is
obtained from Monte-Carlo simulations.  In either case, an analytic
continuation is not possible without further information or
assumptions. In particular, while the Euclidean correlator can be
expanded in powers of the strong coupling, the parametric dependence
of the viscosity (\ref{eq:etasPT}) makes it clear that some form of
resummation of the perturbative series is required. In general it is
not known how to implement the resummation in this way, and an
effective kinetic theory treatment has been employed instead
in~\cite{Arnold:2000dr}.  In the representation (\ref{eq:GErho}),
which is the form most commonly employed when the Euclidean correlator is
determined by Monte-Carlo simulations, it is clear that the smoothness
of the kernel implies that the fine features of $\rho(\omega)$
(particularly on scales $\Delta\omega\lesssim T$) are encoded in the
correlator in a way which is numerically very suppressed.

\begin{figure}
\centerline{\includegraphics[angle=0,width=0.45\textwidth]{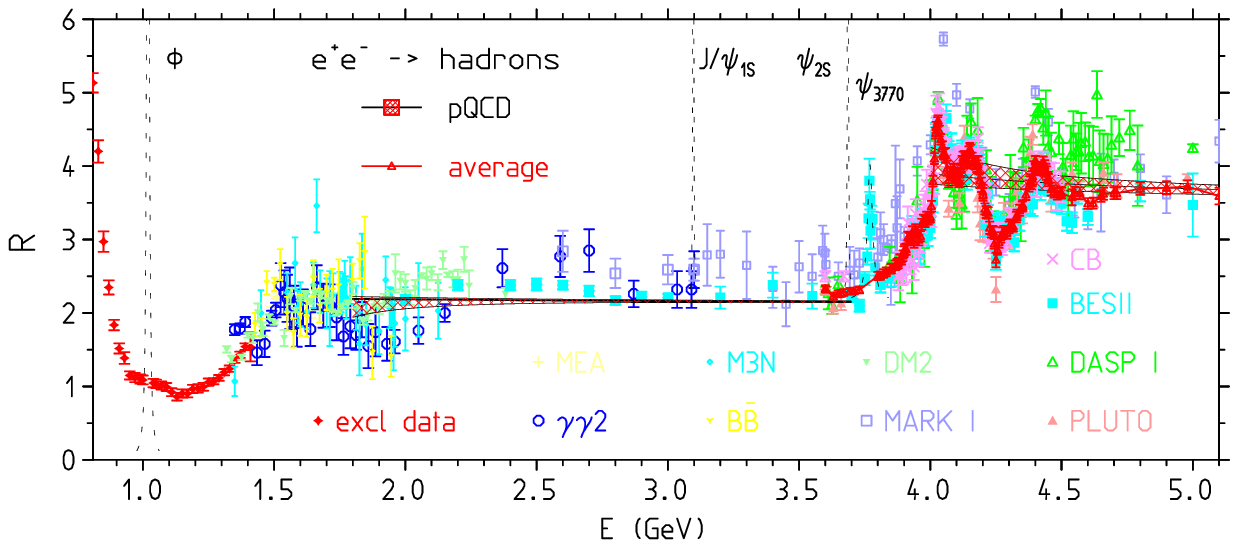}}
% \centerline{\includegraphics[angle=0,width=0.49\textwidth]{FIGS/Pi-bernecker.ps}}
\centerline{\includegraphics[angle=0,width=0.41\textwidth]{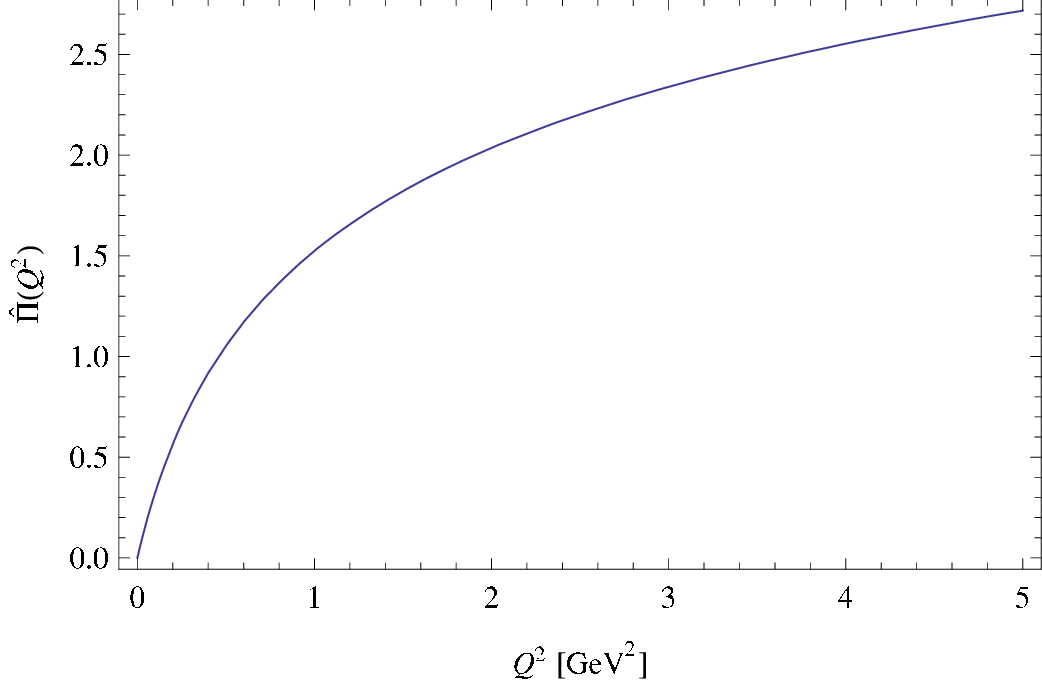}~~~~}
\caption{The $R$ ratio, \eq(\ref{eq:R}), determined experimentally 
(Fig. from~\cite{Jegerlehner:2009ry}), and
the subtracted vacuum polarization $\widehat\Pi(Q^2)\equiv \Pi(0)-\Pi(Q^2)$ 
in the space-like domain calculated using 
\eq(\ref{eq:Pidisprel},\ref{eq:R})~\cite{Bernecker}.}
\label{fig:vacpol}
\end{figure}

We wish to illustrate the latter point somewhat more concretely.
In the limit of zero-temperature, the kernel in (\ref{eq:GErho}) simply becomes $e^{-\omega t}$,
so that the Euclidean correlator is simply the Laplace transform of the 
spectral function. Since Lorentz symmetry is restored in the limit $T\to0$, 
the problem is simplified by going to four-momentum space.
We will consider the case of the electromagnetic current correlator, 
$j_\mu(x) = \frac{2}{3}\bar u\gamma_\mu u -\frac{1}{3} \bar
d \gamma_\mu d -\frac{1}{3} \bar s\gamma_\mu s+\dots$
Current conservation and Lorentz invariance implies that a single 
function of four-momentum squared determines all correlation functions 
of the electromagnetic current,
\be  
{\int} \ud^4x \; \<j_\mu(x)j_\nu(0)\>\;  e^{iq\cdot x} = 
% \Pi_{\mu\nu}(q) = 
(q_\mu q_\nu- q^2 g_{\mu\nu})\,\Pi(q^2).
\ee
In Euclidean four-momentum space, the spectral representation of the
vacuum polarisation $\Pi(Q^2)$ reads
\be
\Pi(0)-\Pi(q^2) = {q^2}\int_0^\infty \ud s \frac{ \rho(s)}{s(s+q^2)}.
\la{eq:Pidisprel}
\ee
Via the Optical Theorem, the spectral density is accessible to experiments
\be
\pi\rho(s) = \frac{s}{4\pi\alpha(s)} \sigma_{\rm tot}(e^+ e^-\!\!\to{\rm hadrons})
\equiv \frac{\alpha(s)}{3\pi} R_{\rm had}(s),~~~~\;
\la{eq:R}
\ee
where $\alpha$ is the QED coupling and 
the first equality holds up to QED corrections\footnote{
Neglecting the muon mass and up to QED radiative corrections,
the quantity $R_{\rm had}$ can be expressed as
$ R_{\rm had}(s)\approx \frac{\sigma(e^+e^-\to {\rm hadrons})}
                       {\sigma(e^+e^-\to \mu^+\mu^-)} $, 
which is convenient from an experimental point of view~\cite{Jegerlehner:2009ry}.}.
To illustrate the different character of a retarded correlator in the
time-like and the space-like region, we reproduce
in \fig\ref{fig:vacpol} the quantity $R_{\rm had}$, as determined by experiments
spanning several decades, as well as the corresponding vacuum polarization
$\Pi(Q^2)$ in the space-like region.
While many vector resonances show up in $R_{\rm had}$ (and therefore
the spectral function), the vacuum polarisation in the space-like
domain is a very smooth function. In the absence of auxiliary information,
extracting the existence, position and width 
of each resonance from $\Pi(Q^2)$ in the space-like region 
is therefore a numerically ill-posed problem.

Returning to the physics of the quark-gluon plasma, it is not known
precisely how far up in temperatures the sharp features of the vacuum
spectral function depicted in \fig(\ref{fig:vacpol}) survive.
However, in the case of conserved currents, the functional form of the
spectral function in the region $\omega\ll \tau_{\rm max}^{-1}$, where
$\tau_{\rm max}$ is the longest relaxation time in the system, is
completely determined by the continuous global symmetries of the
theory\footnote{At a phase transition, additional slow modes
associated with an order parameter (possibly of a discrete symmetry)
may play an important role.}. The effective theory that describes this
kinematical regime is hydrodynamics. This aspect is reviewed in
section (\ref{sec:kubo}).  The universality of these features means
that they can be used as an ansatz for the spectral function in the
small-$\omega$ region.

Beyond these universal features, the stronger assumption that the
conserved charges are carried by quasiparticles leads to additional
distinctive features of the spectral function around $\omega=0$. The
latter can be predicted in detail by using kinetic theory. However,
thanks to the AdS/CFT correspondence, we now know of at least one
theory where these features are absent (the strongly coupled ${\cal
N}=4$ SYM theory).  We invite the reader to compare figures
(\ref{fig:rho1212-kin}, black dashed curve) and
(\ref{fig:rho1212-teaney}, red curve), which illustrate the difference
between the functional form of the shear stress spectral function in a
kinetic description (Boltzmann equation) and in a theory which
obviously does not admit any quasiparticles.  The most important
qualitative question about the quark-gluon plasma is whether, at the
temperatures explored in heavy-ion collisions, it contains
quasiparticles as assumed in kinetic theory. The alternative
possibility is that there are instead no poles near the real axis in
the retarded Green's function, other than those dictated by the
symmetries of the theory; this possibility is realized in the strongly
coupled SYM theory.  To answer the question from first principles, we
need to study the spectral functions of conserved currents in the
small-frequency regime, in other words we need to study the transport
coefficients.

This review is organized as follows.  We start in section
(Sec.~\ref{sec:analy}) with a review of the analytic properties of thermal
correlators, detailing in particular the relation between Euclidean
and real-time correlators and giving the general tensor structure of
correlators in the channels of interest (Sec.~\ref{sec:tensor}). We
also describe the derivation of Kubo formulae and of the hydrodynamic
predictions for spectral functions (Sec.~\ref{sec:kubo}), as well as
the perturbative and kinetic theory treatments (Sec.~\ref{sec:boltz}).
These results are contrasted with those obtained in strongly coupled
theories via the AdS/CFT correspondence (Sec.~\ref{sec:adscft}).
An overview of the state-of-the-art in lattice calculations of 
Euclidean correlators is given in Sec.~(\ref{sec:latt}), as well as of 
the scaling of their computational cost;
we discuss the electric conductivity, the heavy-quark diffusion
and shear \& bulk viscosities.
The analytic continuation problem is discussed from a numerical perspective
in Sec.~(\ref{sec:EC}), where we show the advantages and disadvantages
of linear and non-linear methods.
We finish with a summary in Sec.~(\ref{sec:concl}).

There are a number of subjects closely related to those treated in
this review, which are only briefly mentioned here.  The extraction of
a shear viscosity estimate from heavy-ion data is reviewed
in~\cite{Teaney:2009qa}.  The thermodynamic properties of the
quark-gluon plasma~\cite{Aoki:2006br,Aoki:2009sc,Borsanyi:2010bp,Bazavov:2009zn,Cheng:2009zi}
have been reviewed recently~\cite{DeTar:2009ef}.
The behavior of spatial correlation lengths is discussed
in~\cite{Laermann:2003cv,DeTar:2009ef}. For recent developments in
hydrodynamics we refer to~\cite{Romatschke:2009im}.  For introductions
to relativistic hydrodynamics and the gauge/gravity duality,
see~\cite{Son:2008zz,Gubser:2009md,CasalderreySolana:2011us}. Finally,
the physics of shear viscosity and its perturbative calculation are
succinctly reviewed in~\cite{Moore:2004kp}.

% \newpage 

\section{Definition and Properties of Thermal Correlators\label{sec:analy}}
% and \cite{RefJ}
% {\sc Definitions. Completion end of November}

% \section{An Overview of Lattice Calculations\la{sec:latt}}

In this section we review the different correlators relevant to
calculations of transport properties. We introduce the Minkowski-space
correlators and the Euclidean-space correlators, and describe how they
are interrelated. Their spectral representation is given, from which
the dispersion relation expressing the retarded correlator in terms of
the spectral function is given. Finally, we summarize the tensor
structure of the vector current and energy-momentum tensor
correlators.  In order to facilitate the initiation of the reader not
accustomed to finite-temperature field theory, we begin with a
reminder on vacuum correlators and then generalize the definitions for
a thermal ensemble.

\subsection{Vacuum correlation functions\la{sec:corfunc}}

We start with vacuum correlation functions, working in the metric
({\small$-+++$}).  The time-evolution operator 
\be U(t) \equiv e^{-iH  t} 
\ee 
is used to define the operators in the Heisenberg picture,
\be 
A(t) \equiv (U(t))^{-1} A(0) U(t). 
\ee 
We recall the time-ordering operation for operators, 
\be 
T\big(A(t_1) B(t_2)\big)
\equiv A(t_1)B(t_2)\,\theta(t_1-t_2) \pm
B(t_2)A(t_1)\,\theta(t_2-t_1), 
\ee 
where the $+$ ($-$) sign applies to
bosonic (fermionic) operators.  Time-ordered correlation functions,
\ba
% G_T^{AB}(t)&\equiv& \<0 | T\big(A(t) B(0)\big)|0\>,
% \\
G_T^{AB}(\omega)&\equiv& \int_{-\infty}^\infty {\ud t} \,
e^{i\omega t} \,
\<0 | T\big(A(t) B(0)\big)|0\>,
% G_T^{AB}(t),
\la{eq:G_T}
\ea
have simple Lorentz invariance properties.
For instance, the `Feynman' propagator of a free scalar field reads
\be
 \<0 | T\big(\hat\phi(t,\bx) \hat\phi(0)\big)|0\>
= -i{\int} \frac{\ud^{D}p}{(2\pi)^{D}} 
  \frac{e^{ipx}}{p^2+m^2-i\epsilon}.
\ee
Up to the factor $i$, the Euclidean space propagator has the same
expression, but the scalar products are to be interpreted with the
({\small $++++$}) metric, and the $i\epsilon$ prescription is
unnecessary, since the integration region is well separated from the
poles $p_0^2+\bp^2=-m^2$.

The commutator $ [A(t),B(0)] $ 
vanishes outside the light-cone in a relativistic causal theory.
In the free scalar field theory,
\be
i\theta(x_0) \<0 |[\hat\phi(x), \hat\phi(0)] |0\>
= {\int} \frac{\ud^{D}p}{(2\pi)^{D}} 
  \frac{e^{ipx}}{(p_0+i\epsilon)^2-\bp^2-m^2}.
\ee
The correlator
\be
G_R^{AB}(\omega) = i\int_0^\infty \ud t\, e^{i\omega t} 
\<0| [A(t),B(0)] |0\>,
\ee
will play an important role in the following.  % finite temperature.
In the vacuum % (and in the case $B=A^\dagger$) 
it is related to the time-ordered correlator (\ref{eq:G_T}), which
is familiar from time-dependent perturbation theory, through
\be
G_R^{AB}(\omega) = \left\{ 
\begin{array}{l@{\quad}l}
i G_T^{AB}(\omega)                               & \re \omega>0 \\ 
\big(i G_T^{B^\dagger A^\dagger}(\omega)\big)^*  &  \re \omega<0.
\end{array}\right.
\ee

Time-ordered products of fields appear when expressing the
time-evolution operator, and in particular the $S$-matrix, in terms of
the interaction-picture Hamiltonian. It is then natural to work with
time-ordered propagators.  Retarded correlation functions occur
typically when coupling a field to an external source, which is
switched on only for a finite time interval. A classic example is the
calculation of the bremsstrahlung emitted by an electron passing near
a nucleus. In the following we will deal mainly with this second type
of situations, where the response comes however from a thermal medium
rather than a single particle.

\subsection{Finite-temperature correlation functions}

In order to describe quantum statistical properties, the vacuum
expectation values of operators are replaced by averages over an
ensemble of states described by a density matrix $\hat\rho$.
The finite-temperature, equilibrium density matrix is 
\be
\la{eq:densmat}
\hat\rho \equiv \frac{1}{Z} e^{-\beta H}. %% cite Parisi book
\ee
with $Z$ such that $\tr\{\hat\rho\}=1$.
At finite temperature, the Wightman correlation functions are defined as
\ba
G_>^{AB}(t)&\equiv& \tr\{ \hat\rho A(t) B(0) \}\,,
\la{eq:G>}
\\
G_<^{AB}(t)&\equiv& \tr\{ \hat\rho B(0) A(t) \}\,.
\ea
Obviously, due to time-translation invariance of the equilibrium density matrix,
\be
% G_>^{BA}(t) = G_<^{AB}(-t).
G_<^{AB}(t) = G_>^{BA}(-t) .
\la{eq:swap}
\ee
The reality property
\be
% G_>^{AA^\dagger}(t) = G_>^{AA^\dagger}(-t)^*
G_>^{A^\dagger B^\dagger}(t) = G_>^{BA}(-t^*)^*
\la{eq:reality}
\ee
is also easily proven.
Last but not least, the definition (\ref{eq:G>})
implies the Kubo-Martin-Schwinger (KMS) relation,
\be
G_>^{AB}(t) =  G_>^{BA}(-t-i\beta).
\la{eq:KMS}
\ee
In fact, it can be shown that this property uniquely characterizes the 
equilibrium density matrix. %  $\rho \equiv \frac{1}{Z} e^{-\beta H}$. %% cite Parisi book
An obvious consequence of \eq(\ref{eq:KMS}) is that the thermal correlation 
function of an anticommutator is periodic in the imaginary time direction.

The expectation value of a commutator,
\be
G^{AB}(t) = i \tr\{\hat\rho [A(t),B(0)]\} = i\left(G_>^{AB}(t) - G_<^{AB}(t)\right),
\la{eq:G}
% = \int_{-\infty}^{\infty} \ud\omega\; e^{-i\omega t} \rho_{AB}(\omega).
\ee
is physically important. The commutator vanishes outside the light-cone, 
a reflection of the causality of the theory.
Using \eq(\ref{eq:swap}) and (\ref{eq:reality}), one obtains the properties
\ba
G^{AB}(-t) &=& -G^{BA}(t), % ,\qquad \forall A,B;
\la{eq:Gisodd}
\\
G^{A^\dagger B^\dagger}(t) &=& G^{AB}(t^*)^*.   % , \qquad A=A^\dagger.
\la{eq:Gisreal}
\ea

The Euclidean correlator is defined as
\be
G_E^{AB}(t) = G_{>}^{AB}(-it).
% \< A(-it)\, B(0)\>.
\la{eq:GE}
\ee
As a special case of (\ref{eq:KMS}), it obeys the relation
\be
G_E^{BA}(\beta-t) = G_E^{AB}(t)\,.
\ee

The Fourier transform of $G^{AB}(t)$ defines the so-called 
spectral function~\footnote{We use a small `hat' on the density matrix
to distinguish it from the spectral function.}
\be
\rho^{AB}(\omega)= \frac{1}{2\pi i}
\int_{-\infty}^{+\infty} \ud t\; e^{i\omega t}\; G^{AB}(t).
\la{eq:rhodef}
\ee
It enjoys the symmetry properties
\ba
\rho^{AB}(-\omega) &=& -\rho^{BA}(\omega),
\\
\rho^{A^\dagger B^\dagger}(\omega) &=& \rho^{BA}(\omega)^*.
\ea
The integral tranform over the positive half-axis
\be
G_R^{AB}(\omega) = \int_0^\infty \ud t\, e^{i\omega t} G^{AB}(t)
\ee
is analytic in the half complex plane $\im(\omega) > 0$.
It is called the retarded correlator and plays a central role in linear response theory, 
as we shall see shortly.
The Euclidean correlator can be expressed as a Fourier series
on the interval $0\leq t<\beta$,
\ba
G_E(t) &=& T\sum_{\ell\in{\mathbb{ Z}}} G_E^{(\ell)}\, e^{-i\omega_\ell t}\,,
\\
G_E^{(\ell)} &=&  \int_0^\beta  dt e^{i\omega_\ell t} G_E(t)\,,
\la{eq:GEell}
\ea
where $\omega_\ell= \omega_M\cdot \ell = 2\pi T\,\ell$ and we have dropped 
the label specifying the operators $A, B$.

For real $\omega$ and in finite volume, 
$\rho$ is a distribution, namely a discrete sum of delta functions.
However, in the infinite-volume limit of an interacting theory, 
it is typically a smooth function, except possibly at a finite
number of points. The most useful relations among the correlators 
are those that survive the infinite-volume limit.

Using the KMS relation (\ref{eq:KMS}), one easily shows 
\ba
{\int_{-\infty}^\infty} \frac{\ud t}{2\pi} \, e^{i\omega t} \<\{A(t),B(0)\}_\pm\>
 &=& (1\pm e^{-\beta\omega}) \cdot
\\
&& \cdot{\int_{-\infty}^\infty} \frac{\ud t}{2\pi} \, e^{i\omega t} G_>^{AB}(t),
\nonumber
\ea
whence it follows that the spectral function is related to the 
anticommutator-defined correlator via
\be
\rho^{AB}(\omega) = \tanh\big({\txts\frac{\beta\omega}{2}}\big) \, 
\int_{-\infty}^\infty \frac{\ud t}{2\pi} \, e^{i\omega t} \<\{A(t),B(0)\}_+\>.
\la{eq:fluc-dissip}
\ee
This relation is known as the fluctuation-dissipation theorem.
Since the retarded propagator $G_R^{AB}$ can be expressed in terms 
of $\rho^{AB}$ via a dispersion relation, it so follows that 
the knowledge of the anticommutator-defined correlator also 
determines $G_R^{AB}$.
% \subsection{Relations among correlators}

\subsection{Spectral representation and relations between correlators}

Using (\ref{eq:Gisodd}) and (\ref{eq:Gisreal}), one shows that 
\be
\rho^{AB}(\omega)=\frac{1}{2\pi i} 
\left(G_R^{AB}(\omega)- G_R^{B^\dagger A^\dagger}(\omega)^*\right).
\ee
In particular, for $B=A^\dagger$, 
the spectral function is identically related 
to the imaginary part of the retarded correlator,
\be
\rho^{AA^\dagger}(\omega) = \frac{1}{\pi} \im G^{AA^\dagger}_R(\omega)\in \mathbb{R}\,.
% \qquad A=A^\dagger.
\la{eq:rho-ImGR}
\ee
% For the rest of this subsection, we assume $B=A^\dagger$.

The spectral representation of the correlators is useful 
to prove further relations between the Euclidean and 
real-time correlators. 
Inserting two complete sets of energy eigenstates
in the definitions of $G(t)$ and $G_E(t)$, one obtains
\ba
G^{AB}(t) &=& \frac{2i}{Z}
\sum_{m,n} A_{mn}B_{nm} e^{-\beta (E_n+E_m)/2} 
\\
&& \qquad\quad\cdot \sinh({\txts\frac{\beta E_{nm}}{2}}) e^{-iE_{nm}t},
\nn
G_E^{AB}(t) &=& \frac{1}{Z} \sum_{n,m} A_{mn} B_{nm}
        e^{-\beta E_m} e^{-E_{nm} t} \,,
\la{eq:G_E1}
\ea
where we employ the notation
\be
E_{nm}=E_n-E_m\,,\qquad A_{nm} = \< n | A(t=0) | m\>\,.
\ee
The expression for the retarded correlator reads
\ba\la{eq:G_R2}
G^{AB}_R(\omega) 
&=& \frac{2}{Z}\sum_{n,m} \frac{-A_{mn}B_{nm}}{\omega-E_{nm}} 
\\
&& \qquad~~ \cdot\; e^{-\beta(E_n+E_m)/2} \sinh({\txts\frac{\beta E_{nm}}{2}}) \,,
\nonumber
\ea
and the spectral function for $\omega\in\mathbb{R}$ takes the form
\ba\la{eq:rho}
\frac{\rho^{AB}(\omega)}{2 \sinh\frac{\beta\omega}{2}}
&=& \frac{1}{Z}\, \sum_{m,n}  A_{mn} B_{nm} 
\\
&& \qquad~~\cdot\; e^{-\beta (E_n+E_m)/2} \delta(\omega-E_{nm})\,.
\nonumber
\ea
For $B=A^\dagger$, this formula can be obtained by taking the imaginary part of 
$G_R(\omega+i\epsilon)$ and using  the standard representation
of the delta function,
$\delta(\omega)=\frac{1}{\pi} \frac{\epsilon}{\omega^2+\epsilon^2}$.
After some algebraic manipulations,
one finds for the Fourier coefficients (\ref{eq:GEell})
% (assuming again $A=\pm A^\dagger$)
\be
G_E^{(\ell)} =  \frac{2}{Z}\sum_{m,n} 
e^{-\beta(E_n+E_m)/2}\sinh(\beta E_{nm}/2)
\frac{ A_{mn}B_{nm}}{-i\omega_\ell+E_{nm}}.
\la{eq:G_El}
\ee

\begin{figure}
\centerline{\includegraphics[width=0.37\textwidth]{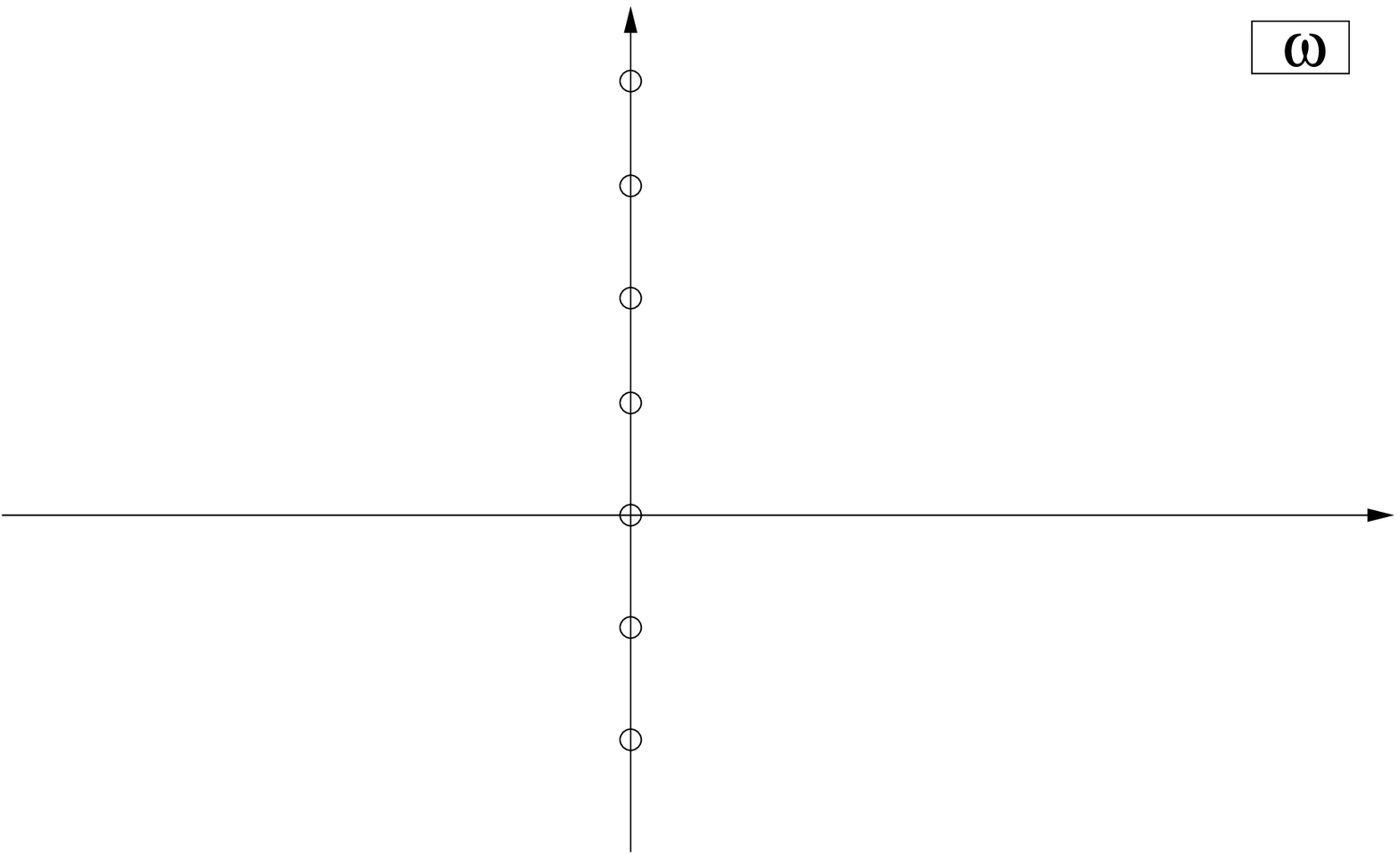}}
\centerline{\includegraphics[width=0.45\textwidth]{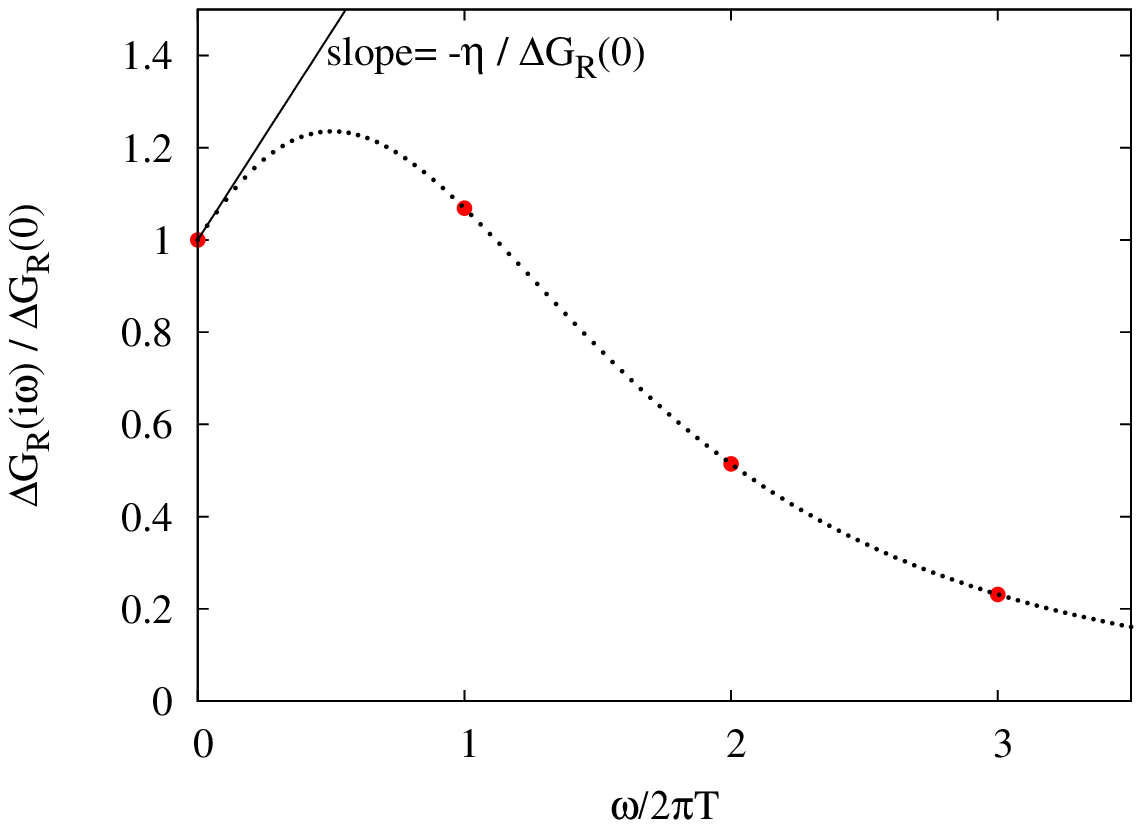}~~~}
\caption{\underline{Top}: Analytic structure of the retarded correlator $G_R(\omega)$
in the complex $\omega$ plane. At the circled values of $\omega$ on the 
positive imaginary axis, it coincides with the Euclidean correlator (\eq\ref{eq:l.neq.0}).
\underline{Bottom}: 
At the origin, the odd derivatives of $G_R$ along the imaginary axis coincide
with the corresponding derivatives of $\im G_R$ along the real axis. 
These odd derivatives are given by transport coefficients (section \ref{sec:kubo}).
The figure illustrates
a case where one subtraction has been made to make the dispersion integral converge.
}
\label{fig:GR_complex}       
\end{figure}

Formulae (\ref{eq:G_R2}) and (\ref{eq:G_El}) show in particular that 
\be
G_R(i\omega_\ell) = G_E^{(\ell)} \,, \quad
\ell\neq 0\,.
\la{eq:l.neq.0}
\ee
Thus the frequency-space Euclidean correlator is the analytic continuation 
of the retarded correlator; the situation in the complex $\omega$ plane
is illustrated in \fig(\ref{fig:GR_complex}).
For $\ell=0$, one has to remember that the retarded correlator
was initially defined for $\im(\omega)>0$. Therefore its value 
at the origin can only be defined as 
$\lim_{\epsilon\to0_+}G_R(i\epsilon)$.
This value differs from $G^{(0)}_E$ if
the spectral function is singular at the origin.
% the correlator $G(t)$ does not vanish  sufficiently fast when $t\to+\infty$. 
Specifically, 
\be
\frac{\rho(\omega)}{\omega}=A\delta(\omega)+{\rm finite} 
~~~\Rightarrow~~~ 
G_E^{(0)} - \lim_{\epsilon\to0_+} G_R(i\epsilon) = A.
\la{eq:l.eq.0}
\ee 
Finally, it is easy to verify, using \eq(\ref{eq:G_E1}) and (\ref{eq:rho}),
that the configuration-space Euclidean correlator can be obtained 
from $\rho$ via
\ba\la{eq:ClatRho}
&& G^{AB}_E(t) + G_E^{AB}(\beta-t)
= 
\\
&& \qquad \int_{-\infty}^\infty \!\!\!\!\ud\omega \rho^{AB}(\omega)
\frac{\cosh\omega(\frac{\beta}{2}-t)}{\sinh\beta\omega/2},
\nonumber
\\
&& G^{AB}_E(t) - G_E^{AB}(\beta-t)
\\
&& =  \qquad\int_{-\infty}^\infty\!\!\!\! \ud\omega \rho^{AB}(\omega)
\frac{\sinh\omega(\frac{\beta}{2}-t)}{\sinh\beta\omega/2}.
\nonumber
\ea
These formulae show that the part of the Euclidean correlator
that is even (odd) around $t=\beta/2$ is in correspondence with 
the odd (even) part of the spectral function.
In particular, $B=A$ implies that the Euclidean correlator is symmetric
around $t=\beta/2$ and that the spectral function is an odd function,
while $B=A^\dagger$ implies that both $G_E^{AA^\dagger}(t)$ and 
$\rho^{AA^\dagger}(\omega)$ are real. In most cases of interest, 
$B=A=A^\dagger$, so that both properties are satisfied.

A simple calculation based on the explicit expression
for the spectral density (\ref{eq:rho}), 
or alternatively a contour integration,
then leads to the Kramers-Kronig relation
\be
\int_{-\infty}^{+\infty}d\omega\, \rho(\omega)
\frac{\omega}{\omega^2+\omega_{\rm I}^2}
=   G_R(i\omega_{\rm I})\,,
\qquad\forall \omega_{\rm I}>0.
\la{eq:KK-main}
\ee
Starting from (\ref{eq:KK-main}), it is easy to 
derive by recursion the following equalities,
\ba
(-1)^n\frac{d^{2n}}{d\omega_{\rm I}^{2n}}
  G_R(i\omega_{\rm I}) &\!=\!& 
\int_{-\infty}^{\infty} d\omega \rho^{(2n)}(\omega)\,
\frac{\omega}{\omega^2+\omega_{\rm I}^2}\,,
\nn
(-1)^{n+1}\frac{d^{2n+1}}{d\omega_{\rm I}^{2n+1}}
  G_R(i\omega_{\rm I})
&\!=\!& 
\int_{-\infty}^\infty d\omega \rho^{(2n+1)}(\omega) 
\frac{\omega_{\rm I}}{\omega^2+\omega_{\rm I}^2}
\nn
&\stackrel{\omega_I\to0}{=}& \pi \rho^{(2n+1)}(\omega=0)\,.
\la{eq:dGR}
\ea
Thus at the origin, the even derivatives of the retarded
correlator along the imaginary axis are given by
integrals over the spectral density, while the odd
derivatives are equal to the corresponding 
derivatives of the spectral function, see the bottom panel of 
\fig(\ref{fig:GR_complex}).

\subsection{Properties of the Euclidean correlator}

We have seen that in configuration space, the Euclidean correlator is
defined as the analytic continuation of the Wightman correlator, while
in frequency space it turns out to be the analytic continuation of the
retarded correlator. Here we discuss further relations between Euclidean and 
real-time correlators. 

In our subsequent considerations, it will prove useful to express the
Euclidean correlator in terms of the real-time correlator $G(t)$, in
order to analyze the contribution of hydrodynamic modes.  
The relation between the Euclidean correlator and the real-time correlator 
is 
\be
G_E(t) = T\int_0^\infty \!\!\!\ud u\, G(u)\, \frac{\sinh(2\pi T u)}{\cosh(2\pi Tu)-\cos(2\pi Tt)}.
\la{eq:GEG}
\ee
The moments
\ba
\frac{d^{2n}}{dt^{2n}}G_E(t) &\stackrel{t=\frac{\beta}{2}}{=}&\!\! \int_0^\infty\!\! \ud\omega
\frac{\omega^{2n}\rho(\omega)}{\sinh\frac{\beta\omega}{2}}
\la{eq:moments}\\
&=& (-1)^n \!\int_0^\infty \!\!\!\ud t \, G(t) \,
\frac{d^{2n}}{dt^{2n}}\tanh\left({\pi T t}\right) 
\nonumber
\ea
are thus expressed in terms of integrals of $G(t)$.

The positivity of $\rho(\omega)/\omega$
implies the positivity of certain linear combinations of 
the Euclidean correlator. For instance,
\ba
G_E^{(0)} -G_E^{(\ell)}
&=&\omega_\ell^2
\int_{-\infty}^\infty \frac{\ud\omega}{\omega}
\frac{\rho(\omega)}{\omega^2+\omega_\ell^2}\,,
\la{eq:lincomb1}
\\
\la{eq:lincomb2}
3G_E^{(0)} - 4G_E^{(\ell)} + G_E^{(2\ell)}
&=& 12\omega_\ell^4\int_{-\infty}^\infty \frac{\ud\omega}{\omega}
\\
&& ~ \cdot\;
\frac{\rho(\omega)}{(\omega^2+\omega_\ell^2)(\omega^2+4\omega_\ell^2)}\,.
\nonumber
\ea
We note that relations (\ref{eq:GEG}) and (\ref{eq:lincomb1}--\ref{eq:lincomb2}) 
only hold if the integrals converge in the ultraviolet. In the cases of interest,
at least one subtraction (for example between two different temperatures) is necessary.

In a number of contexts, it is natural to study the difference between
the finite-temperature and the zero-temperature spectral density. For
that purpose it is convenient to define the so-called reconstructed
Euclidean correlator by
\be 
G_E^{\rm rec}(t,T;T') {\equiv} \int_0^\infty d\omega \rho(\omega,T')
\frac{\cosh\omega(\frac{\beta}{2}-t)}{\sinh \omega \beta/2} \,.
\la{eq:Grec1-main}
\ee
The identity ($0\leq t\leq \beta$)
\be
\frac{\cosh\omega(\frac{\beta}{2}-t)}{\sinh \omega \beta/2}
= \sum_{m\in{\bf Z}} e^{-\omega|t+m\beta|}
\ee
allows us to derive in particular the exact relations
\ba
G_E^{\rm rec}(t,T;0) &=& \sum_{m\in{\mathbb Z}} G_E(|t+m\beta|,T=0)\,,
\la{eq:Grec2}
\\
G_E^{\rm rec}(t,T;{\txts\half} T) &=& 
G_E(t,{\txts\half} T) + G_E(\beta-t, {\txts\half} T)\,.
\ea

\subsection{Tensor Structure of Thermal Correlators\label{sec:tensor}}
% {\sc Vector current correlators.  Heavy-quarks. Energy-momentum tensor correlators.}

The tensor structure of current correlators and energy-momentum tensor
correlators was worked out systematically in~\cite{Kovtun:2005ev}.  We
simply report the results in the notation of section (\ref{sec:corfunc}). 
Setting 
\be
A(t) \leftarrow {\txts\frac{1}{\sqrt{V}}\int} \ud\bx\, e^{-i\bk\cdot\bx}\, J^\mu(t,\bx)
\ee
and defining $B(t)$ analogously in terms of $J^\nu$, we denote their
retarded correlator by $G_R^{\mu\nu}(k)$, with $k=(k_0,\bk)$.  Due to
conservation of the current,
\be
k_\mu G_R^{\mu\nu}(k) = 0.
\ee
Then one defines the tensors
\ba
P_{\mu\nu}&=& \eta_{\mu\nu}-\frac{k_\mu k_\nu}{k^2}\,;\quad
P_{\mu\nu} = P_{\mu\nu}^L + P_{\mu\nu}^T\,,
\\
P_{00}^T &=& 0,\quad P_{0i}^T=0\quad P_{ij}^T = \delta_{ij}-\frac{k_i k_j}{\bk^2}.
\ea
The current-current correlator then has the generic tensor structure
\be
G_{R,\mu\nu}(k)= P_{\mu\nu}^T \,\Pi^T(k_0,\bk^2) + P_{\mu\nu}^L \,\Pi^L(k_0,\bk^2).
\la{eq:GRmunu}
\ee
Thus there are two independent functions of two variables $(k_0,\bk^2)$
that characterize all possible current correlators at finite
temperature.  When $\Pi^T=\Pi^L=\Pi$, one recovers the familiar vacuum
form of the correlators, which are characterized by a single function of
$k^2$.

For the energy-momentum tensor, one proceeds similarly. Setting 
\be
A(t) \leftarrow {\txts\frac{1}{\sqrt{V}}\int} \ud\bx\, e^{-i\bk\cdot\bx}\, T^{\mu\nu}(t,\bx)
\ee
and defining $B(t)$ analogously in terms of $T^{\alpha\beta}$, we denote their
retarded correlator by $G_R^{\mu\nu\alpha\beta}(k)$, with $k=(k_0,\bk)$.  
At zero temperature and in $D$ spacetime dimensions,
\ba
G_R^{\mu\nu,\alpha\beta}(k)&=&
P^{\mu\nu}P^{\alpha\beta}G_B(k^2) + H^{\mu\nu,\alpha,\beta}G_S(k^2),
\la{eq:GTzero}\\
H^{\mu\nu,\alpha,\beta} &=& 
\half(P^{\mu\alpha} P^{\nu\beta} + P^{\mu\beta}P^{\nu\alpha})
-{\txts\frac{1}{D-1}} P^{\mu\nu}P^{\alpha\beta}.~~~~\;
\ea
The tensor $H$ is the polarization tensor of a spin-two particle.
The correlators of the energy-momentum tensor  are thus characterized by
two independent functions.
At finite temperature, the tensor $H$ splits into three independent tensors,
which one may choose as follows
\ba
H_{\mu\nu,\alpha,\beta} &=& S_{\mu\nu,\alpha\beta} + Q_{\mu\nu,\alpha\beta} 
+ L_{\mu\nu,\alpha\beta},
\\
S_{\mu\nu,\alpha\beta}&=& \half\big[ P^T_{\mu\alpha}P^L_{\nu\beta}
+P^L_{\mu\alpha} P^T_{\nu\beta} + P^T_{\mu\beta} P^L_{\nu\alpha}
+ P^L_{\mu\beta}P^T_{\nu\alpha}\big],~~~~~
\\
Q_{\mu\nu,\alpha\beta}&=& {\txts\frac{1}{D-1}}\big[
{\txts(D-2)} P^L_{\mu\nu}P^L_{\alpha\beta}
+{\txts\frac{1}{D-2}} P^T_{\mu\nu}P^T_{\alpha\beta} 
\\
&&~~~~~~~
-(P^T_{\mu\nu} P^L_{\alpha\beta} + P^L_{\mu\nu} P^T_{\alpha\beta})\big].
\nonumber
\ea
The general tensor form of the correlators in a rotationally invariant 
average over states is 
\ba\la{eq:GRmunurhosig}
&& G_{R,\mu\nu,\alpha\beta}(k) =
 \big(P^T_{\mu\nu}P^T_{\alpha\beta} 
+\half(P^T_{\mu\nu}P^L_{\alpha\beta} \!+\! P^L_{\mu\nu}P^T_{\alpha\beta}) \big) C_T(k_0,\bk^2)
\nn
&& ~~~~~~~~~~~~~~~ + \big( P^L_{\mu\nu}P^L_{\alpha\beta}
+\half(P^T_{\mu\nu}P^L_{\alpha\beta} \!+\! P^L_{\mu\nu}P^T_{\alpha\beta}) \big) C_L(k_0,\bk^2)
\nn
&& +S_{\mu\nu,\alpha\beta}G_1(k_0,\bk^2) + Q_{\mu\nu,\alpha\beta}G_2(k_0,\bk^2)
+L_{\mu\nu,\alpha\beta} G_3(k_0,\bk^2).
\nonumber
\ea
There are thus five independent functions that specify all correlators 
of $T_{\mu\nu}$. The function $G_1$ describes the shear channel,
$G_3$ describes the tensor channel, while $G_2$, $C_L$ and $C_T$ describe
the sound channel and the bulk channel.
When $C_L=C_T$ and $G_1=G_2=G_3$, one recovers the zero-temperature form
(\ref{eq:GTzero}).

\section{Retarded correlators: physical significance and analytic calculations}

In the previous section, we introduced the definitions of the
Minkowski and Euclidean correlators, and described their basic
analytic and tensorial structure.  Here, using linear response theory,
we want to show why the retarded correlator $G_R(\omega,\bq)$ is
physically important.  We then go through the most important analytic
techniques to calculate the thermal spectral functions in different
regimes.  We describe how hydrodynamics makes a prediction for the low
frequency and momentum dependence of $G_R(\omega,\bq)$, and thereby
establish the Kubo formulae which allow one to compute transport
coefficients \emph{ab initio} (section \ref{sec:kubo}). When the
hydrodynamic regime arises from the many-body dynamics of weakly
interacting quasiparticles, additional structures can be predicted in
the spectral functions using kinetic theory
(sections \ref{sec:boltz}, \ref{sec:lo-etazeta}).  We also review the
predictions of perturbation theory and the operator product expansion
at high frequencies in section (\ref{sec:pt}).  Some important results
on spectral functions obtained using AdS/CFT results are summarized in
(\ref{sec:adscft}). Thermal sum rules provide useful constraints on
the spectral functions, and those are discussed in section
(\ref{sec:sumrules}).  Finally, certain aspects of spectral functions
in finite volume (where lattice QCD simulations take place) are
pointed out in section (\ref{sec:finivol}).

%and the transport properties are determined by 

% A particular way in which the hydrodynamic regime can be reached is
% via the regime of kinetic theory, and its transport coefficients

\subsection{Linear response theory}

% {Physical significance of the correlators}

When studying dynamical properties of a thermal medium, it is natural
to start with small amplitude departures from equilibrium.  The linear
response of the system to `slow' perturbations is sensitive to the
first order transport coefficients, as well shall see in the next
section.

We assume that the time-dependent perturbation of the system by an
external classical field $f(t)$ coupling to operator $B$ is described
by a Hamiltonian of the form
\be
H_f(t) = H - f(t) B(t),
\la{eq:Hpert}
\ee
$H$ being the Hamiltonian of the unperturbed system.
The perturbation leads to a `response' of physical quantities, i.e.~a change 
in their expectation values with respect to the unperturbed ensemble.
The evolution equation of an operator $A$ is given by
\be
i\frac{\partial}{\partial t}A(t) = -[H_f(t),A(t)].
\ee
% The time dependence of operators is now given by 
% \ba
% A(t) &=& U_f(t) A(0) (U_f(t))^{-1},
% \\
% U_f(t) &=& P\;\exp\{ i{\txts\int_0^t} \ud t'\,H_f(t')\}.
% \ea
One then finds that to linear order in $f$, the expectation value of $A$ 
in the perturbed system minus its unperturbed value is~\cite{Kapusta:2006pm}
\ba
\delta\<A(t)\>&\equiv& \<A(t)\>_f - \<A(0)\>
\la{eq:linresp1}
\\
&=& \!\int_{-\infty}^t \!\ud t' G^{AB}(t-t') f(t')+ {\rm O}(f^2).
\nonumber
\ea
Equation (\ref{eq:linresp1}) is the master formula of linear response
theory. It shows that the correlator $G^{AB}$ determines
the response of an observable $A$ to a time-dependent external field
that couples to $B$.

% In frequency space, if
% $f(t) = \int_{-\infty}^\infty \ud\omega\,e^{-i\omega t} \tilde f(\omega)$
% and similarly for $\delta \<A(t)\>$, 
% \be
% \delta \<\tilde A(\omega)\> =  G_R^{AB}(\omega)\; \tilde f(\omega)~~~({\rm if}~f({ t<0})=0).
% \la{eq:linresp2}
% \ee
% Equations (\ref{eq:linresp1}) and (\ref{eq:linresp2}) are the master 
 %formulas of linear response theory. They show that 
% the retarded correlator $G_R^{AB}$ determines the response 
% of an observable $A$ to a time-dependent external field that couples to $B$.

A source term of the form
\be
f(t) = e^{\epsilon t} \theta(-t) f_0
\la{eq:adiabpert}
\ee
is often adopted to study how the system relaxes back to equilibrium
after having been perturbed adiabatically.
The static susceptibility is defined as the expectation value of $A$ at $t=0$,
\be
\delta\<A(t=0)\>_f = \chi^{AB}_s \, f_0.
\la{eq:chi_stat1}
\ee
From (\ref{eq:linresp1}), it follows that 
\be
\chi^{AB}_s = \int_0^\infty \ud t\, e^{-\epsilon t} \,G^{AB}(t)\,
= G^{AB}_R(i\epsilon).
\la{eq:chi_stat2}
\ee
Looking back at \eq(\ref{eq:l.eq.0}), the zero-frequency Euclidean correlator
is thus equal to the static susceptibility, barring any delta function 
at the origin in the spectral function.

Integrating both sides of (\ref{eq:linresp1}), $\int_0^\infty
\ud\omega\,e^{i\omega t}(.)$, one obtains for the adiabatic
perturbation (\ref{eq:adiabpert}) 
\be G_R^{AB}(\omega)f_0 = \<\delta
A(0)\>_f + i\omega \int_0^\infty \ud t \, e^{i\omega t} \<\delta
A(t)\>_f.  
\la{eq:retresp} 
\ee 
This formula shows that the relaxation
of observable $A$ back to its equilibrium value determines the
retarded correlator $G_R^{AB}(\omega)$. Since the late-time relaxation
is described by hydrodynamic evolution, this equation can be exploited
to obtain a prediction of the small-$\omega$ functional form of $G_R$,
thereby establishing the famous Kubo formulae for the transport
coefficients (section (\ref{sec:kubo})).

In the special case where $B$ is a conserved charge, i.e.~$[H,B]=0$, 
the situation described by $\epsilon=0$
simply corresponds to looking at the system with a small `chemical potential' $f_0$.
Then $\delta\<A(0)\>$ is given by the thermodynamic derivative
\be
\delta\<A(0)\> = \beta f_0 \left(\<B(0) A(0)\>-\<A\>\<B\>\right),
\ee
which, by comparison with (\ref{eq:chi_stat2}), shows that 
it is natural to assign to the retarded correlator at the origin the value 
\be
G_R^{AB}(0) = \beta \left(\<A(0) B(0)\>-\<A\>\<B\>\right).
\ee

% REMARK ON THE LIMIT $\epsilon\to0$ ???
% In most cases, the limit $\epsilon\to0$ of this expression is smooth.
% However, in the case where $B=A$ is a conserved charge, i.e.~$[H,A]=0$, 
% the situation described by $\epsilon=0$
% simply corresponds to looking at the system with a small `chemical potential' $f_0$.
% Then $\delta\<A(0)\>$ is given by the thermodynamic derivative
% \be
% \delta\<A(t)\> = \beta f_0 \<A(0) A(t)\>.
% \ee

% \subsection{Derivation of Kubo formulae}

\subsection{Hydrodynamic predictions and Kubo Formulas\label{sec:kubo}}

There are recent excellent reviews on the modern point of view on
hydrodynamics~\cite{Romatschke:2009im,Teaney:2009qa,Son:2008zz}.
Therefore our goal here is simply to describe the important
steps that lead to the Kubo formulae.
Hydrodynamics can be thought of as a low-energy effective theory:
\bi
\item 
it makes predictions for the low-momentum behavior of correlation functions.
\item it has unknown `low-energy constants', which are called the 
transport coefficients in this context: they are the coefficients 
of a derivative expansion.
\item their values can be determined by a matching procedure with 
the underlying quantum field theory.
\item the number of coefficients grows with the order of the expansion.
\ei

For illustration we choose as a simple example the case of 
the diffusion of a massive particle species in a thermal medium.
In the hydrodynamic treatment, the conservation of the particle number 
is expressed by the classical equation
% The particle current is conserved, 
\be
\partial_t n + \nabla \cdot\boldsymbol{j} = 0.
\la{eq:jcons}
\ee
Second, a phenomenological `constitutive equation' must be introduced to
close the system of equations for $(n,\bj)$.  To leading order in
gradients, it takes the form of Fick's law,
\be
\bj = -D\nabla n.
\la{eq:Fick}
\ee
The conjugate variable to particle density is the chemical potential.
The particle number can be perturbed by 
\ba
H_{\mu} &=& H - {\txts\int} \ud\bx\, \mu(t,\bx) n(t,\bx),
\\
\mu(t,\bx) &=& \mu(\bx)\,e^{\epsilon t} \theta(-t),
\nonumber
\ea
(see \eq(\ref{eq:Hpert}) with $f\leftarrow \mu$ and $B\leftarrow n$).

% Fourier transforming with respect to the spatial coordinates and Laplace transforming in time, 
% $\tilde n(t,\bx) = \int \frac{\ud\bk}{(2\pi)^3} \, e^{i\bk\cdot \bx}\, n(t,\bk)$,
We define
\ba
\tilde n(\omega,\bk)= \int_0^\infty e^{i\omega t} \int \ud\bx \,e^{-i\bk\cdot\bx}
n(t,\bx)
\la{eq:nFourier}
\\
\mu(\bk) = \int \ud\bx\, e^{-i\bk\cdot\bx} \,\mu(\bx).
\ea
The conservation equation (\ref{eq:jcons}) becomes, with the help of 
Fick's law, the diffusion equation 
\be
\partial_t n(\bx) = D\nabla^2 n(\bx).
\la{eq:diff_eq}
\ee
Upon Fourier transformation, the solution reads
\be
\tilde n(\omega,\bk) = \frac{n(0,\bk)}{-i\omega +D\bk^2}.
\ee
The initial condition $n(0,\bk)$ is determined by the static susceptibility
(see \eq(\ref{eq:chi_stat1},\ref{eq:chi_stat2})),
 \ba
 n(0,\bx) &=& {\txts\int_0^\infty} \ud t\, e^{-\epsilon t} {\txts\int} \ud\bx' 
 \,G^{nn}(t,\bx-\bx')\,\mu(\bx'),~~~
 \\
 G^{nn}(t,\bx) &=& i \<[n(t,\bx),n(0)]\>,
 \ea
 or equivalently,
\ba
n(0,\bk) &=& \chi_s(\bk) \mu(\bk),
\\
\chi_s(\bk) &=& {\txts\int_0^\infty} \ud t \, e^{-\epsilon t}
            \, G^{nn}(t,\bk),
\\
G^{nn}(t,\bk) &=& {\txts\int} \ud\bx e^{-i\bk\cdot\bx} G^{nn}(t,\bx).
\ea
This completes the prediction for the long-wavelength, 
late-time evolution of the particle number density.
In Fourier space, the relaxation rate of a particle density 
perturbation of wave-vector $\bk$ is encoded in the position of a 
pole  in the lower-half of the complex $\omega$ plane, namely 
$\omega_{\rm pole}(\bk^2) = -i D\bk^2$.
The residue of the pole is given by the initial condition 
of the relaxation process, which is determined by the static 
susceptibility.

We can now straightforwardly apply the general formula (\ref{eq:retresp}).
In spatial Fourier space, this equation applies for each wave-vector $\bk$.
The result is
\ba
G_R^{nn}(\omega,\bk) &=&
\frac{1}{\mu(\bk)}\left(n(0,\bk) + i\omega \tilde n(\omega,\bk)\right)
\\
&=& \chi_s(\bk)  + i\omega\, \frac{\chi_s(\bk)}{-i\omega+D\bk^2}
\\
&=& 
% \left[ 
\frac{(D\bk^2)^2 + i\omega\, D\bk^2}
{\omega^2+(D\bk^2)^2} 
% {\omega^2+(D\bk^2)^2}
% \right]
\,\chi_s(\bk).
\la{eq:G_Rnn}
\ea
In particular, the imaginary part yields the spectral function,
\be
\frac{\rho^{nn}(\omega,\bk)}{\omega} = 
\frac{\chi_s(\bk)}{\pi} \frac{ D\bk^2}{\omega^2+(D\bk^2)^2}.
\la{eq:rho_nn}
\ee
This equation describes how the transport properties are encoded 
in the low-frequency and low-momentum part of the spectral function.
The Kubo formula is usually expressed in terms of the 
longitudinal part of the current correlator. The latter is related 
to the density correlator by the conservation equation,
$\rho_L(\omega,\bk) = \frac{\omega^2}{\bk^2} \rho^{nn}(\omega,\bk)$
(see \eq\ref{eq:GRmunu}),
\be
\frac{\rho_L(\omega,\bk)}{\omega} =
\frac{\chi_s(\bk)}{\pi} \frac{ D\omega^2}{\omega^2+(D\bk^2)^2},
\la{eq:rhoLhydro}
\ee
implying in particular the Kubo formula
\be
D\chi_s^N = \pi\lim_{\omega\to0}\lim_{\bk\to0} \frac{\rho_L(\omega,\bk)}{\omega}.
\la{eq:kubo}
\ee
Here $\chi_s^N = \beta \int \ud\bx \, n(t,\bx) n(0)$ is the particle 
number susceptibility.

\subsubsection{Electric conductivity\la{sec:elec_cond}}

The correlators of the electromagnetic current are of particular 
phenomenological interest. We will therefore study the operator
\be
A(t) = \int \ud\bx \, e^{-i\bk\cdot \bx}\, j^{\rm em}_{\mu}(t,\bx).
\ee
In QCD, the explicit expression of the electromagnetic current is 
\be
j^{\rm em}_\mu(x) = {\txts\frac{2}{3}} \bar u \gamma_\mu u 
 -{\txts\frac{1}{3}} \bar d \gamma_\mu d 
- {\txts\frac{1}{3}} \bar s \gamma_\mu s + \dots
\ee
In the long-wavelength limit the current-current correlator is given by
the diffusion equation, as described in the previous section, however
here we wish to include the effects of coupling the current to an external
electromagnetic field, $(\boldsymbol{E},\boldsymbol{B})$, and to promote
the treatment to second order accuracy -- we follow the treatment of~\cite{Hong:2010at}.
For a linearized theory invariant under parity, Fick's law generalizes to
\be
\boldsymbol{j} = -D \nabla n + \sigma \boldsymbol{E} 
  - (\sigma\tau_J) \partial_t \boldsymbol{E}
+\kappa_B \nabla\times\boldsymbol{B}.
\la{eq:jconstitut}
\ee
However, a perturbation of the form $\mu(x)+A_0(x)=0$ does not affect the 
system at all, since both $\mu$ and $A_0$ couple 
to the charge density in the Hamiltonian. 
Therefore only the gradient of this combination can appear in the constitutive
equation (\ref{eq:jconstitut}), and we conclude
\be
\chi_s D = \sigma.
\ee
The conservation law $\partial_t n+\nabla\cdot\boldsymbol{j}=0$ 
can be solved for $n(\omega,\bk)$ in the presence of a sinusoidal electric field.
Teaney and Hong~\cite{Hong:2010at} find (for $\bk=(0,0,k)$)
\ba
G_R^{zz}(\omega,\bk) &=& D\chi_s\;\frac{\omega^2 +i\tau_J\omega^3}{-i\omega+D\bk^2},
\\
G_R^{xx}(\omega,\bk) &=& iD\chi_s\omega -D\chi_s\tau_J \omega^2 +\kappa_B\bk^2.
\ea
From here, one can write out Kubo formulae for the second-order coefficients $\tau_J$ 
and $\kappa_B$, 
\ba
D\chi_s \tau_J &=&  -\frac{\partial}{\partial\omega^2} G_R^{xx}(\omega,\bk)_{\omega=\bk=0}\,,
\\
\kappa_B &=&   \frac{\partial}{\partial k^2} G_R^{xx}(\omega,\bk)_{\omega=\bk=0}\,.
\ea
The coefficient $\kappa_B$ can thus be extracted directly by
differentiating the static Euclidean correlator with respect to $\bk$
at $\omega=0$.  An interesting fact about $\kappa_B$ is that it
vanishes in a theory whose transport properties are described by the
Boltzmann equation~\cite{Hong:2010at}. Thus measuring a non-vanishing
$\kappa_B$ would exhibit a shortcoming of kinetic theory. The
parameter $\kappa_B$ is dimensionless and based on the constitutive
equation for the current, it describes how a flux of particles
$\int\ud\boldsymbol{\sigma}\cdot \boldsymbol{j}$ appears through a surface 
around which there is a non-vanishing circulation of a magnetic field, 
$\oint \ud\boldsymbol{\ell}\cdot \boldsymbol{B}$.
Finally, using the conservation law, the spectral function for the charge
density $n$ is modified from \eq(\ref{eq:rho_nn}) to
\be
\frac{\rho^{nn}(\omega,\bk)}{\omega} = 
\frac{\chi_s}{\pi} \frac{ D\bk^2(1+\tau_JD\bk^2)}{\omega^2+(D\bk^2)^2}.
\la{eq:rho_nnb}
\ee

We conclude this section by noting that the vector spectral function
determines the emission rate of photons by the thermal medium.
Denoting the spectral function of components $\mu$ and $\nu$ of the
electromagnetic current by $\rho_{\mu\nu} $, the real photon
emissivity of the quark-gluon plasma at temperature $T$ is related to
the vector spectral function through (see~\cite{Arnold:2001ba} and
Refs. therein)
\be
\frac{\ud\Gamma_\gamma}{\ud^3\bk} = 
\big({\txts\sum_{\rm f}} Q_{\rm f}^2\big)\frac{\alpha_{\rm em}}{2\pi} n_B(\omega,T)\, 
\left.\frac{\rho_\mu^\mu(\omega,\bk,T)}{\omega}\right|_{\omega=|\bk|}\;,
\la{eq:gam-emissiv}
\ee
where $Q_{\rm f}$ are the fractional charges of the quarks. 
Thus the conductivity $\sigma$ determines in particular the emission of 
soft photons.
Similarly, the thermal production rate of dilepton pairs
of invariant mass $M^2=\omega^2-\bk^2$ reads
\be
\frac{\ud N_{\ell_+ \ell_-}}{\ud\omega\ud\bk^3} =
\big({\txts\sum_{\rm f}} Q_{\rm f}^2\big) \,
\frac{\alpha_{\rm em}^2}{3\pi^2} \, n_B(\omega) \,
\frac{\rho_\mu^\mu(\omega,\bk,T)}{\omega^2-\bk^2}.
\ee

% However, since in QCD every flavor number is separately conserved, we
% will also consider for instance the charm current, $j_\mu^{\rm c}=
% \bar c\gamma_\mu c$.  The electric conductivity is given by the Kubo
% formula
% \be
% \sigma(T) = \frac{\pi}{3} 
% \left.\frac{d\rho_i^i(\omega,\boldsymbol{0},T)}{d\omega}\right|_{\omega=0}\,.
% \ee

\subsubsection{Kubo formulae for the transport of energy and momentum}

\begin{figure}
\centerline{\includegraphics[width=0.45\textwidth]{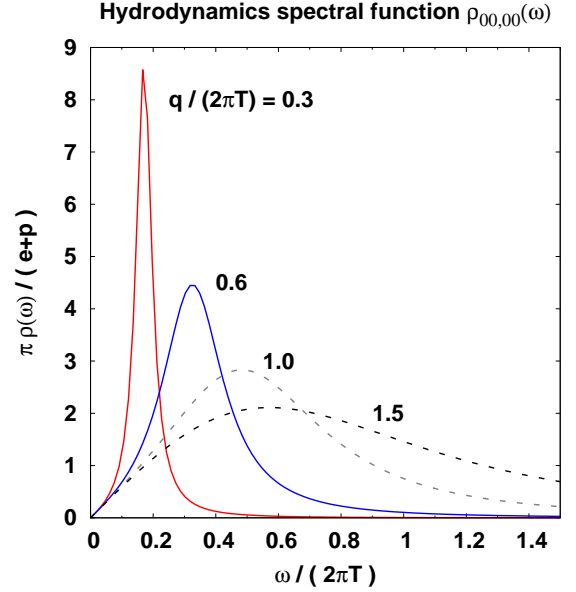}}
\caption{The sound channel spectral function: correlator of 
$T_{00}$ at finite spatial momentum $\bq$. This is the functional form
predicted by hydrodynamics at small momenta and frequencies, for
$c_s^2=\frac{1}{3}$ and $T\Gamma_s=\frac{1}{3\pi}$. For comparison 
with \fig(\ref{fig:sf-sound-AdSCFT}), the dashed curves are displayed, 
although they correspond to momenta well beyond the validity of hydrodynamics.}
\label{fig:sf-sound-hydro}
\end{figure}

In hydrodynamics the transport of energy and momentum is determined by
the correlation functions of the energy-momentum tensor $T_{\mu\nu}$.
It is a symmetric, $T_{\mu\nu}=T_{\nu\mu}$, and conserved operator,
$\partial_\mu T_{\mu\nu}=0$.  The constituent equation, which is the
analogue of Fick's law (\ref{eq:Fick}), reads
\ba
T^{\mu\nu} &=& (e+P)u^{\mu}u^{\nu} + Pg^{\mu\nu}
\la{eq:constit}
\\
&-& \eta P^{\mu\alpha}P^{\nu\beta}
\big(  \partial_\alpha u_\beta + \partial_\beta u_\alpha 
     - {\txts\frac{2}{3}} g_{\alpha\beta} \partial\cdot u \big)
-\zeta P^{\mu\nu} (\partial\cdot u).
\nonumber
\ea
where $u_\mu u^\mu = -1$.  The energy-momentum conservation equations,
together with \eq(\ref{eq:constit}), determine the evolution of
the velocity and temperature fields, $u_\mu(x)$ and $T(x)$.  
To study the linearized modes of these
equations, one writes $\boldsymbol{u}_\perp$, $\boldsymbol{u}_{||}$ and $e$ in
Fourier components as in (\ref{eq:nFourier}), and 
decomposes the velocity field into a longitudinal and a
transverse part with respect to $\bk$, 
$\boldsymbol{u}_{||}$ and $\boldsymbol{u}_{\perp}$.  The retarded correlator
for the momentum density, $\boldsymbol \pi=(e+p) \boldsymbol{u}$ is then obtained
along the same lines as \eq(\ref{eq:G_Rnn}). We refer to appendix C 
of~\cite{Teaney:2006nc} for the details and quote the leading-order result
\ba
\!\!\!\! G_R^{\pi_\perp\pi_\perp}(\omega,\bk) &=&
 \frac{\eta \bk^2}{-i\omega + \frac{\eta \bk^2}{e+p}}\,,
\\
\!\!\!\! G_R^{\pi_{||}\pi_{||}}(\omega,\bk) &=&
\frac{i\omega \Gamma_s\bk^2 - (c_s\bk)^2}
     {\omega^2 -(c_s\bk)^2 + i\omega\Gamma_s\bk^2}\,.
\ea
respectively for the transverse and longitudinal momentum density, 
where 
\be
\Gamma_s = \frac{\frac{4}{3}\eta + \zeta}{e+p}
\ee
determines the damping of sound waves.
The spectral functional is obtained by taking the imaginary part,
\ba
\!\!\!\! \frac{\rho^{\pi_\perp\pi_\perp}(\omega,\bk)}{\omega} &=&
\frac{1}{\pi} \frac{\eta \bk^2}{\omega^2+(\frac{\eta \bk^2}{e+p})^2}\,,
\la{eq:ImGRshear}
\\
\!\!\!\! \frac{\rho^{\pi_{||}\pi_{||}}(\omega,\bk)}{\omega} &=&
\frac{e+p}{\pi} \frac{\omega^2\Gamma_s\bk^2}
     {(\omega^2-c_s^2\bk^2)^2+ (\omega \Gamma_s\bk^2)^2} \,.
\la{eq:ImGRsound}
\ea
The first line corresponds to the shear channel, the second to the
sound channel.  The former admits a diffusive pole, similar to the
pole appearing in the vector current correlator. The latter contains a
pole at $\omega=\pm c_s k - \half i\Gamma k^2$, which corresponds to
the propagation of sound waves with velocity $c_s$ with an attenuation
$\propto e^{-\frac{1}{2}\Gamma_s k^2 t}$ for a plane wave of
wavelength $\lambda = \frac{2\pi}{k}$. The sound channel spectral
function is displayed for two different values of the wavelength
in \fig(\ref{fig:sf-sound-hydro}).  Using the momentum conservation
equations (see \ref{eq:GRmunurhosig}) and denoting the spectral
function of $T_{\mu\nu}$ and $T_{\rho\sigma}$ by
$\rho^{\mu\nu,\rho\sigma}$, one then obtains the Kubo formulae,
for $\bk=(0,0,k)$,
\ba
\eta(T) &=& \pi \lim_{\omega\to0} \frac{\rho^{13,13}(\omega,{\bf 0},T)}{\omega},
\la{eq:Kubo-eta}
\\ % \qquad\qquad
\zeta(T) &=& \frac{\pi}{9} \sum_{i,j=1}^{3}
\lim_{\omega\to0} \frac{\rho^{ii,jj}(\omega,{\bf 0},T)}{\omega} .
\la{eq:Kubo-zeta}
\ea
In numerical practice, it is convenient if $\rho(\omega)/\omega$ behaves
smoothly at $\omega\to0$. It is therefore important to study the
spectral function in detail in this
region~\cite{Romatschke:2009ng,Meyer:2010ii}.
One finds that 
\be
\frac{1}{9}\sum_{i,j=1}^3\frac{\rho_{ii,jj}(\omega,{\bf 0},T)}{\omega}
= \frac{\zeta}{\pi}+(e+p)c_s^2\delta(\omega)+\dots
\ee
contains a delta function at the origin (see section \ref{sec:sumrules}).
When studying the bulk (scalar) channel it is then preferable 
to work instead with the operator 
\be
T^{kk} + 3 c_s^2 T^{00}, 
\ee
written here in Euclidean metric,
whose spectral function at $\bk=0$ is identical to the spectral function 
of $T^{kk}$, except for not having the delta function at the origin.

The most general constitutive equation for $T^{\mu\nu}$ has been
worked out to second order in gradients of the velocity fields for a
conformal theory~\cite{Baier:2007ix}, and more recently even in the
non-conformal case~\cite{Romatschke:2009im}. In the conformal case,
there are five new transport coefficients, denoted
$\tau_\Pi, \kappa, \lambda_1, \lambda_2, \lambda_3$.  These
second-order transport coefficients can be obtained from Kubo
formulae,
\ba
\kappa &=& -\frac{\partial^2}{\partial k^2} G_R^{12,12}(\omega,\bk)_{\omega=k=0}
\la{eq:kubo_kappa}
\\
\eta\tau_\Pi -\half \kappa &=& \frac{1}{2} 
\frac{\partial^2}{\partial \omega^2}G_R^{12,12}(\omega,\bk)_{\omega=k=0}.
\la{eq:kubo_tauPi}
% \frac{\partial^2}{\partial \bk^2}G_R^{12,12}(\omega,\bk) \big)
\ea
still with the convention $\bk=(0,0,k)$.  A physically important effect
is that at second order, the speed of sound receives a
wavelength-dependent correction~\cite{Baier:2007ix}, $\omega = c_s(q) q$ with
\be
% \omega = c_s(q) q,\qquad
c_s(q)=c_s \left\{1 + \frac{\Gamma_s}{2}q^2
\left(\tau_\Pi-\frac{\Gamma_s}{4c_s^2}\right)+{\rm O}(q^4)\right\}\,.
\la{eq:vsq}
\ee
This provides a way to determine the coefficient $\tau_\Pi$
if the leading-order speed of sound $c_s^2=\frac{\partial p}{\partial e}$ 
and $\Gamma_s$ are known.
The coefficient $\kappa$ can be extracted directly from the $\omega=0$ Euclidean correlator\footnote{See 
\eq(\ref{eq:l.eq.0}), where in view of (\ref{eq:delta1313}) 
no subtlety occurs when taking the limit $\omega\to0$. Note however that a quartic 
divergence must be subtracted from $G_R^{12,12}$.}.
The coefficients $\lambda_{1,2,3}$ influence the hydrodynamic evolution only at
quadratic order in the amplitude of the perturbation. Kubo formulae have
been derived for these second-order coefficients as well~\cite{Moore:2010bu}.
They involve three-point functions of the energy-momentum tensor. 
Among them we want to point out the expression for $\lambda_3$, 
which like $\kappa$ can be extracted directly from a Euclidean correlator,
\be
\lambda_3 = 6 \lim_{p,q\to0} \frac{\partial^2}{\partial q\partial k} G_E^{11,01,01}(q,k),
\ee
where $\bq$ and $\bk$ are aligned along the $\hat3$-direction.
Both $\kappa$ and $\lambda_3$ have a finite limit when the coupling constant goes to 
zero~\cite{Baier:2007ix,Moore:2010bu}.

The physical interpretation of the second order coefficients is not always straightforward.
The coefficients $\kappa$ and $\lambda_3$ have been interpreted as a 
thermodynamic response to vorticity in the fluid 
velocity field~\cite{Romatschke:2009kr,Moore:2010bu}.
The coefficient $\tau_\Pi$ has the unit of time and 
from its contribution to the constituent equation of $T_{\mu\nu}$, 
it implies a retardation in the response of the fluid to an change in the background metric.
A relaxation time $\tau_R$ in the sense of the 
Israel-Stewart formalism is reflected in a pole in the retarded correlator 
at $\omega_{\rm pole } = -i \tau_R^{-1}$.
Whether the retarded correlator actually admits such a pole cannot be answered 
within hydrodynamics, instead it must be answered at 
a more microscopic level, see the discussion in~\cite{Denicol:2011fa} and Refs. therein.

\subsection{Quasiparticles and the Boltzmann Equation\label{sec:boltz}}

In addition to the pole structure of certain spectral functions
predicted by hydrodynamics, additional characteristic features show up
if the conserved charges are transported by quasiparticles whose mean
free path is long compared to the thermal scale.  This is the subject
of this section, where we first consider the technically simpler case
of a diffusing heavy quark, and then discuss the transport of quark
number, energy and momentum by the constituents of the quark-gluon
plasma.

\subsubsection{Diffusion of a heavy quark}

The diffusion of a heavy quark (one has in mind the charm or bottom)
in the quark-gluon plasma is characterized by a time scale $M/T^2$
which is long compared to the thermal time scale of $1/T$.  For this
reason it is widely believed that a classical Langevin equation should
appropriately describe the thermalization of heavy
quarks~\cite{Moore:2004tg}.  The equations of motion for the latter
are
\ba 
\frac{d\bx}{dt} = \frac{\bp}{M},
&\quad & \frac{d\bp}{dt} = \boldsymbol{\xi}(t) - \eta \bp(t), 
\\ \<\xi^{i}(t) \xi^{j}(t')\> &=&
\kappa \delta^{ij} \delta(t-t').  
\la{eq:xixi}
\ea 
For a given $\xi(t)$, the equation is easily solved to give
\be
\bp(t) = e^{-\eta t} \big[\bp(0) +
 {\txts\int_0^t} \ud s \boldsymbol{\xi}(s) e^{\eta s}\big].
\ee
Then 
\be
\lim_{t\to\infty} \< p_i(t) p_j(t)\> = \frac{\kappa}{2\eta} \delta_{ij}.
\ee
The equipartition of energy requires $\frac{\bp^2}{2M}$ to be $\frac{3}{2}T$ in
equilibrium. The drag and fluctuation coefficients are then related by the
fluctuation-dissipation relation (established by Einstein in 1905),
\be
\eta = \frac{\kappa}{2MT}.  
\la{eq:Einstein}
\ee 
Furthermore, the mean square distance is also easily worked out, 
\ba
{\txts\frac{1}{3}}\<\bx^2(t)\> &=& \frac{\kappa}{\eta^2M^2} 
\big[t-{\txts\frac{1}{\eta}}(1-e^{-\eta t}) -
{\txts\frac{1}{2\eta}}(1-e^{-\eta t})^2 \big]
\nn
&& +\frac{\bp^2(0)}{3M^2\eta^2} (1-e^{-\eta t})^2.
\la{eq:x2langevin}
\ea
For large times, the first term dominates.
Recall that the diffusion equation (\ref{eq:diff_eq}) yields
a density of particles $n(t,\bk)= n(0,\bk) e^{-D\bk^2 t}$, or 
$n(t,\bx)\propto e^{-\frac{\bx^2}{4Dt}}$ for an initial distribution 
localized at the origin. Therefore the mean square radius of the
particles after time $t$ is $\frac{1}{3}\<\bx^2\> = 2Dt$, and comparing 
with (\ref{eq:x2langevin}), one finds that the Langevin process leads 
at late times to a diffusion with diffusion coefficient
\be
D = \frac{T}{M\eta}.
\ee
For a thermal initial distribution of momenta, $\<p_i(0) p_j(0)\>=MT\delta_{ij}$, 
expression (\ref{eq:x2langevin}) simplifies to
\be
{\txts\frac{1}{3}}\<\bx^2(t)\> = 
2D\big[ t  - {\txts\frac{1}{\eta}}(1-e^{-\eta t})\big]
\la{eq:x2lange2}
\ee
This equation describes both the early-time directed motion,
$ {\txts\frac{1}{3}}\<\bx^2(t)\> = \frac{1}{3} v^2 t^2$, ${\txts\frac{1}{3}}v^2=\frac{T}{M}$, 
and the late-time diffusive motion, ${\txts\frac{1}{3}}\<\bx^2(t)\> = 2Dt$~\cite{Petreczky:2005nh}.
% This relation ensure that the heavy quark eventually thermalizes.

\begin{figure}
\centerline{\includegraphics[angle=0,width=0.45\textwidth]{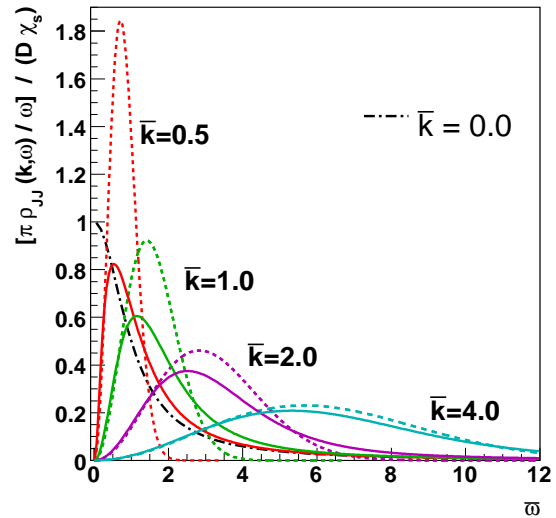}}
\caption{Heavy-quark spectral function of the longitudinal current correlator
as predicted by the Langevin equation~\cite{Petreczky:2005nh}
as a function of a scaled frequency $\bar\omega \equiv \omega D(M/T)$
for various values of a scaled momentum $\bar{\bk}\equiv \bk DM/T$. 
The solid lines are the predictions of the Langevin equations,
while the dotted lines are the predictions of the free theory.
The dash-dotted line shows the $\bk= 0$ result of the Langevin equation.}
\label{fig:chinn}
\end{figure}

We follow the treatment of~\cite{Petreczky:2005nh} to establish 
the connection between the distribution of heavy quarks after a time
$t$ and the retarded correlator of the heavy quark current.
Let $P(t,\bx)$ be the probability that a heavy quark starts at the
origin at $t = 0$ and moves a distance $\bx$ over a time $t$.
If the distribution of heavy quarks at time zero is $N(0,\bx)$, 
at time $t$ it will be given by the convolution
\be
N(t,\bx) = \int \ud\bx' \, P(t,\bx-\bx')\, N(0,\bx'),
\ee
or equivalently 
\be
N(t,\bk) = P(t,\bk) \,N(0,\bk).
\ee
Applying the general rule (\ref{eq:retresp}), we thus find 
\be
G^{nn}_R(\omega,\bk) = \chi_s(\bk) % \frac{N(0,\bk)}{\mu(\bk)}
\left[ 
1+i\omega \int_0^\infty \ud t \, e^{i\omega t}\, P(t,\bk)
\right]
\la{eq:G_HQ}
\ee
If one assumes the noise to be Gaussian distributed, 
then the probability distribution $P(t,\bx)$ is
Gaussian~\cite{Petreczky:2005nh}, with a width given
by \eq(\ref{eq:x2lange2}), and therefore so is $P(t,\bk)$.  The static
susceptibility $\chi_s(\bk)$ can be obtained by noting that the
initial phase-space distribution of the heavy quarks is described by
$f(\bx,\bp,t=0)=e^{\beta(\mu(\bx)-M-\bp^2/2M)}$.  Taking into account
both quarks and antiquarks, the static susceptibility is independent
of $\bk$ to linear order in the perturbation $\mu(\bx)=\mu_0+\delta\mu(\bx)$, 
\be
\chi_s(\bk) = \frac{\nu}{T}\Big[\frac{MT}{2\pi}\Big]^{3/2}\,e^{-\beta M}\cosh{\beta\mu_0},
\la{eq:chi_HQ}
%% check the factor 1/T with Derek
\ee
where $\nu=4N_c$ is a multiplicity factor.
\eq(\ref{eq:G_HQ}) and (\ref{eq:chi_HQ}) thus provide all the ingredients
to compute the kinetic theory prediction of $G_R$ at low frequencies and 
momenta. \fig(\ref{fig:chinn}), from~\cite{Petreczky:2005nh}, 
displays the spectral function of the longitudinal current correlator
($\rho_L(\omega,\bk) = \frac{\omega^2}{\pi k^2}\im G^{nn}_R(\omega,\bk)$, 
denoted by $\rho_{JJ}$ on the figure).
In particular, the $\bk=0$ spectral function takes the form
\be
\frac{\rho_L(\omega,\boldsymbol{0})}{\omega}= 
\frac{\chi_s}{\pi}\frac{T}{M}\,\frac{\eta}{\omega^2+\eta^2}.
\la{eq:rho_L}
\ee
By comparison with the hydrodynamic prediction (\ref{eq:rhoLhydro}), 
we see that the kinetic treatment predicts not just the intercept of $\rho_L/\omega$, 
but also its local functional form, a Lorentzian. 
The width has the interpretation of an inverse relaxation time.
It is a long-known fact that kinetic theory establishes a connection between the 
diffusion coefficient $D=\frac{T}{M\eta}$ and the relaxation time $\eta^{-1}$.
Already Drude's classical description of the electric conductivity in metals
around 1900 showed that $\sigma/(\tau n)=\frac{e^2}{m}$, where $n$ is the density of 
electrons and $m$ their mass.

The spectral structure that is obtained from the Langevin equation is
expected to arise for a sufficiently heavy diffusing particle.
Conversely, the presence of a transport peak allows one
to \emph{define} the quantities appearing in the Langevin equation
directly from the spectral function~\cite{CaronHuot:2009uh}. The
effective mean-square velocity is then given by
\be
{\txts\frac{1}{3}}\<\boldsymbol{v}^2\> \equiv \frac{1}{\chi_s}\int_{-\Lambda}^{\Lambda} 
 \frac{\ud\omega}{\omega} \rho_L(\omega),
\ee
where $\Lambda$ is a cutoff that separates the scale $\eta$ from 
the correlation time of the medium (which is typically of order $T$, or $gT$ at weak coupling).
The `kinetic mass' $M_{\rm kin}$ is further defined so as to satisfy the 
equipartition theorem, $M_{\rm kin}\<\boldsymbol{v}^2\>=3T$.
Finally, the momentum diffusion coefficient $\kappa(M)$ can be defined as 
\be
\kappa(M)= \frac{2\pi M_{\rm kin}^2}{\chi_s} 
\;\omega\rho_L(\omega)\Big|_{\eta\ll |\omega| \ll \Lambda}.
\la{eq:kappa_M}
\ee
If one inserts the spectral function (\ref{eq:rho_L}) derived from the
Langevin equation, one recovers Einstein's relation
(\ref{eq:Einstein}). A weak-coupling calculation shows that while
$\kappa(M)$ and $D$ are only weakly dependent on $M$, the drag
coefficient $\eta\sim T^2/M\times $ a power of the coupling constant
is parametrically small compared to the medium time-scale.  This
justifies \emph{a posteriori} the assumption that the transport peak
is narrow.  The momentum diffusion coefficient $\kappa$ converges to a
finite value when the static limit $M\to\infty$ is taken.  Note
that although the peak in $\rho_L(\omega)$ becomes arbitrarily
narrow in that limit, the limit $M\to\infty$ cannot be interchanged with 
the limit $\omega\to0$ in the definition of $\kappa$.

The formula (\ref{eq:kappa_M}) can be understood heuristically as
follows~\cite{CaronHuot:2009uh}: $\rho_L$ is the spectral function for
the operator $j_k$ at vanishing spatial momentum. Classically, if the
quark is heavy, this operator measures a single quark's velocity,
$v_k=\int \ud\bx \,j_k$.  Therefore the operator
$M\int \ud\bx\,\frac{\ud j_k}{\ud t}$ measures the force acting on the
quark. We now recall that $\kappa$ measures the size of the force
exerted by the medium on the quark, \eq(\ref{eq:xixi}).  We then have
to evaluate the classical force-force correlator, whose quantum analog
is the symmetrized correlator.  Combining these ingredients, one
reaches
\ba \la{eq:jkjk}
\kappa &=&\frac{\beta}{6}\sum_{k=1}^3 
\lim_{\omega\to0} \Bigg[
\lim_{M\to\infty} \frac{M_{\rm kin}^2}{\chi_s}
\int_{-\infty}^{\infty}\!\!\ud t\, e^{i\omega(t-t')} 
\\ &&~~~~~~~~~~~~~
\int\!\!\ud\bx 
\big\<\big\{\frac{\ud\hat j_k(t,\bx)}{\ud t},
              \frac{\ud\hat j_k(t',\boldsymbol{0})}{\ud t'}
\big\}\big\> \Bigg].
\nonumber
\ea
A second, rigorous derivation was given for this formula 
in~\cite{CaronHuot:2009uh}.
Through the fluctuation-dissipation formula (\ref{eq:fluc-dissip}),
the symmetrized correlator can be expressed in terms of the
corresponding spectral function, which then leads back to formula
(\ref{eq:kappa_M}) in the heavy-quark limit. As one might expect
based on the classical argument laid out above, 
the correlator (\ref{eq:jkjk}) can be expressed in the Heavy-Quark 
Effective Theory (HQET) as a force-force correlator, where the leading 
force is the chromo-electric force $g\boldsymbol E$,
\ba
\kappa &=&\frac{\beta}{6}\sum_{k=1}^3
\lim_{M\to\infty} \frac{1}{\chi_s}\int \ud t\ud\bx\,
\Big\< \Big\{ \big(\hat\phi^\dagger gE^k\hat\phi-
\hat\theta^\dagger gE^k\hat\theta\big)(t,\bx),\,
\nn && ~~~~~~~~~~~~~~~~~~
\big(\hat\phi^\dagger gE^k\hat\phi-
\hat\theta^\dagger gE^k\hat\theta\big)(0,\boldsymbol{0})\Big\}\Big\>.
\ea
where $\theta$ and $\phi$ are two-component spinors of HQET and the
$\omega\to0$ limit has now been taken.  The corresponding Euclidean
correlator reads, after evaluating the fermion line contractions,
\be
G^{\rm HQET}_E(t) = 
\frac{\Big\<\re\tr \big( U(\beta,t)gE_k(t,\boldsymbol{0})
                   U(t,\boldsymbol{0})gE_k(0,\boldsymbol{0})\big)\Big\>}
{ -3\; \< \re\tr U(\beta,0)\>},
\la{eq:G_HQET}
\ee
where the color parallel transporters $U(t_2,t_1)$ 
in the fundamental representation are propagators of static quarks.
In particular the denominator of (\ref{eq:G_HQET}) is the Polyakov loop.
The momentum diffusion coefficient is given by the low-frequency limit
of the corresponding spectral function via \eq(\ref{eq:ClatRho}),
\be
\kappa = \lim_{\omega\to0} \frac{2\pi T}{\omega} \rho^{\rm HQET}(\omega).
\la{eq:kappa_HQET}
\ee
The obvious advantage of this formulation is that the large scale $M$
has disappeared from the problem. The spectral function $\rho^{\rm HQET}$
has been studied in detail at next-to-leading order in perturbation
theory~\cite{Burnier:2010rp}.  Remarkably, even in the weak-coupling
limit, the function is smooth as small frequencies. This is in
contrast with the narrow transport peaks that are found at
weak-coupling in e.g.~the shear channel. This property is a clear
advantage for numerical studies of the spectral function, as we will 
see in section (\ref{sec:EC}).

\subsubsection{Kinetic theory in other channels}

When studying the diffusion of a light quark number, or the transport
of transverse momentum, the applicability of kinetic theory is not
\emph{a priori} guaranteed. It is only valid if the medium admits 
quasiparticles that are sufficiently long-lived. In the heavy-quark
diffusion case, the existence of a quasiparticle is guaranteed by the
separation of scales $M/T$. Typically, although perhaps not
necessarily, quasiparticles arise in weakly coupled systems. In
$\phi^4$ theory, the quasiparticle carries the same quantum numbers as
the particles at zero-temperatures. At sufficiently high temperatures
in QCD, due to asymptotic freedom, quasiparticles are expected to
emerge, with the quantum numbers of quarks and gluons.  At low
temperatures, well below the crossover region, we expect the
well-known hadron resonances to be good quasiparticles. In that regime
the parameter $1/N_c$ can be used to provide an expansion around the
non-interacting hadron gas ($N_c$ is the number of colors).

The physical picture one has in mind is that of a particle propagating
freely until it makes a `collision' with a constituent of the medium,
upon which its momentum or even its nature may be changed. The
collision generally has to be described by a quantum-mechanical matrix
element. The propagation of the quasiparticle between collisions can
be treated as classical, provided the time until the next collision 
is long enough that any coherence effects can be neglected. The evolution 
of the quasiparticle can then be written as a convolution of transition 
probabilities, rather than amplitudes. In this approximation the evolution
is described by the Boltzmann equation for the probability distribution 
$f(\bx,\bp,t)$.

\begin{figure}
\centerline{\includegraphics[angle=0,width=0.40\textwidth]{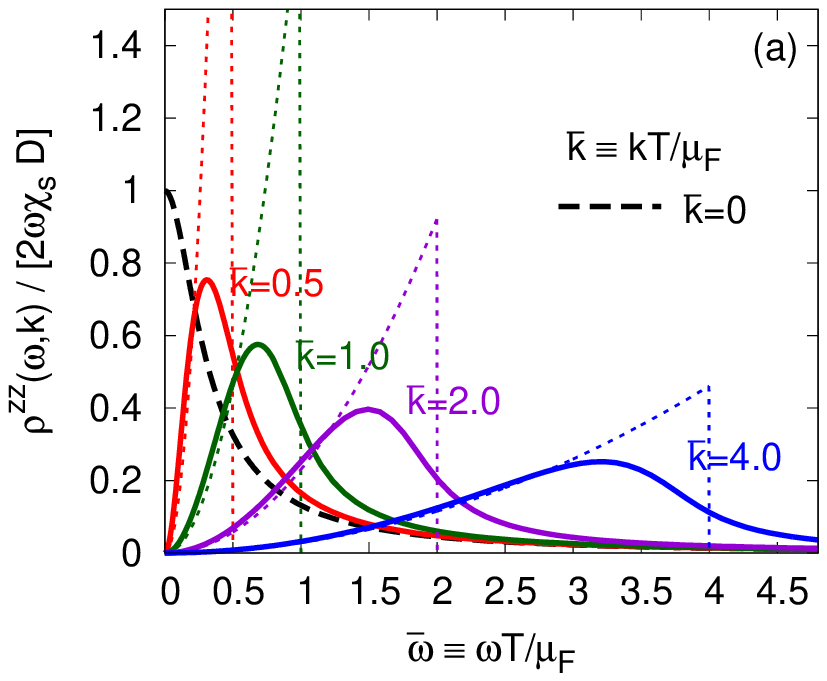}}
\centerline{\includegraphics[angle=0,width=0.40\textwidth]{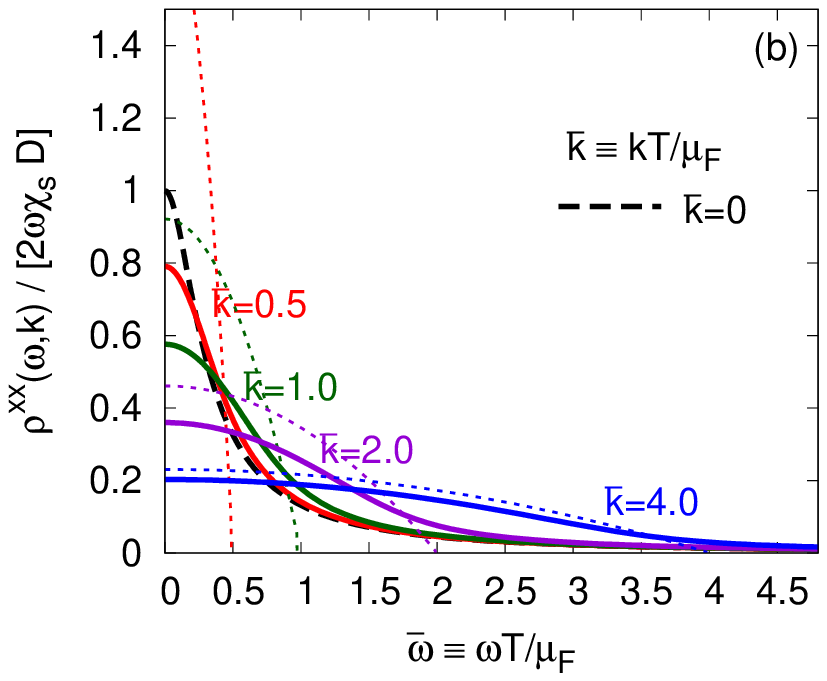}}
\caption{Kinetic theory prediction for the vector current correlator in
quenched QCD for massless quarks~\cite{Hong:2010at}.  
The momentum is directed in the $z$
direction, $\bk=(0,0,k)$, and $\mu_F=g^2C_Fm_D^2\log(T/m_D)/8\pi$ is
the drag coefficient of a quark in leading-log approximation. 
The thick dashed curve corresponds to $\bk=0$. The thin
dashed lines show the result of the free Boltzmann 
equation~\cite{Aarts:2005hg,Hong:2010at}. Note that the normalization of $\rho$ 
differs from our convention by a factor $2\pi$.}
\label{fig:rho-diff}
\end{figure}

To illlustrate these general remarks, we review a recent kinetic
theory treatment of the diffusion of \emph{light} quarks~\cite{Hong:2010at}.
To treat this problem one needs the Boltzmann equation in the presence of an external field.
Recalling that the Lorentz force on a charged particle is $F^i= Q {F^i}_\mu v^\mu$,
the Boltzmann equation (for a single flavor charged under the externally applied gauge field) reads
\be
\frac{1}{E_{\bp}}\left(p^\mu\partial_\mu 
+ Q F^{\mu\nu}p_\nu \frac{\partial}{\partial p_\mu}\right)
f^a = C^a[f,\bp].
\ee
Here $Q$ denotes the electric charge of the quark and below we will employ the notation
\ba
n^B(\omega) = \frac{1}{e^{\beta\omega}-1},
\\
n^F(\omega) = \frac{1}{e^{\beta\omega}+1},
\ea
often with the understanding of an ultrarelativistic dispersion relation,
i.e. $n_p^F\equiv n^F(\omega=|\bp|)$.
Specializing to a spatial gauge potential, $A_\mu=(0,\boldsymbol{A})$, and linearizing 
the Boltzmann equation yields in Fourier space $(\omega,\bk)$
\ba
&& (-i\omega +i \boldsymbol{v}_{\bp}\cdot\bk) \delta f^a(\omega,\bk) 
\\
&& \qquad -i\omega n(1 - n) Q A_i\frac{p^i}{E_{\bp}} = C^a[\delta f,\bp].
\nonumber
\ea
Since the coupling of the gauge field is linear in the charge, it only influences
the difference of quarks and antiquarks,
\ba
\la{eq:Boltz}
&&(-i\omega +i \boldsymbol{v}_{\bp}\cdot\bk) \delta f^{s-\bar s}(\omega,\bk) 
\\
&&\qquad
-i\omega n(1-n) 2Q A_i \frac{p^i}{E_{\bp}} = C^{s-\bar s}[\delta f,\bp].
\nonumber 
\ea
At this point one needs to specify the collision kernel $C^{s-\bar s}$.
The quarks and gluons contribute a `loss' term (1st line) 
and a `gain' term~\cite{Arnold:2000dr,Hong:2010at},
\ba
\!\!\!(C^{q}_{qg}-C^{\bar q}_{qg}) &=& 
-2\gamma \frac{n_p^F(1+n_p^B)}{p} [\chi^q(\bp)-\chi^{\bar q}(\bp)]
\\
&& + \frac{2\gamma}{\xi_{BF}} \frac{n_p^F(1+n_p^B)}{p}\int \frac{\ud\bk}{(2\pi)^3}
\nonumber
\\
&& \quad \frac{n_k^F(1+n_k^B)}{k}\,[\chi^q(\bk)-\chi^{\bar q}(\bk)],
\nonumber
\\
\delta f(\bp) &=& n_p^F(1-n_p^F)\chi(\bp),
\\
\gamma &\equiv &  \frac{g^4C_F^2\xi_{BF}}{4\pi}\log\left({\frac{T}{m_D}}\right),
\\
\xi_{BF} &\equiv& \int \frac{\ud\bk}{(2\pi)^3} \frac{n_k^F(1+n_k^B)}{k}= \frac{T^2}{16}.
\ea
Hong and Teaney solved the Boltzmann equation (\ref{eq:Boltz}) by expanding functions 
of momenta in real-valued linear combinations of spherical harmonics. The radial 
part of the momenta is discretized, and the problem is reduced to inverting
a large matrix.
The out-of-equilibrium expectation value of the current is determined from $\delta f$ via
\be
\<J_i\>_A = Q\nu_s\int \frac{\ud\bp}{(2\pi)^3} \frac{p_i}{E_{\bp}} \delta f^{s-\bar s},
\ee
and from there the retarded correlator is obtained via the frequency version of the 
linear response formula (\ref{eq:linresp1}),
\be
\<J^\mu(\omega,\bk)\>_A = G_R^{\mu\nu}(\omega,\bk) A_\nu(\omega,\bk).
\ee
The result is shown in \fig(\ref{fig:rho-diff}).
In the rescaled variables, the functional form of the curves corresponding 
to different momenta are similar to those obtained for heavy quarks, 
\fig(\ref{fig:chinn}).

\begin{figure}
\centerline{\includegraphics[angle=0,width=0.40\textwidth]{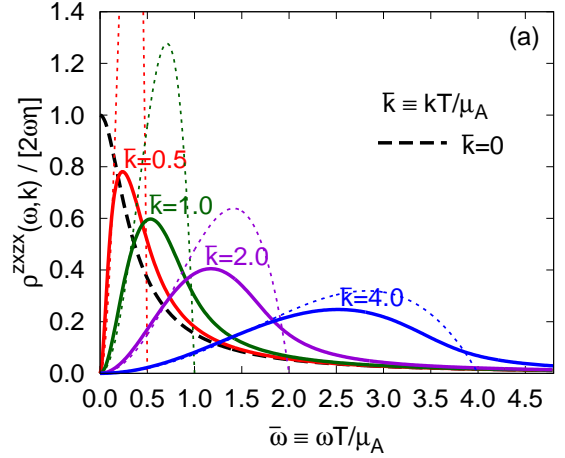}}
\caption{Spectral function of $T_{xy}$ in the shear channel,
as predicted by the Boltzmann equation~\cite{Hong:2010at} in leading log approximation.
The thick dashed curve corresponds to $\bk=0$, and $\mu_A=g^2C_A m_D^2\log(T/m_D)/8\pi$.
The thin dashed curves represent the free theory result~\cite{Meyer:2008gt,Hong:2010at}.
The normalization of $\rho$ differs from our convention by a factor $2\pi$.
}
\label{fig:rho1212-kin}
\end{figure}

A similar approach was used in~\cite{Hong:2010at} to determine the
correlators of $T_{\mu\nu}$. Now the perturbing field is gravitational
rather than electromagnetic, and consequently the Boltzmann equation
has to be generalized to non-flat background metrics.  Another
important aspect of the calculation is the selection of appropriate
boundary conditions. Hong and Teaney argue that the appropriate
boundary condition for $\chi(\bp)$ is a Dirichlet boundary condition
at $\bp=0$. The reason is that the rate of soft gluon emission is
parametrically high compared to the transport time scales.  At high
momenta the solution that grows exponentially is discarded by
implementing the appropriate finite-difference scheme at the highest
discretized momentum. The result for the shear channel is shown
in \fig(\ref{fig:rho1212-kin}). By the Kubo formula
(\ref{eq:Kubo-eta}), the shear viscosity in leading logarithmic
approximation can be read off from the intercept of the $\bk=0$ curve
(parametrically, $\eta/s \sim [g^4 \log(1/g)]^{-1}$).

\subsection{Leading order results for the shear and bulk viscosities\la{sec:lo-etazeta}}

The conductivity, as well as the shear and bulk viscosities, have been
obtained beyond the leading-logarithmic order, namely to full leading
order in the gauge coupling $g$~\cite{Arnold:2003zc,Arnold:2006fz}.
The method follows the steps sketched in the previous section, namely
a Boltzmann equation is written down and subsequently linearized.  In
addition to the ingredients necessary for a leading-logarithmic
accuracy, the collision kernel must be treated more carefully at low
momenta, and the Landau-Pomeranchuk-Migdal (LPM) effect, which
involves collinear emissions and their destructive interferences, must
be taken into account~\cite{Moore:2004kp}.

\begin{figure}
\centerline{\includegraphics[angle=0,width=0.40\textwidth]{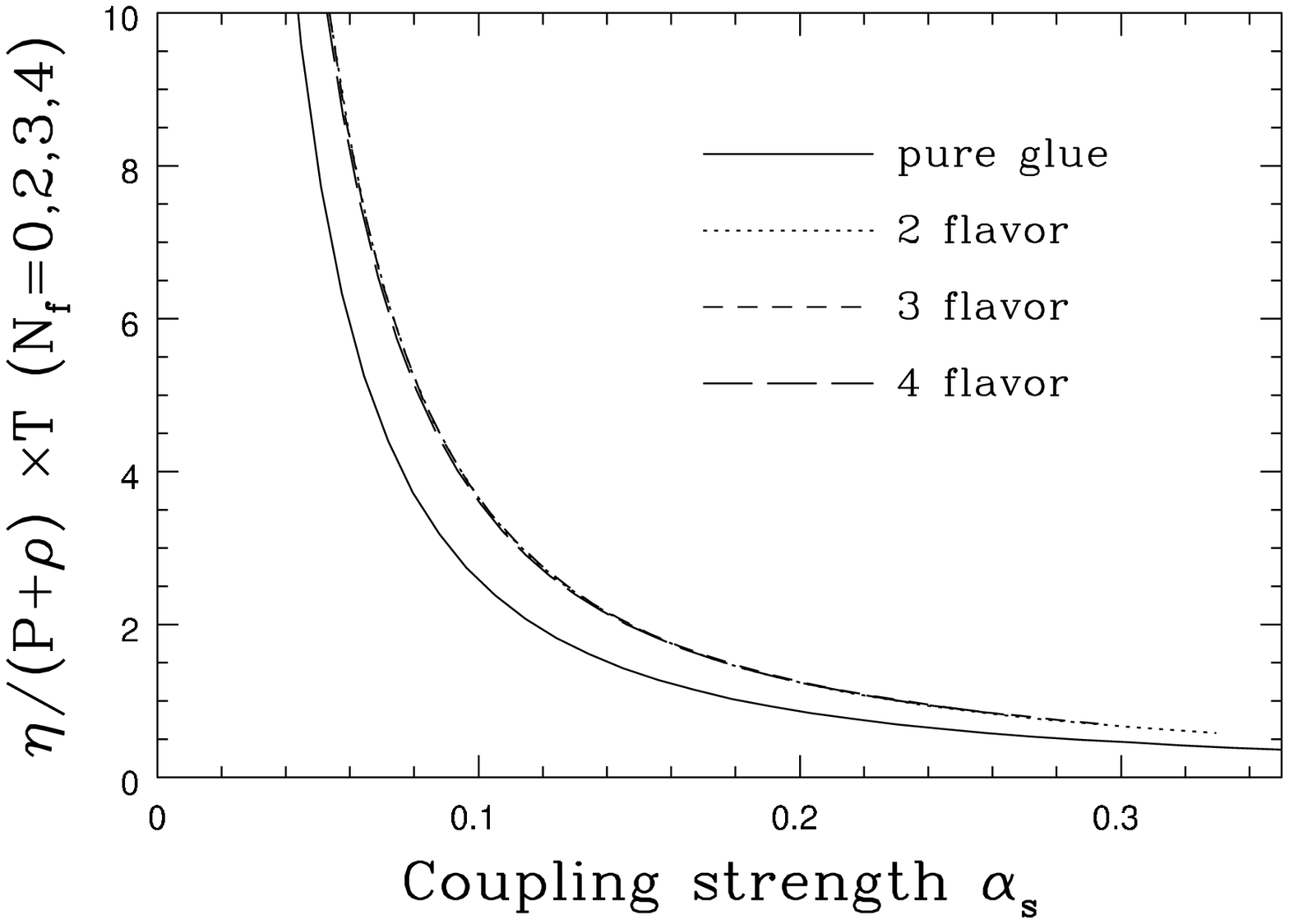}}
\centerline{\includegraphics[angle=0,width=0.40\textwidth]{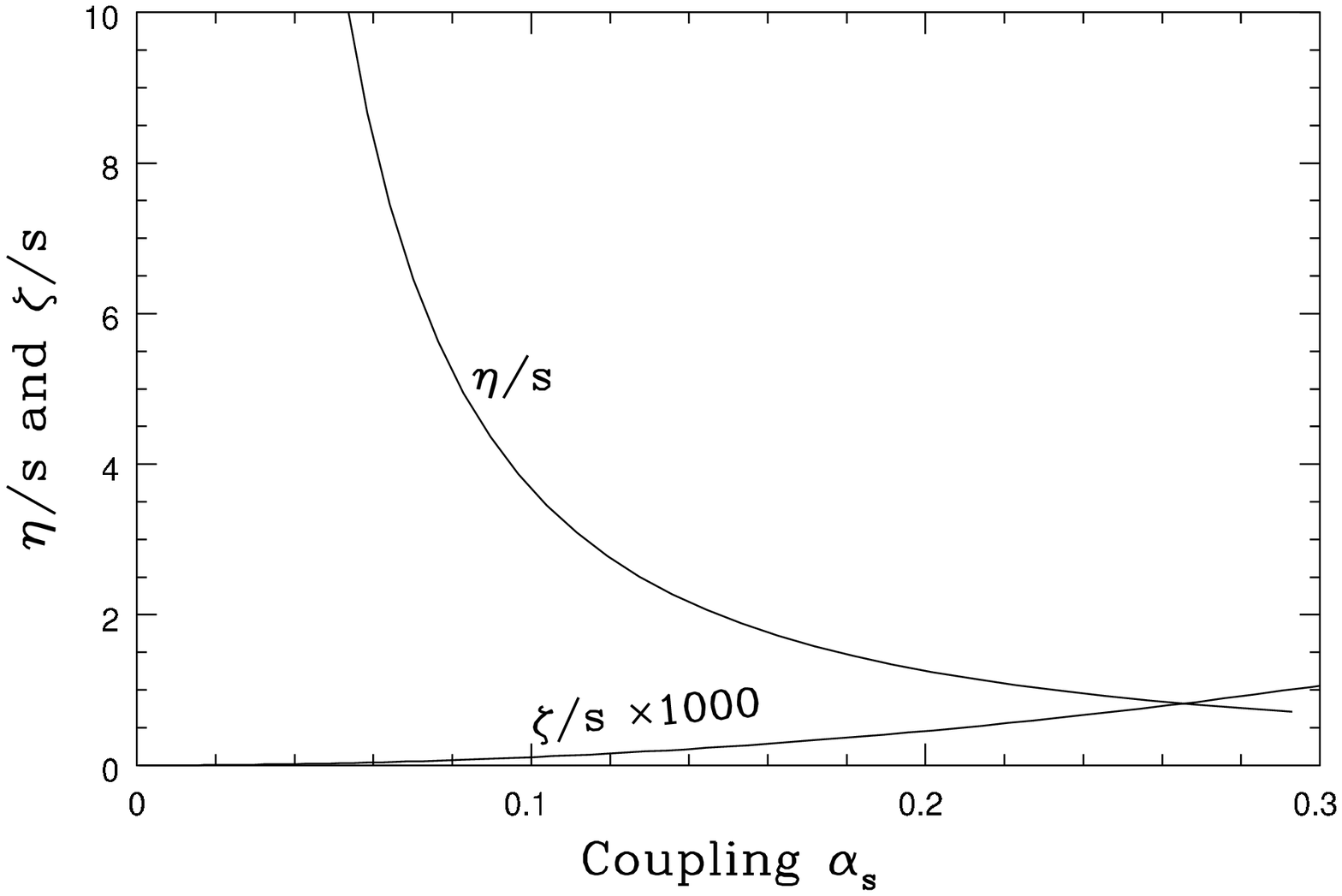}}
\caption{The shear ($\eta$) and bulk ($\zeta$) viscosities scaled by the 
entropy density, in full leading order of perturbation theory
(\cite{Moore:2004kp,Arnold:2003zc} and \cite{Arnold:2006fz}
respectively).}
\label{fig:etazeta-pert}
\end{figure}

Arnold, Moore and Yaffe (AMY) solved these linear integral equations using a
variational approach (a method introduced earlier
in~\cite{Baym:1990uj}).  Numerical results for the shear and bulk
viscosity are displayed in \fig(\ref{fig:etazeta-pert}).  The results
are divided by the entropy density, effectively normalizing them by the
number of degrees of freedom. The ratio $\eta/s$ is about $30\%$ lower
in the pure gauge theory than in $N_{\rm f}=3$ QCD, primarily because
the larger color charge carried by the gluons makes them equilibrate
faster.

Numerically, $\eta/s$ is about 2.0 for $\alpha_s=0.15$, but the
accuracy of the calculation deteriorates for larger values of the
coupling. In the weak coupling regime, the bulk viscosity, of order
$g^4$ in the gauge coupling, turns out to be negligible compared to
the shear viscosity, see the bottom panel
of \fig(\ref{fig:etazeta-pert}).

Recently, the treatment of the inelastic process $gg\leftrightarrow ggg$
in the shear viscosity calculation 
has been scrutinized by Xu and Greiner (XG)~\cite{Xu:2007ns}. 
They showed that at moderate values of $\alpha_s$
this process can numerically dominate the elastic $gg \leftrightarrow gg $ 
scattering due to soft, not necessarily collinear emissions. 
The calculation was revisited very recently 
by Chen et al.~\cite{Chen:2010xk} without relying on the 
soft gluon bremsstrahlung approximation. 
The inclusion of gluon emissions appears
to lower somewhat the ratio $\eta/s$, but a full next-to-leading 
order calculation has not been performed yet in a non-Abelian gauge theory.

\subsection{Perturbative predictions for spectral functions and operator production expansion\la{sec:pt}}

\begin{figure}
\centerline{\includegraphics[width=0.35\textwidth,angle=-90]{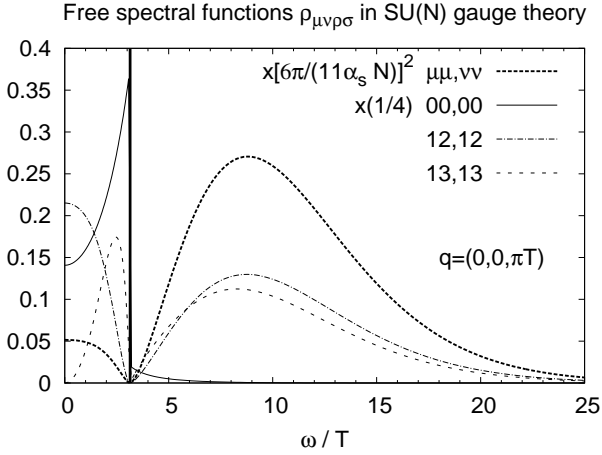}~~~~~~}
\caption{Non-interacting spectral functions in the SU($N$) gauge theory.
More precisely, $q=\pi T$ and the functions plotted are
$\frac{1}{d_AT^4}[\rho(\omega,{\bf q},T)/\tanh(\omega/2T)-\rho(\omega,{\bf q},0)]$
for four different channels. The sound channel $\rho_{00,00}$ admits 
a discontinuity at threshold, $\omega=q$.}
\label{fig:sfq2}       
\end{figure}

We begin with the energy-momentum tensor correlators. 
At frequencies $\omega\gg T$, straightforward perturbation 
theory is expected to apply.
Here we give some results for the SU($\Nc$) gauge theory
in the limit of zero coupling.
In the non-interacting limit, gluodynamics is of course 
a scale-invariant theory. There are thus three independent 
correlators of the stress-energy tensor, which were 
calculated in~\cite{Meyer:2008gt}. 
To describe them, we assume in the following expressions
that $\omega,q>0$, $\bq=(0,0,q)$.
We define the functional 
\ba
{\cal I}[P] &=&
\theta(\omega-q) \int_0^1  dz \frac{P(z)\,
\sinh\frac{\omega}{2T}}{\cosh\frac{\omega}{2T}-\cosh \frac{q z}{2T}}
\\
&+& \theta(q-\omega) \int_1^\infty dz \frac{P(z)\,
\sinh\frac{\omega}{2T}}{\cosh\frac{qz}{2T}-\cosh \frac{\omega}{2T}}.
\nonumber 
\ea
Thus ${\cal I}[P]$ is a function of $(\omega,q,T)$.
Then the spectral functions read, respectively in the 
shear channel, the sound channel and the tensor channel,
\ba
\rho_{13,13}(\omega,q,T) &=& {\txts\frac{d_A   }{8\,(4\pi)^2}}
\; \omega^2(\omega^2-q^2)\;  {\cal I}[1-z^4]\,,
\la{eq:sfqfirst}
\\
\rho_{00,00}(\omega,q,T) &=&  
{\txts\frac{d_A }{4\,(4\pi)^2}}\;q^4 \;  
{\cal I}[(1-z^2)^2]\,,
\\
\rho_{12,12}(\omega,q,T) &=& {\txts\frac{d_A}{32\,(4\pi)^2}}\; 
(\omega^2-q^2)^2 \; {\cal I}[1+6z^2+z^4],~~~~
\ea
where $d_A \equiv \Nc^2-1$.
In addition, the correlator of the trace anomaly reads
\ba
\rho_{\theta,\theta}(\omega,q,T) &=&  \!\!
{\txts\left(\frac{11\alpha_sN_c}{6\pi}\right)^2
\frac{d_A }{4\,(4\pi)^2}}\; (\omega^2-q^2)^2 
\la{eq:sfelem}\\
% {\cal I}([1], \omega, q, T)\,,
&& \Big[-\frac{\omega}{q} \theta(q-\omega)
\,+\, \frac{2T}{q}\,
\log\frac{\sinh(\omega+q)/4T}{\sinh{|\omega-q|/4T}}\Big].
\nonumber
\ea
For $P$ a polynomial, the quantity $I[P]$ can be expressed in terms of
polylogarithms~\cite{Meyer:2008gt}. The correlators of $T_{\mu\nu}$
can be expressed in terms of polylogarithms up to order 5.  The
spectral functions are displayed for $q=\pi T$
in \fig(\ref{fig:sfq2}).  Using expressions of this kind, 
the second-order transport coefficient
$\kappa$ can be extracted via the Kubo formula
(\ref{eq:kubo_kappa}), yielding~\cite{Romatschke:2009ng},
after subtraction of the vacuum contribution,
\be
\kappa = \frac{d_A}{18}T^2 = \frac{5s}{8\pi^2T}.
\la{eq:kappa_freeSUN}
\ee

Even in the non-interacting limit, the functional form of the spectral
functions is relatively complicated, due to the three scales
$\omega,q,T$ involved in the problem. The expressions simplify
significantly when the spatial momentum $q$ is set to zero, for
instance (without summation over the indices $\mu,\nu$)
\ba
\rho_{\mu\nu,\mu\nu}(\omega,0,T) 
&=&  \frac{A_{\mu\nu} ~\omega^4 }{\tanh\quarter \omega \beta} 
     + B_{\mu\nu} \beta^{-4}~\omega \delta(\omega) , ~~~\la{eq:tl}
\la{eq:rho_q.eq.0}
\\
A_{12,12} = A_{13,13} &=& \frac{1}{10} \frac{d_A}{(4\pi)^2},
\\
B_{12,12} =B_{13,13}  &=& \left(\frac{2\pi}{15}\right)^2 ~ d_A ;
\\
A_{\theta} &=&  d_A \left(\frac{11\alpha_sN}{3(4\pi)^2}\right)^2,
\\
B_{\theta} &=& 0.
\ea
The $\delta$-function at the origin corresponds to the fact that
gluons are asymptotic states in the free theory and implies an
infinite viscosity.

Next-to-leading order corrections are known in some cases.  In the
bulk channel $\rho_{\mu\mu,\nu\nu}$, they have been calculated
recently~\cite{Laine:2010tc}, see \eq(\ref{eq:rho_laine}).  In the
shear channel, the $\omega^4$ term in the large-frequency behavior of
the spectral function $\rho_{13,13}(\omega,{\bf 0},T)$ is corrected by
the factor
$\left[1-\frac{5\alpha_s\,N_c}{9\pi}\right]$~\cite{Kataev:1981gr,Pivovarov:1999mr}.

Expressions similar to \eq(\ref{eq:sfqfirst}) were obtained earlier by
Aarts and Martinez Resco for the free spectral functions of
dimension-three quark bilinears~\cite{Aarts:2005hg} in terms of
polylogarithms. In this case polylogarithms up to order 3 appear.  At
$q=0$, the expressions again simplify considerably. For the vector
current, the result is for one flavor (see for instance
~\cite{Karsch:2000gi,Petreczky:2005nh})
\ba
\rho_{00}(\omega,0,T) &=&  -\chi_s \omega \delta(\omega),
\\
\sum_{i=1}^3 \rho_{ii}(\omega,0,T) &=&   \chi_s \<v^2\>\omega \delta(\omega) 
\\
& +& \frac{N_c}{4\pi^2} \theta(\omega-2m) \sqrt{1-\big(\frac{2m}{\omega}\big)^2} ~~
\nn &&
\Big(1+ \half \big(\frac{2m}{\omega}\big)^2 \Big) \omega^2 \tanh\frac{\omega}{4T}.
\nonumber
\ea
The susceptibility and average velocities have
simple expression in the massless and in the heavy-quark limits,
\ba
\chi_s,~\<v^2\> &=& \left\{ \begin{array}{l@{\quad}l@{\qquad}l} 
\frac{N_c}{3}T^2, & 1  & m=0  \\
\frac{4N_c}{T}\Big[\frac{mT}{2\pi}\Big]^{\frac{3}{2}}\,e^{-\beta m} ,& \frac{3T}{m}  & m \gg T.
\end{array}\right.~~~~
\ea

At high frequencies $\omega\gg T$, the leading power in $\omega$ of
the spectral function gets modified by radiative corrections. In this
frequency regime, they are independent of the temperature and can be
calculated in the vacuum. Famously, the vector-current spectral
function $\rho_{ii}$ is then modified by a factor
$(1+\frac{\alpha_s}{\pi})$~\cite{Jegerlehner:2009ry}. The quark number
susceptibility $\chi_s$ also receives calculable corrections at high
temperature, $\chi_s = \frac{\Nc}{3}T^2 (1- \frac{g^2(T)}{2\pi^2})$ in
the massless case \cite{Vuorinen:2002ue}. The broadening of the delta
function in $\rho_{ii}$ into a peak of width on the order of the
inverse relaxation time however requires either an explicit
resummation procedure or a kinetic theory treatment, see section
(\ref{sec:boltz}).

\subsection{Thermal Correlators from the Gauge/Gravity Correspondence\label{sec:adscft}}

In this section we summarize some of the results on spectral functions
that have been obtained in the strong coupling regime using the
holographic principle.  We will restrict ourselves to presenting some
of the spectral functions of the ${\cal N}=4$ SYM theory.  Our main
goal here is to contrast their analytic structure with the analytic
structure found at weak coupling using kinetic theory. For the
interested reader, the method of computing spectral functions based on
the holographic
principle~\cite{Son:2002sd,Policastro:2002se,Policastro:2002tn,Kovtun:2006pf,Teaney:2006nc}
has been reviewed
in~\cite{Son:2007vk,Gubser:2009md,CasalderreySolana:2011us}.

The spectral function of the R-charge correlator at vanishing spatial
momentum, remarkably, can be written in terms of elementary
functions~\cite{Myers:2007we},
\be
\la{eq:N4vc}
\rho(\omega) = \chi_s \,\left(\frac{\omega}{2\pi T}\right)^2
\frac{\sinh\frac{\omega}{2T}}{\cosh\frac{\omega}{2T}-\cos\frac{\omega}{2T}}\,.
\ee
The static susceptibility is given by $\chi_s= \frac{N_c^2T^2}{8}$, and the 
diffusion constant $D=\frac{1}{2\pi T}$ can be read off from this formula 
by using the Kubo formula, \eq(\ref{eq:kubo}).
This function has a series of poles along the diagonals in the lower half 
of the $\omega$ complex plane, 
\be
\omega_p = n\cdot 2\pi T (\mp 1-i).
\ee

\begin{figure}
\centerline{\includegraphics[angle=0,width=0.40\textwidth]{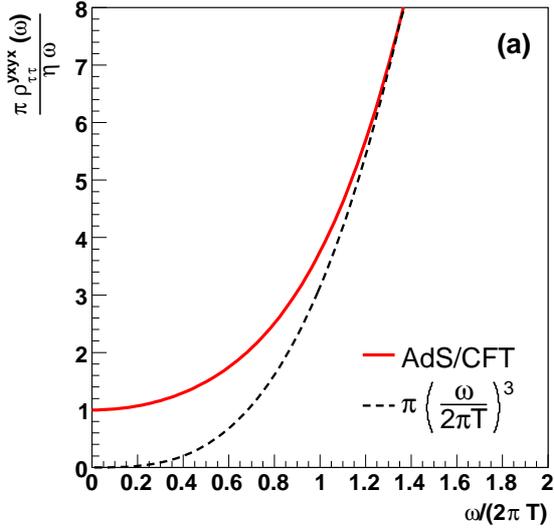}}
\caption{Spectral function of shear stress at strong coupling in the ${\cal N}=4$ 
SYM theory at vanishing spatial momentum. The dashed curve shows the zero temperature
result (\ref{eq:symT.eq.0}) normalized by the same factors. Due to a non-renormalization
theorem in these channels, the zero temperature spectral densities in the free and interacting
theories are equal~\cite{Freedman:1998tz,Chalmers:1998xr}. 
At finite temperature the kinetic theory peak observed in \fig(\ref{fig:rho1212-kin}, $\bk=0$ curve) 
does not exist in the strongly interacting theory. Reproduced from~\cite{Teaney:2006nc}.}
\label{fig:rho1212-teaney}
\end{figure}

The shear channel spectral function ($\rho(\omega)/\omega$) displayed
in \fig(\ref{fig:rho1212-teaney})~\cite{Teaney:2006nc,Kovtun:2006pf}
smoothly interpolates between the vacuum spectral function
\be
\rho^{\rm SYM}_{12,12}(\omega,q=0,T=0)=\frac{\Nc^2}{64\pi^2}\omega^4
\la{eq:symT.eq.0}
\ee
at large frequencies and a finite value at $\omega=0$, which yields
the shear viscosity $\eta=\frac{\pi}{8}\Nc^2 T^3$ via the Kubo formula
(\ref{eq:Kubo-eta}). Combining this with the entropy density of $s
= \frac{\pi^2}{2}\Nc^2T^3$, one obtains the ratio $\eta/s=1/4\pi$
(see \eq(\ref{eq:etasSYM}) in the Introduction).

Numerically no particular structure or excitation is visible in the
spectral function at any frequency.  This is in stark contrast with
the peak structure predicted at the origin by the Boltzmann equation,
see \fig(\ref{fig:rho1212-kin}). One concludes from figure
(\ref{fig:rho1212-teaney}) that the conformal plasma of the ${\cal
N}=4$ SYM theory does not admit quasiparticles. There is no separation
between the transport time scale $\tau_\Pi$ and the thermal scale $\pi
T$.  In fact, the second-order transport coefficients were
calculated in the strongly coupled ${\cal N}=4$ SYM
theory~(\cite{Baier:2007ix}, see also \cite{Natsuume:2007ty} for $\tau_\Pi$), 
with the result
\ba
\tau_\Pi &=& \frac{2-\log 2}{2\pi T}, 
\la{eq:tauPiSYM}
\\
\kappa &=& \frac{\eta}{\pi T} = \frac{s}{4\pi^2 T}.
\la{eq:kappaSYM}
\ea

\begin{figure}
\centerline{\includegraphics[angle=0,width=0.45\textwidth]{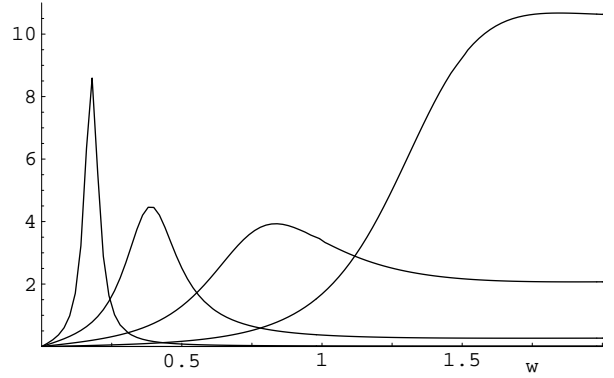}}
\caption{Spectral function of the energy density at spatial momenta
obtained from the AdS/CFT correspondence for the ${\cal N}=4$ SYM
theory.  The quantity displayed is $\pi\frac{\rho(\omega)}{e+p}$    % $2\rho/(\pi d_A T^4)$ 
in infinite-coupling, large-$N_c$ SYM theory, computed by AdS/CFT
methods.  The $x$-axis variable is {\tt w}$\,=\omega/2\pi T$ and the curves
correspond to $q/2\pi T=0.3,0.6,1.0$ and 1.5.  Figure reproduced
from~\cite{Kovtun:2006pf}.}
\label{fig:sf-sound-AdSCFT}
\end{figure}

The sound channel spectral function, namely the spectral function for
the energy density $T_{00}$ at finite spatial momentum, is diplayed
in \fig(\ref{fig:sf-sound-AdSCFT}). Comparison
with \fig(\ref{fig:sf-sound-hydro}) shows that in this strongly
coupled theory, hydrodynamics provides a good description of the
spectral function up to momenta as large as $\pi T$.  In a
weakly-coupled theory, this range of validity is parametrically smaller.

Assuming the holographic principle to hold for other theories as well,
non-conformal theories can also be studied by these methods. In the
past few years, a program has been undertaken to tune the properties
of the `bulk' theory to reproduce the thermodynamic properties of QCD
on the `boundary', and then predict those properties such as transport
coefficients that are difficult to extract from first
principles~\cite{Gursoy:2007cb,Gubser:2008yx,Gursoy:2010fj}. These
models can give qualitative insight into whether the shape of the
spectral functions displayed in \fig(\ref{fig:rho1212-teaney}) and
(\ref{fig:sf-sound-AdSCFT}) are robust again modest modifications of
the theory. A longer-term goal could be to understand the crossover
between the functional form observed at weak coupling and that found
at strong coupling.

We remark that in the `AdS/QCD' models the shear viscosity to entropy
density ratio remains $1/4\pi$ at all temperatures, which shows that
the breaking of conformal invariance is not of the most general
form. A bound that is found to be observed by these
theories~\cite{Gursoy:2010fj} is Buchel's conjectured
bound~\cite{Buchel:2007mf},
\be
\frac{\zeta}{\eta} \geq 2 \left(\frac{1}{3} - c_s^2  \right).
\ee
It implies that the bulk viscosity must become at least comparable to
the shear viscosity when the speed of sound becomes small, as happens
for instance at certain phase transitions.

\subsection{Thermal Sum Rules\label{sec:sumrules}}

\begin{figure}
\centerline{\includegraphics[angle=0,width=0.35\textwidth]{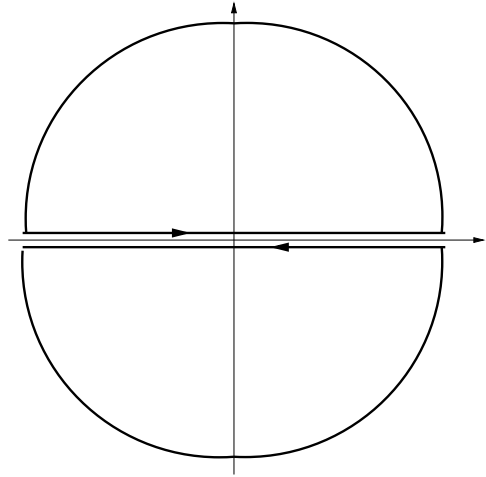}}
\caption{Contour chosen in the derivation of thermal sum rules, 
see \eq(\ref{eq:general-sumrule}).}
\label{fig:contour}
\end{figure}

We briefly remind the reader of the derivation of sum rules.  The
first step is to derive a dispersion
relation~\cite{Weinberg:1995mt}. For this purpose it is useful to
introduce the function
\be
{\cal G}(\omega) = \left\{ 
\begin{array}{l@{\quad}l} G_R(\omega) \equiv 
\int_0^\infty \ud t \, e^{i\omega t}\, G(t)  &  \im(\omega)>0 \\ 
G_A(\omega) \equiv
-\int_{-\infty}^0 \ud t \, e^{i\omega t}\, G(t) & \im(\omega)<0
\end{array}
\right.
\ee
It is analytic in $\omega$ everywhere except along the real axis.  We
now apply Cauchy's theorem for a contour ${\cal C}$ which is the sum of
two large semi-circles, as depicted on \fig(\ref{fig:contour}).  In
order for the contributions at infinity to vanish, the theorem is
applied to ${\cal G}(\omega)/P(\omega)$, where $P(\omega)$ is a
polymonial of sufficiently high order $n$ with simple zeros at
$\omega=\omega_\nu$. The contour integral then yields
\be
\frac{1}{2\pi i} \oint_{\cal C} \frac{{\cal G}(z) \ud z}{(z-\omega)P(z)}
= \frac{{\cal G}(\omega)}{P(\omega)} + 
\sum_\nu \frac{{\cal G}(\omega_\nu)}{(\omega_\nu-\omega)P'(\omega_\nu)},
\la{eq:contour}
\ee
By construction, the contribution of the semi-circles vanishes in the
limit of infinite radius. Now recalling the definition of the
spectral function, the left-hand side is equal to 
\be
\int_{-\infty}^\infty \frac{\ud z \, \rho(z)}{(z-\omega)P(z)}.
\ee
Therefore, noticing that ${\cal G}(\omega)=G_R(\omega)$ for $\im(\omega)>0$, 
we obtain
\be
G_R(\omega) = P(\omega) \int_{-\infty}^\infty \frac{\ud z \,\rho(z)}{(z-\omega)P(z)} 
+ Q(\omega),
\la{eq:general-sumrule}
\ee
where 
\be
Q(\omega) \equiv -P(\omega)\sum_\nu \frac{{\cal G}(\omega_\nu)}{(\omega_\nu-\omega)P'(\omega_\nu)}.
\ee
is a polynomial of degree $n-1$.  In practice it means that $n$
constants have to be calculated before formula
(\ref{eq:general-sumrule}) determines the retarded correlator For
instance, for the vector correlator, \eq(\ref{eq:Pidisprel}), one
subtraction is sufficient to make the dispersion integral converge. We
reserve the name sum rule to the case where $G_R(\omega)$ can be
calculated in an independent way, for instance in terms of
thermodynamic potentials.  \eq(\ref{eq:general-sumrule}) can then be
interpreted as a global constraint on the spectral function
$\rho(\omega)$, which may help to parametrize it in a consistent way.

Since the derivation of a sum rule involves integrating the spectral
function over all frequencies, it is necessary to know its large
frequency behavior in order to ensure the integral's UV convergence.
This leads us to the subject of the operator product expansion (OPE).
In thermal field theory, the spectral functions typically rise at
large frequencies with a positive power of frequency.  The simplest
subtraction to implement is then to take the difference between two
temperatures. This reduces the asymptotic power by four units, due to
the absence of gauge invariant operators of lower dimension (chiral
symmetry implies that the operator $\bar\psi\psi$ appears accompanied
by a power of the quark mass).

Correlation functions deep in the Euclidean region admit an
operator-product expansion. Through the dispersion relation, this
leads to a large frequency expansion for the spectral
function~\cite{Huang:1994vd}.  Leading-order results for the OPE of
current and energy-momentum tensor correlators were obtained
in~\cite{CaronHuot:2009ns}.  Recently, a full two-loop calculation of
the correlator of $\theta\equiv T_{\mu\mu}$ has been
performed~\cite{Laine:2010tc} in Yang-Mills theory (see also the
closely related work~\cite{Laine:2010fe}).  In addition to giving the
radiative corrections to the leading $\omega^4$ power, this
calculation provides the Wilson coefficients of the $\theta$ and
$\theta_{00}\equiv T_{00}-\frac{1}{4}\theta$ operators in the operator
product expansion of the trace anomaly correlator. We quote the
$\omega\gg \pi T$ result,
\ba
 \frac{\rho_\theta(\omega)}{4c_\theta^2 } & =
 & \frac{d_A \omega^4}{(4\pi)^2} \,\tilde\gamma_\rmi{$\theta;\!
 1$}(\omega)
\\ &&
 \; - \; {2(e+p)} 
 \,\tilde\gamma_\rmi{$\theta;e+p$}(\omega)
~~~\la{eq:rho_laine}\nn
&&
 \; - \; 2 (e-3p)
 \,\tilde\gamma_\rmi{$\theta;e-3p$}(\omega)
 \; + \; \rmO\biggl( \frac{T^6}{\omega^2} \biggr)
 \;, \nn \la{rho_theta} 
\tilde\gamma_\rmi{$\theta;\! 1$}(\omega) & = & 
 g^4 + \frac{g^6\Nc}{(4\pi)^2}
 \biggl( \frac{22}{3}\ln\frac{\bmu^2}{\omega^2} + \frac{73}{3} \biggr) 
\\ &&  + \rmO(g^8)
 \;, \la{tgamma_theta_unit} \nn
 \tilde\gamma_\rmi{$\theta;e+p$}(\omega) & = & 
 \frac{11g^6 \Nc}{(4\pi)^2} + \rmO(g^8)
 \;, \la{tgamma_theta_ep} \\[2mm] 
 \tilde\gamma_\rmi{$\theta;e-3p$}(\omega) & = & 
 g^4 + \rmO(g^6)
 \;, \la{tgamma_theta_emtp} % \\[2mm] 
\ea
where $c_\theta=-\frac{b_0}{2}-\frac{b_1}{4} g^2$, $b_0 =
\frac{11\Nc}{3(4\pi)^2}$ and $b_1 = \frac{34\Nc^2}{3(4\pi)^4}$, and
$g^2$ is the dimensionless renormalized coupling constant, evaluated
at the $\msbar$ scheme renormalization scale $\bmu$ ($ \mu^{2} =
\bmu^{2} {e^{\gammaE}}/{4\pi} $).  Expressions such as
\eq(\ref{eq:rho_laine}) are very useful to constrain the form of the
spectral function at high frequencies, particularly since the
thermodynamic potentials $e$ and $p$ can be evaluated
non-perturbatively.  In particular, if the vacuum correlator is
subtracted from the thermal correlator, then the leading asymptotic
behavior is determined by two linear combinations of energy density
and pressure. This has recently been exploited to constrain the
spectral function $\rho_\theta$~\cite{Meyer:2010ii}, at which time
however only the leading order behavior of the Wilson coefficients was
known~\cite{CaronHuot:2009ns}.

The calculation shows that (a) the scalar and pseudoscalar correlators
do not have contact terms (the correlators vanish in the limit
$\omega\to\infty$) and (b) the integral of $\Delta\rho/\omega$ is
UV-finite.  Using these facts, one can derive a sum rule in the
SU($\Nc$) Yang-Mills theory~\cite{Ellis:1998kj,Kharzeev:2007wb}.
First, the following expression
\be
G_{E,\theta}^{(0)}(T) -G_{E,\theta}^{(0)}(T=0)
= T^5\frac{\partial}{\partial T} \frac{e-3p}{T^4}\,,
\la{eq:srE}
\ee
is obtained for the $(\omega,\bq)=0$ Euclidean correlator,
where $e$ and $p$ are the energy density and pressure of the 
finite-temperature system. 
We now convert identity (\ref{eq:srE}) into a
sum rule for the bulk spectral function
with the goal to constrain the latter.
Combining equations 
(\ref{eq:srE})  and (\ref{eq:GErho}), we obtain
\be
\int_{-\infty}^\infty  \frac{d\omega}{\omega} 
[\rho_{\theta}(\omega,T) -\rho_\theta(\omega,0)] = 
T^5 \partial_T \frac{e-3p}{T^4} \,.
\la{eq:srtheta}
\ee

Next, we separate an infrared contribution of the form $\omega\delta(\omega)$ in
$\rho_\theta$ from the rest of the spectral
function~\cite{Romatschke:2009ng}.  Starting from
\eq(\ref{eq:ImGRsound}), we recall that for any smooth function such
that $\int_{-\infty}^\infty f(x)dx=1$,
$\frac{1}{\epsilon}f(x/\epsilon)$ provides a representation of the
delta function. Here we will exploit this fact for the function
\be
f(x) = \frac{1}{\pi(b^2-1)}\left[
\frac{x^4}{(x^2-b^2)^2+x^2}-1\right]
\ee
and $b=c_s/(\Gamma_s k)$ and $\epsilon = \Gamma_s k^2$.
In this way one finds that, in the sense of distributions, 
\be
\la{eq:rho3333}
\lim_{{\bk}\to0}
\frac{\rho_{33,33}(\omega,{\bk},T)}{\omega} =
\frac{e+p}{\pi} \Gamma_s  + (e+p)c_s^2\delta(\omega) + \dots
\ee
In the shear channel on the other hand, 
\ba
\la{eq:delta1313}
\frac{\rho_{13,13}(\omega,{\bk},T)}{\omega}
&=& \frac{\eta}{\pi} \frac{\omega^2}
       {\omega^2+\left(\frac{\eta k^2}{e+p}\right)^2}
\\
&\stackrel{{\bk}\to0}{\sim}& \frac{\eta}{\pi} 
-\frac{\eta^2 k^2}{e+p}\delta(\omega)
\nonumber
\ea
and the delta function is suppressed by a power of ${\bk}^2$.
Thus combining the shear and sound channels,
\ba
\frac{1}{9}\frac{\rho_{ii,kk}(\omega,{\bf 0},T)}{\omega}
&=& \frac{\rho_{33,33}(\omega,{\bf 0},T)}{\omega}
 - \frac{4}{3}  \frac{\rho_{13,13}(\omega,{\bf 0},T)}{\omega}
\nn
&=& \frac{\zeta}{\pi}+(e+p)c_s^2\delta(\omega)+\dots
\ea
Using the exact relations
\ba
\rho_{00,00}(\omega,{\bf 0},T) 
&=& \frac{e+p}{c_s^2} \omega \delta(\omega)\,,
\\
\rho_{00,kk}(\omega,{\bf 0},T) 
&=& -3(e+p) \omega \delta(\omega)\,,
\ea
one finds
\ba\la{eq:separ}
\frac{\rho_\theta(\omega,{\bf 0},T)}{\omega}
% &\equiv& \frac{\rho_{\mu\mu,\nu\nu}(\omega,{\bf 0},T)}{\omega}
% \\
&=& \frac{9\zeta}{\pi} 
+ \frac{e+p}{c_s^2}(3c_s^2-1)^2\delta(\omega)
+\dots
\nonumber
\ea
We introduce the operator
\be
{\cal O}_\star = L^{-3/2}
\int d^3{\bx}\,\left(T_{\mu\mu}(x)+(3c_s^2-1)T_{00}(x)\right)\,.
\la{eq:Ostar}
\ee
Its spectral function, which we denote by $\rho_\star$,
satisfies
\be
\rho_{\theta}(\omega,{\bf 0},T)
= \rho_{\star}(\omega,{\bf 0},T)
+ \frac{e+p}{c_s^2}(3c_s^2-1)^2\,\omega\,\delta(\omega)
\la{eq:separ2}
\ee
and, in view of \eq(\ref{eq:separ}), 
is free of the delta function at the origin.

We are now ready to turn \eq(\ref{eq:srtheta}) into a sum rule for
$\rho_\star$, the spectral function of the operator ${\cal O}_\star$
(\ref{eq:Ostar}), which is smooth at the origin.  Using
\eq(\ref{eq:separ2}), one finds
\be
2\int_0^\infty \frac{d\omega}{\omega}
[\rho_{\star}(\omega,T)
 -\rho_{\star}(\omega,0)]
= 3(1-3c_s^2)(e+p) -4(e-3p)\,.
\la{eq:bsr1}
\ee
\eq(\ref{eq:srtheta}) was already obtained in~\cite{Kharzeev:2007wb},
but the presence of the $\omega\delta(\omega)$ term in $\rho_\theta$
was missed.  In~\cite{Romatschke:2009ng}, a finite spatial momentum
${\bk}$ was used as an infrared regulator and the contribution of
the zero-frequency delta function correctly identified for the first
time.  If one insists on expressing \eq(\ref{eq:bsr1}) in terms of
$\rho_\theta$, one should strictly write $\lim_{\epsilon\to0}
\int_{-\infty}^\infty \frac{\omega \,d\omega}{\omega^2+\epsilon^2}
\Delta\rho_\theta(\omega,T) = 3(1-3c_s^2)(e+p) -4(e-3p)$.

In the shear channel, a similar sum rule applies in the Yang-Mills
theory, as first pointed out in~\cite{Romatschke:2009ng}.
It was later also studied in lattice regularization~\cite{Meyer:2010gu}.
There it was given in the form
\ba
\la{eq:shear-sr}
\int_{-\infty}^\infty \frac{\ud\omega}{\omega}\,
\Delta\rho_{12,12}(\omega,{\bp=0},T)
&=& 
\frac{2}{3} e(T)
\\
&& -\lim_{\omega\to\infty}\Delta G(\omega,T)\,,
\la{eq:conti-sr}
\nonumber
\ea
where
\ba
\la{eq:defDG}
&&\Delta G(\omega,T) \equiv 
\int \ud^4x\, e^{i\omega x_0} \; \Big\<   -  {\txts\frac{2}{3}} T_{00}(x)\,T_{00}(0)~~
\\
&& \qquad + {\txts\frac{1}{4}} (T_{11}(x)-T_{22}(x))(T_{11}(0)-T_{22}(0))
\Big\>_{T-0}.
\nonumber
\ea
was proved to be finite.  The form (\ref{eq:shear-sr}) is equivalent
to the form given in~\cite{Romatschke:2009ng}.  The sum rule remains
incomplete, because a finite, but yet undetermined contribution of the
trace anomaly can affect $\Delta G(\omega\to\infty,T)$.  The
determination of its OPE coefficient requires a two-loop calculation.
To our knowledge, this contact term can not be determined by Ward
identities.

Another important example is the difference of thermal vector and
axial-vector current correlators, which generalizes the Weinberg sum
rules of the sixties.  The difference of the corresponding spectral
functions obeys sum rules $\forall \bk$ in the chiral
limit~\cite{Kapusta:1993hq},
\ba
\int_0^\infty \frac{\ud\omega \,\omega}{\omega^2-\bk^2}
\,\left(\rho^V_L(\omega,\bk) - \rho^A_L(\omega,\bk)\right) &=& 0.
\la{eq:sr-av1}
\\
\int_0^\infty {\ud\omega}\,{\omega}
\,\left(\rho^V_L(\omega,\bk) - \rho^A_L(\omega,\bk)\right) &=& 0.
\la{eq:sr-av2}
\\
\int_0^\infty {\ud\omega}\,{\omega}
\,\left(\rho^V_T(\omega,\bk) - \rho^A_T(\omega,\bk)\right) &=& 0.
\la{eq:sr-av3}
\ea
A remarkable point is that the difference of the vector and
axial-vector spectral functions falls off rapidly at large
frequencies. In the OPE analysis, the correlators first differ
by a four-quark operator, which means that the difference of spectral functions
falls off as $\omega^{-4}$ at high frequencies.

\subsection{Correlators in Finite-volume\label{sec:finivol}}

All the analytic treatments covered in this section are performed in
infinite spatial volume. However, the Monte-Carlo calculations
reviewed in section (\ref{sec:latt}) are necessarily performed in
finite volume $V=L^3$.  The boundary conditions are normally taken to
be periodic.  As is manifest from the formal spectral expression
(\ref{eq:rho}) of the spectral density, the latter is a distribution,
which consists of a collection of delta functions with a weight given
by the matrix elements of the current under investigation between
energy eigenstates. The spectral weight at the origin $\omega=0$ is
thus determined by transition matrix elements of the current between
states whose energy differs only infinitesimally. Symmetry
considerations thus do not result in as stringent restrictions as at
zero temperature.  For instance, for any energy eigenstate above the
two-particle threshold, there typically exist energy eigenstates whose
energy differs only by an amount of order $1/L$ and which are
connected to it by a non-vanishing matrix element.

Clearly one cannot literally apply the Kubo formula to the
finite-volume spectral function, since this would given a undefined
value. The situation is similar in this respect to the discussion of
the spectrum of the Dirac operator. In that context, the spectral
density at threshold in infinite volume is equal to the chiral
condensate $-\<\bar\psi\psi\>$. In finite volume, the Dirac spectrum
is discrete and therefore it is necessary to send the volume to
infinity (before setting the quark mass $m$ to zero) in order to
extract the value of the chiral condensate. However, if the density is
given by
\be
\rho(\lambda,m) = \sum_{k=1}^\infty \delta(\lambda-\lambda_k),
\ee
the difficulty can be dealt with by working with the integrated
spectral density~\cite{Giusti:2008vb},
\be
\nu(M,m) = V \int_{-\Lambda}^\Lambda \ud\lambda\, \rho(\lambda,m),
\qquad \Lambda = \sqrt{M^2-m^2}.
\ee
A finite spectral density manifests itself by a linearly rising
$\nu(M,m)$ as a function of $M$ on a scale much larger than the
typical eigenvalue interspacing, but much smaller than the scale on
which a curvature arises. This behavior is readily seen in simulations
of QCD in large but finite volume~\cite{Giusti:2008vb}.
In the case at hand  of thermal spectral functions of gauge
invariant operators, a natural way to deal with the singular nature of
the spectral functions is to work with the coordinate-space retarded
correlator $G^{AB}(t)$. For $B=A=A^\dagger$, inverting the Fourier
transform (\ref{eq:rhodef}) yields
\be
G(t) = 2\int_0^\infty \ud\omega\, \sin(\omega t) \, \rho(\omega).
\la{eq:Gt}
\ee

Consider for illustration the quantum Hamiltonian of a free particle coupled to an assembly of 
harmonic oscillators ($[q,p]=i$, $[q_n,p_m]=i\delta_{nm}$) \cite{Kac,Gardiner},
\be
\hat H = \frac{\hat p^2}{2m} + \half \sum_n\{ (\hat p_n-\kappa_n \hat q)^2+ \omega_n^2 \hat q_n^2\}.
\ee
The Heisenberg equations of motion $\dot {\hat Y} = i[\hat H,\hat Y]$ are easily obtained
(first for the annihilation operators $\hat a_n\equiv \frac{\omega_n \hat q_n + i\hat p_n}{\sqrt{2\omega_n}}$, 
then for $\hat p$ and $\hat q$) and cast in the form
\ba
\dot {\hat q}(t) &=& \hat p(t) / m \,,
\\
\dot {\hat p}(t) &=& \hat \xi(t) -\int_0^t \ud s\; f(t-s) \;\; \dot \hat\!\! q(s) \la{eq:Langevin2} \,,
\\
\hat \xi(t) &\equiv &  \sum_n \kappa_n \frac{\omega_n}{2i} 
\Big[ \hat a_n(0) e^{-i\omega_n t} - \hat a_n^\dagger(0) e^{i\omega_n t} \Big],
\\
f(t) &=& \sum_n \kappa_n^2 \cos\omega_n t.
\ea
The operator equations for $\hat q$ and $\hat p$ resemble formally the classical Langevin
equation with a memory kernel given by $f(t)$.  
The commutator-defined correlator of the operator $\hat \xi(t)$, which
plays the role of the random force term, is given by
\be
G^{\xi\xi}(t) \equiv i[\xi(t),\xi(0)] = - \frac{df(t)}{dt}
\ee
and, based on \eq(\ref{eq:Gt}), the spectral function by
\be
\frac{2\rho(\omega)}{\omega} = \sum_n \kappa_n^2 \delta(\omega-\omega_n).
\ee
It has the expected form for a spectral function (\eq\ref{eq:rho}).

Imagine now that the frequencies $\omega_n$ correspond to differences of 
photon or phonon modes in a box of size $L$,
\be
\omega_n = 2\pi n /L, \qquad n\in\mathbb{Z},
\ee
and that 
\be
\kappa_n^2 = \frac{1}{L} \frac{2\Gamma \chi}{\Gamma^2+\omega_n^2},
\ee
then one finds
\be
G^{\xi\xi}(t) = -\chi \Gamma \frac{\sinh\Gamma(t-L/2)}{\sinh\Gamma L/2}\,.
\la{eq:Gxixi-fv}
\ee
In the limit of infinite volume, linear response predicts that 
the retarded correlator falls off exponentially 
and the spectral function has the form of a Lorentzian\footnote{Non-linear 
effects in general dominate at large times with 
$G(t) \sim t^{-3/2}$. The cross-over time from the exponential to the power-law 
behavior is however parametrically 
of order $N_c^2$ in SU($N_c$) gauge theories~\cite{Kovtun:2003vj}.},
\ba
G^{\xi\xi}(t)  &=& 
 \chi \Gamma e^{-\Gamma |t|}
 \la{eq:Gxixi-L}
\\
\rho^{\xi\xi}(\omega) &=& \frac{\chi}{\pi} \frac{\Gamma\omega}{\Gamma^2+\omega^2}.
\la{eq:rhoxixi-L}
\ea
The lesson we draw from this example is that in infinite volume, the
parameters of a transport peak could be read off the asymptotic
$t\to\infty$ behavior of $G^{\xi\xi}(t)$.  In finite volume on the
other hand, \eq(\ref{eq:Gxixi-fv}), these parameters have to be read
off for $t\ll L$, but large enough for contributions from higher
frequencies to have died off. The window of real-time in which the
transport properties can be extracted thus only exists for a
sufficiently large volume.

To conclude with this example, when $\Gamma\to\infty$,
$\rho(\omega)= \frac{\chi}{\pi} \frac{\omega}{\Gamma}$, the memory
function becomes a delta function, $f(t) = {2\gamma}\delta(t)$,
$\gamma\equiv\frac{\chi}{\Gamma}$ and \eq(\ref{eq:Langevin2}) takes
the form of the memory-free Langevin equation~\cite{Gardiner}.  If one
then attributes a thermal distribution to the occupation numbers of
the photonic energy levels, $\<\hat a_n^\dagger \hat a_m\> = \delta_{nm}
n_B(\omega_n)$, $n_B(\omega)=(e^{\beta\omega}-1)^{-1}$, one recovers
familiar results (see section \ref{sec:hq}) such as
\be
\frac{1}{2m}\<\hat p^2(t)\> \stackrel{t\to\infty}{=} 
\frac{\gamma}{\pi m}\int_0^\infty \ud\omega\, 
\frac{\omega n_B(\omega)}{(\gamma/m)^2+\omega^2} + \dots
\ee
where the dots represent a temperature-independent term.
For $\gamma/m\to0$, the first term reproduces the equipartition result $\half k_BT$.

\begin{figure}
\centerline{\includegraphics[width=0.35\textwidth,angle=-90]{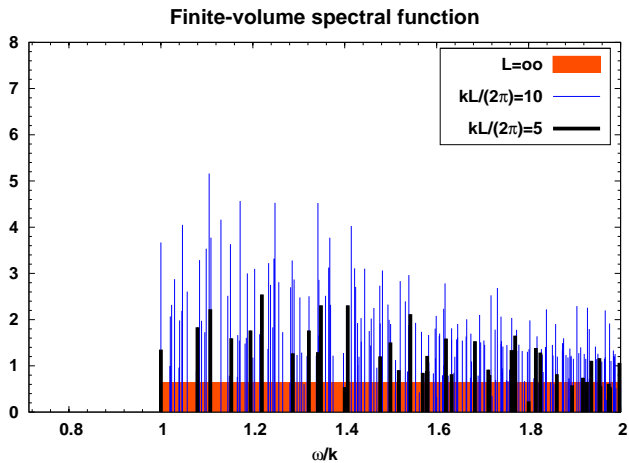}}
% \centerline{\includegraphics[width=0.45\textwidth]{FIGS/}}
\caption{Comparison of spectral functions in finite and infinite volume, in a 
free (3+1)-dimensional massless scalar field theory at zero temperature. The infinite-volume
spectral function is a step function at threshold, $\omega=|\bk|$. Two different
volumes are shown.}
\label{fig:fft-illustrn}       
\end{figure}

\begin{figure}
\centerline{\includegraphics[width=0.35\textwidth,angle=-90]{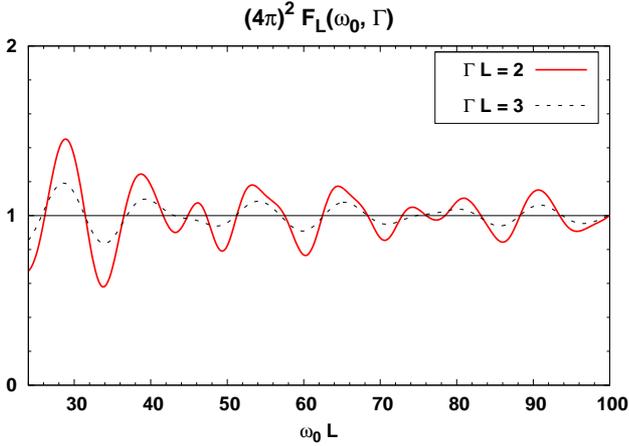}}
% \centerline{\includegraphics[width=0.45\textwidth]{FIGS/}}
\caption{The function $F_L$ as given by \eq(\ref{eq:FL}), normalized to converge
to 1 in the limit $L\to\infty$. It represents the convolution of the 
finite-volume spectral function with a Gaussian of width $\Gamma$ centered at $\omega_0$.
At fixed $L$, increasing the resolution in frequency comes at the cost of enhancing 
the finite-volume effect.
}
\label{fig:FL}       
\end{figure}

To familiarize oneself with the infinite-volume limit further it may
be useful to look at a specific example in 3+1 dimensions.  Consider a
free massless scalar field theory at $T=0$ and the correlator of the
mass operator,
\be
G_E(t,\bk) = \int \ud\bx \, e^{i\bk\cdot\bx}\, \<\phi^2(t,\bx) \,\phi^2(0)\>.
\ee
One easily finds the expressions for the Euclidean correlator
\be
G_E(t,\bk) = \frac{e^{-|\bk|t}}{(4\pi)^2 t}\,,
\ee
and the associated spectral function,
\be
\rho(\omega,\bk) =  \frac{1}{(4\pi)^2}\, \theta(\omega-|\bk|) .
\ee
In a finite periodic box on the other hand,
\be
G_E(t,\bk) = \frac{1}{L^3}\sum_{\bp} \frac{e^{-|\bp|t}}{2|\bp|}\; 
                  \frac{e^{-|\bk-\bp|t}}{2|\bk-\bp|}\,,
\ee
and the corresponding spectral function
\be
\rho(\omega,\bk) = \frac{1}{L^3}\sum_{\bp} \frac{\delta(\omega-|\bp|-|\bk-\bp|)}{4|\bp|\,|\bk-\bp|}
\ee
is, as expected, a collection of delta functions accompanied by a weight factor.
The spectral functions are displayed in \fig(\ref{fig:fft-illustrn}) for illustration.
The `spikes' in the finite-volume spectral function have been drawn with a height
proportional to the weight of the corresponding delta function.
It is clear that the finite-volume spectral function does not converge
towards the infinite spectral function point-by-point in $\omega$, but rather
in the sense of distributions. In infinite volume the convolution
of the spectral function with a Gaussian `resolution function',
\be
F(\omega_0,\Gamma) \equiv \int_0^\infty \ud\omega\, \rho(\omega,\bk=0) 
\frac{e^{-(\omega-\omega_0)^2/2\Gamma^2}}{\sqrt{2\pi}\Gamma}\,,
\la{eq:Fdef}
\ee
amounts to $\frac{1}{4\pi^2}$, for $\Gamma\ll \omega_0$.
In finite volume, on the other hand, the corresponding integral
amounts to (using the Poisson summation formula)
\ba
F_L(\omega_0,\Gamma) &=& 
% &\stackrel{\Gamma L \ll \omega_0/\Gamma}{\simeq}& 
\frac{1}{(4\pi)^2}\sum_{\boldsymbol{m}\in \mathbb{Z}^3}
\frac{2\sin \frac{\omega_0|\boldsymbol{m}|L}{2}}{\omega_0L|\boldsymbol{m}|} 
\la{eq:FL}\\
&& \exp\left(-\frac{\boldsymbol{m}^2L^2\Gamma^2}{8}\right),
\quad \Gamma\ll \omega_0,\;\Gamma^2 L \ll \omega_0.
\nonumber
\ea
Thus one sees that on a resolution scale $\Gamma$ which is not too
small compared to $1/L$, $F_L(\omega_0,\Gamma)$ is a good approximation
to its infinite-volume counterpart. This point was already made a long
time ago in the study of non-relativistic systems~\cite{scalapino}.
The volume thus intrinsically
limits the resolution in $\omega$ at which one can determine
properties of the infinite-volume spectral function with good
accuracy. The simple formula (\ref{eq:FL}), displayed
in \fig(\ref{fig:FL}), also illustrates that the convergence of $F_L$
to $F$ is rather `erratically' dependent on $\omega_0$; for instance,
if $\omega_0$ is a multiple of $2/L$, all contributions of the type
$\boldsymbol{m}=(0,0,m)$ vanish.

% \newpage

\section{An Overview of Lattice Calculations\la{sec:latt}}
% Lattice QCD in a nutshell. Completion end of September.

% \section{An Overview of Lattice Calculations\la{sec:latt}}

In this section we give an overview of the technical aspects of
lattice QCD relevant to the calculations of Euclidean correlators, and
review the latest of these calculations; the discussion of their
analysis in terms of the spectral function is postponed till section
(\ref{sec:EC}). We begin by reminding the reader of the Matsubara
formalism in section (\ref{sec:path-int}). The basics of lattice
regularized QCD are given in section (\ref{sec:actions}). The question
of the renormalization of correlation functions is briefly discussed
in section (\ref{sec:renorm}), and their discretization effects are
investigated perturbatively in section (\ref{sec:latt-pt}). We then
review the lattice calculations of light-quark vector current
correlators, HQET correlators and energy-momentum tensor correlators
(sections \ref{sec:econd} to \ref{sec:EMT} respectively). An important
question for the field is how the accuracy of the calculations scale
with the lattice spacing and volume, and this question is addressed in
section (\ref{sec:ccscaling}). Finally, we briefly mention the
possibility of probing the relaxation of conserved charge densities
through off-diagonal correlators, $G^{AB}$ with $A$ different from
$B$.

\subsection{Path-integral representation of the partition function\la{sec:path-int}}

In this subsection we briefly sketch the origin of the imaginary-time
path-integral formalism of thermal field theory.  To arrive at the
path integral formula for the partition function (see
\eq\ref{eq:densmat}) of a scalar field theory, we evaluate the trace
using the eigenvectors of the field operator
\be 
\hat\phi(0,\bx)|\phi\> = \phi(\bx)|\phi\>. 
\ee 
Then the partition function can be written as
\ba 
Z &=& \tr\{e^{-\beta H}\} = {\txts\int \prod_{\bx}(d\phi_a(\bx))}
\<\phi_a| e^{-\beta H}|\phi_a\> 
\\  &=& {\txts\int
  \prod_{\bx}(d\phi_a(\bx))} \<\phi_a| U(-i\beta) |\phi_a\> \nonumber
\ea 
We can then use the standard path integral formula for the matrix
elements of $U(t)$ (see for instance~\cite{Kapusta:2006pm}),
\ba
&& \<\phi_b|U(t)|\phi_a\> =
{\txts\int_{\phi_a}^{\phi_b} [\ud\phi]} \,\,
\\ && \qquad
\exp\big\{\! -i{\txts \int_0^t \ud t'\int \ud\bx}\,{\cal L}(\partial_t\phi(t',\bx),\phi(t',\bx))\big\}.
\nonumber
\ea
Now setting $t\doteq -i\beta$, $b=a$ and summing over $a$ yields
\ba
&& Z_{\rm boson} = {\txts\int_{\rm periodic}} [d\phi] 
\\ && \quad
\exp\big\{\! -{\txts \int_0^{\beta}\ud t\int \ud\bx}\,
{\cal L}_E(\partial_t\phi(t,\bx),\phi(t,\bx))\big\},
\nonumber
\ea
where
\be
 {\cal L}_E(\partial_t\phi(t,\bx),\phi(t,\bx)) = 
-{\cal L}(i\partial_t\phi(t,\bx),\phi(t,\bx)).
\ee

For fermions, the representation of the partition function 
using Grassmann variables leads to
\ba
&& Z_{\rm fermion} = {\txts\int_{\rm antiperiodic}} [d\psi] 
\\
&& ~~~\exp\big\{\! -{\txts \int_0^{\beta}\ud t\int \ud\bx}\,
{\cal L}_E(\partial_t\psi(t,\bx),\psi(t,\bx))\big\}.
\nonumber
\ea
Without reproducing the proof in detail, we illustrate 
where the antiperiodic boundary condition comes from
in a two-state system, $\{|0\>,|1\>\}$.
Coherent fermionic states are defined by
\be
|\eta\> = e^{-\eta a^\dagger} |0\>,~~~
\<\eta| = \<0| e^{-a \eta^*}.
\ee
where $\eta$ and $\eta^*$ are Grassmann variables.
The definition 
\be
{\txts\int} \ud\eta^*\ud\eta\, \eta\,\eta^* = 1
\ee
furthermore implies the `completeness relation'
\be
{\txts\int} \ud\eta^*\ud\eta\, e^{-\eta^*\eta} |\eta\> \<\eta| = {\bf 1}.
\ee
Some algebraic manipulations then lead to 
\ba
\tr\{e^{-\beta H}\} &=& 
\sum_{0\leq k<2}  \<k|\,e^{-\beta H} |k\> 
% \\ \sum_{0\leq k<2} \<k|{\txts\int} \ud\eta^*\ud\eta\, e^{-\eta^*\eta} \,e^{-\beta H} |k\> 
\\
&= & {\txts\int} \ud\eta^*\ud\eta\, e^{-\eta^*\eta} \< -\eta| e^{-\beta H} | \eta\>,
\ea
where the minus sign in $\<-\eta|$ translates into the antiperiodic
boundary condition.  See for instance the appendix of
Ref.~\cite{Luscher:1976ms} for the generalization to multiple degrees
of freedom.

% after noticing that $\<-\eta| = \<0| + \<1| \eta^*$.

\subsection{Common actions in lattice QCD\la{sec:actions}}

For definiteness we write out the simplest action for lattice
QCD~\cite{Wilson:1974sk}. We restrict ourselves to a four-dimensional
cubic lattice, and the lattice spacing is denoted by $a$.  The unit
vectors $\hat\mu$ are aligned along the four directions of the lattice
of highest symmetry.  The dynamical variables are (a) the `link'
variables $U_\mu(x)\in SU(3)$, which can be thought of as living on
the link from $x$ to $x+a\hat\mu$, and (b) the Grassmann variables
$\psi(x)$, which are Dirac spinors in the fundamental color
representation and live on the lattice sites.  The latter correspond
to the quarks, while the `link variables' are classically related to
the gauge potential by $U_\mu(x) = e^{ia A_\mu(x)}$.

The Wilson gauge action reads
\ba
\!\!\!\!\!\!\!\! S_{\rm g} &=& \frac{1}{g_0^2}\sum_{x,\mu,\nu} \re\tr\{1-P_{\mu\nu}(x)\}\,,
\la{eq:Swg}
\\
\!\!\!\!\!\!\!\! P_{\mu\nu}(x)&=& U_{\mu}(x)U_\nu(x+a\hat\mu)U_\mu(x+a\hat\nu)^{-1}U_\nu(x)^{-1}.
\la{eq:plaq}
\ea
The Wilson fermion action reads
\ba
S_{\rm f}&=& a^4\sum_{x}\bar\psi(x) 
\left(\widetilde\nabla_\mu\gamma_\mu + m + \half a\triangle \right) \psi(x),
\la{eq:Swf}
\ea
where the summation over $\mu$ is understood.
Here $\widetilde\nabla_\mu = \half(\nabla_\mu+ \nabla^*_\mu)$ is the symmetric
covariant derivative and $\triangle = \partial_\mu\partial_\mu^*$ the covariant 
Laplacian,
\ba
\nabla_\mu\psi(x) &=& \frac{1}{a} \big( U_\mu(x)\psi(x+a\hat\mu)-\psi(x)\big),
\\
\nabla_\mu^*\psi(x) &=& \frac{1}{a} \big( \psi(x) - U_\mu(x)^{-1} \psi(x-a\hat\mu)\big).
\ea
In numerical implementations, the hopping parameter $\kappa$, defined by
\be
\kappa^{-1} = 2am + 8,
\ee
is used to parametrize the bare quark mass, and the quark field is
rescaled so as to give the $\bar\psi(x)\psi(x)$ term in
\eq(\ref{eq:Swf}) a unit coefficient.  The discretization errors of
the Wilson action are O($a$), even at the classical level.  Therefore
it is useful to improve the action by adding dimension-five operators
with coefficients tuned to remove the leading cutoff effects on the
spectrum of the theory~\cite{Symanzik:1983dc}. Fortunately, having
made use of the equations of motion, a single operator needs to be
considered, a Pauli term $a c_{\rm sw}\bar\psi(x)
\frac{i}{4}\sigma_{\mu\nu}F_{\mu\nu}(x)\psi(x)$~\cite{Luscher:1996sc}.  The
coefficient $c_{\rm sw}=1+{\rm O}(g_0^2)$ has been determined non-perturbatively for a
wide range of bare couplings in the $N_{\rm f}=0$~\cite{Luscher:1996ug},
2~\cite{Jansen:1998mx} and 3 theories~\cite{Aoki:2005et}.

The Wilson action (improved or not) preserves the vector flavor
symmetry SU($N_{\rm f}$), while the axial symmetry is broken.  The
space-time symmetries are of course broken down to discrete groups on
the lattice.  The discrete symmetries $P$ and $C$ are preserved.  In
general, it is the exact gauge invariance of lattice QCD actions and
the compactness of the gauge degrees of freedom make non-perturbative
calculations possible.

For a brief introduction to Kogut-Susskind (also called staggered)
quarks~\cite{Kogut:1974ag}, see~\cite{Golterman:2009kw}, section
(3.F).  An introduction to
Domain-Wall~\cite{Kaplan:1992bt,Shamir:1993zy} and
Overlap~\cite{Neuberger:1997fp} fermions, which preserve chiral
symmetry on the lattice, is given
in~\cite{DeTar:2009ef,Chandrasekharan:2004cn}.

Since the quark action is quadratic, the Grassmann variables can be
integrated out exactly, yielding in the \emph{numerator} the
determinant of the Dirac operator $D$ (or equivalently, of $Q\equiv
\gamma_5 D= Q^\dagger$).  In the simplest form, the determinant is
then represented as a Gaussian integral over a `pseudofermion' field
$\phi$, which is a c-number field on which the Dirac operator acts in
the same way as on a fermion field. The action reads
\be
S_{\rm g}[U] + \phi^\dagger |Q|^{-N_{\rm f}}\phi . 
\ee
It is manifestly positive-definite for an even number of
mass-degenerate flavors, while the absolute value of $Q$ is taken for
an odd number of flavors. The `Boltzmann' factor $e^{-S}$ is a real,
positive definite integrand in the path integral, and can therefore be
interpreted as a probability distribution for the gauge fields. This
property allows one to apply importance sampling methods to generate
gauge field configuration which dominate the path integral.  The best
known algorithm to generate these configurations is the Hybrid
Monte-Carlo algorithm~\cite{Duane:1987de}. In the past decade many
improvements have been made to its earliest implementation. Of
particular relevance are the more faithful representations of the
quark determinant through several pseudofermions (instead of just
one), where the long wavelength modes are separated from short
wavelength modes. The most widely used realizations of this idea are
the mass preconditioning (MPHMC~\cite{Hasenbusch:2001ne}), the
domain-decomposition (DDHMC~\cite{Luscher:2005rx}) and the rational
HMC (RHMC~\cite{Clark:2006fx}) algorithms.

The calculation of correlation functions amounts to calculating quark
propagators on background gauge fields, i.e. solving the equation
$D[U]\psi=\eta$ for a given spinor field $\eta(x)$ carrying Dirac and
color indices. Much progress has been made in speeding up the solution
of this equation, which results in an improved scaling of the cost
towards the chiral
limit~\cite{Luscher:2003vf,Luscher:2007se,Babich:2010qb}. In the
simplest implementation, gluonic correlators are evaluated by
computing the correlation on the generated ensemble of gauge field
configurations. However, in the absence of fermions the locality of
the action can be exploited to sample the fields more efficiently
(multi-level algorithms~\cite{Luscher:2001up,Meyer:2002cd}).

The `quenched' approximation consists in neglecting the gauge-field
dependence of the fermion determinant altogether. The fermions then
only appear at the level of the correlation functions. As a
consequence, the weighting factor of configurations $e^{-S_g}$ is
local, and a heat-bath update algorithm can be employed, resulting in
an update speedup factor on the order of
100. See~\cite{Luscher:2010ae} for a review of the state-of-the-art in
simulation algorithms.

\subsection{Renormalization of lattice correlation functions\la{sec:renorm}}

Most lattice calculations are restricted to on-shell correlators.  An
on-shell correlator avoids any contact terms in coordinate space, so
that the equations of motion can be used to simplify its
renormalization. In the case of the vector and the axial-vector
current only a finite multiplicative renormalization is then
necessary. For instance, the renormalization factors $Z_A(g_0)$ and
$Z_V(g_0)$ have been determined non-perturbatively in the $N_{\rm
f}=2$ O($a$)-improved Wilson
theory~\cite{DellaMorte:2005rd,DellaMorte:2008xb}.

The renormalization of the lattice energy-momentum tensor is more
complicated already in the continuum, due to the trace anomaly.  In
addition, the lattice regulator breaks continuous translation
invariance, resulting in additional (finite) renormalization effects.
In the continuum, a symmetric and gauge invariant form of the
energy-momentum tensor is
\ba
T_{\mu\nu} \! & \equiv & \!  \theta_{\mu\nu}^{\rm g} +
                  \theta_{\mu\nu}^{\rm f}
               + \quarter\delta_{\mu\nu}(\theta^{\rm g} + \theta^{\rm f}), 
\\
{ \theta_{\mu\nu}^{\rm g} } \! &{=}& \! {
{\txts\frac{1}{4}}\delta_{\mu\nu}F_{\rho\sigma}^a F_{\rho\sigma}^a
   - F_{\mu\alpha}^a F_{\nu\alpha}^a },\nn
\theta_{\mu\nu}^{\rm f} \!&=& \!
\quarter {\txts\sum_f} \bar\psi_f\!\! \stackrel{\leftrightarrow}{D_{\mu}}\!\gamma_{\nu}\psi_f
+ \bar\psi_f\!\! \stackrel{\leftrightarrow}{D_{\nu}}\!\gamma_{\mu}\psi_f
 -{\txts\frac{1}{2}}\delta_{\mu\nu}  \bar\psi_f
\! \stackrel{\leftrightarrow}{D_{\rho}}\! \gamma_{\rho}\psi_f ,\nn
\theta^{\rm g} &=& \beta(g)/(2g) ~ F_{\rho\sigma}^a  F_{\rho\sigma}^a,\quad
\theta^{\rm f} =  {\txts \sum_f} m_f \bar\psi_f \psi_f\,,
\nonumber
\ea
where $\stackrel{\leftrightarrow}{D_{\mu}}=
\stackrel{\rightarrow}{D_{\mu}} - \stackrel{\leftarrow}{D_{\mu}} $.
This form has finite matrix elements between physical states (but
divergences appear in offshell correlation functions).  The term
$\theta^{\rm g}$ is the trace
anomaly~\cite{Adler:1976zt,Fujikawa:1980rc}.  On the lattice, there is
a lot of freedom as to how to discretize these
expressions~\cite{Meyer:2007tm}. An important consideration is that
the traceless part $\theta_{\mu\nu}$ for $\mu=\nu$ belongs to a
three-dimensional irreducible representation of the H(4)
group~\cite{Gockeler:1996mu}, while $\theta_{\mu\nu}$ for $\mu\neq\nu$
belongs to a six-dimensional representation and $\theta$ belongs to
the trivial representation. Because there are no other gauge invariant
operators of dimension four in these representations, their
renormalization is multiplicative.  The renormalization of $\theta$ is
particularly simple when its discretization is based on the action itself, in which case the
renormalization factor is expressed in terms of the lattice beta
function, $dg_0^{-2}/d\log a$.  For $\theta_{\mu\nu}$ ($\mu=\nu$),
the renormalization factor of
$\theta_{\mu\nu}$ for $\mu=\nu$ of the discretization based on plaquettes can be
related to anisotropy coefficients~\cite{Engels:1981qx,Meyer:2007fc}.

A number of quantities of interest are defined through frequency-space
correlation functions, which are not on-shell quantities.  These
include the static susceptibility of the quark number density,
\be
\chi_s(\bk) = \int_0^\beta \ud t \int \ud\bx\, e^{-i\bk\cdot\bx} \, \< n(-it,\bx)\,n(0)\>.
\ee
At $\bk=0$, this corresponds to the derivative of the free energy with
respective to the chemical potential. Another example is the static
susceptibility of the axial current, which was shown to be related to
the finite-temperature dispersion relation of the `pion'
quasiparticle~\cite{Son:2002ci},
\ba
\chi_5 \delta^{ab} &=& \int \ud t\,\ud\bx\,\< A_0^a(-it,\bx)\,A_0^b(0)\>,
\\
A_\mu^a(x) &= & \bar\psi \gamma_\mu\gamma_5 {\txts\frac{\tau^a}{2}} \psi.
\ea
Care has to be taken to renormalize these correlation functions.
Using a conserved current greatly simplifies this task, but this
requires an exact chiral symmetry on the lattice. Formulations
preserving an exact chiral symmetry exist but are computationally
expensive~\cite{Hasenfratz:2002rp}.

We illustrate in the case of the vector current how offshell
correlation functions on the lattice can be given the same structure
as in the continuum~\cite{Gockeler:2003cw}.  By performing a
non-anomalous change of variables in the path integral,
one obtains the identity $\<\delta S {\cal O}\> = \<\delta {\cal O}\>$
for a general operator ${\cal O}$.  We now consider the case of Wilson
fermions for simplicity.  From the transformation
\be
\delta\psi(x) = \omega(x) \psi(x),
\quad 
\delta\bar\psi(x) = -\omega(x) \bar\psi(x),
\ee
one obtains, for ${\cal O}=j_\nu(0)$, 
\ba
\sum_\mu \hat k_\mu \Pi_{\mu\nu}(k)&=& 0, ~~~
(\hat k_\mu= {\txts\frac{2}{a}}\sin(\half ak_\mu)),
\la{eq:kuPimunu}
\\
\Pi_{\mu\nu}(k) &=& a^4\sum_x e^{ik\cdot x}
\<j_\mu(x)\,j_\nu(0)\>
\la{eq:Pimunu_lat}
\\ && 
+ a \delta_{\mu\nu} \<s_\nu(0)\>
\nonumber
\ea
where the conserved vector current and the point-split 
scalar operator take the form
\ba
j_\mu(x) &=& \half \big( 
\bar\psi(x+a\mu) (1+\gamma_\mu)U_\mu^{-1}(x) \psi(x) 
\\ && ~~
-\bar\psi(x) (1-\gamma_\mu)U_\mu(x) \psi(x+a\mu) 
\big).
\nonumber
\\
s_\nu(x) &=& \half \big(
\bar\psi(x+a\hat\nu) (1+\gamma_\nu)U_\nu^{-1}(x) \psi(x)
\\ && ~~
+\bar\psi(x) (1-\gamma_\nu)U_\nu(x) \psi(x+a\nu)
\big).
\nonumber
\ea
Eq.~(\ref{eq:kuPimunu}) corresponds in configuration space to
\be
\sum_\mu \partial_\mu^*\big\<j_\mu(x)
j_\nu(0) + a^{-3}\delta_{\mu\nu}\delta_{x,0}s_\nu(0)\big\> =0\quad  \forall x,
\la{eq:ct}
\ee
which shows that the second term takes care of removing a quadratically
divergent contact term. No summation over $\nu$ is implied in 
\eq(\ref{eq:Pimunu_lat}) and (\ref{eq:ct}).
% which forbids in particular a quadratic divergence in $\Pi_{\mu\nu}$.

% \cite{Meyer:2009jp}
% \cite{Meyer:2007dy}
% \cite{Meyer:2007ic}

\subsection{Lattice perturbation theory\la{sec:latt-pt}}

\begin{figure}
\centerline{\includegraphics[width=0.35\textwidth,angle=-90]{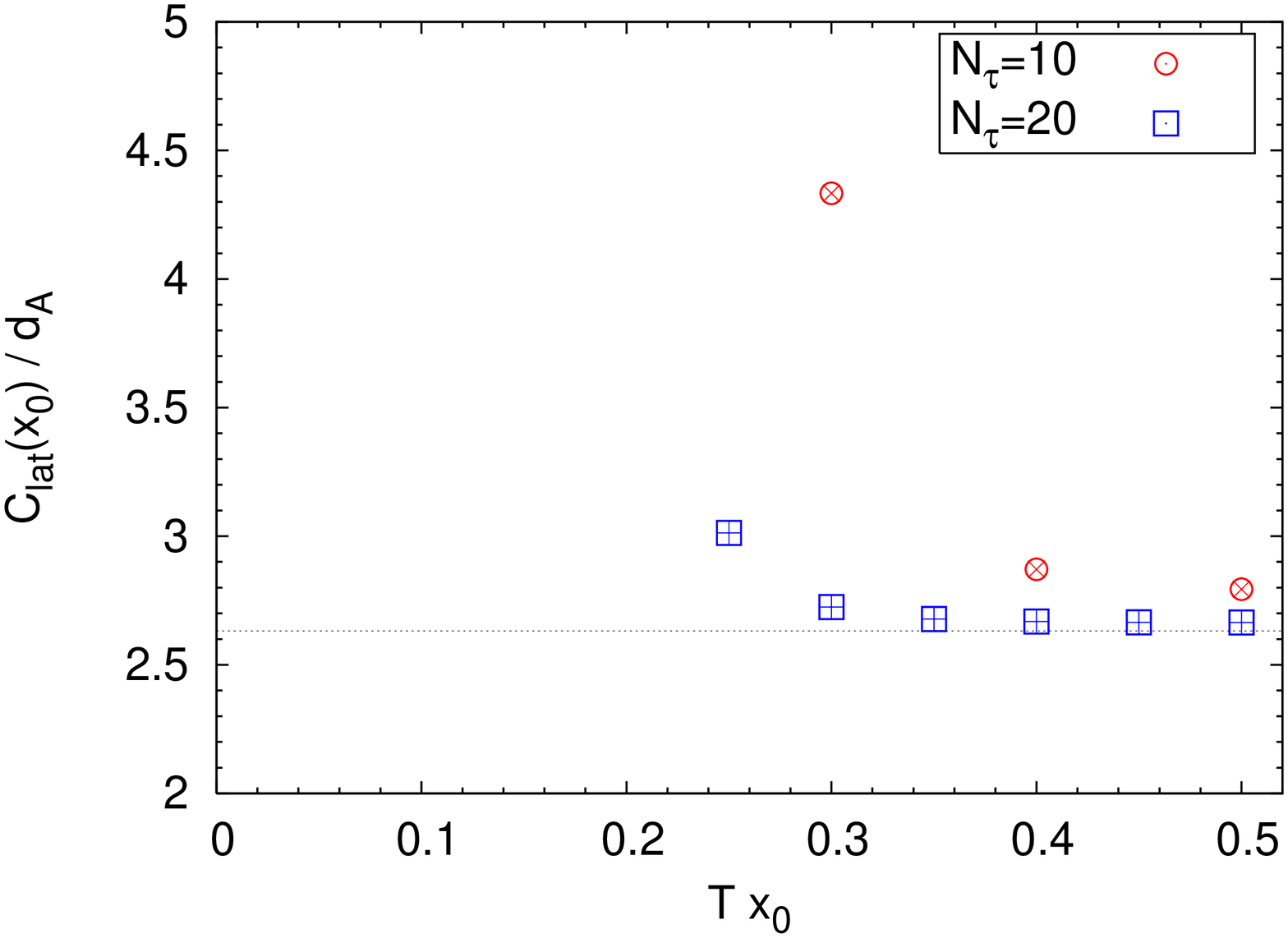}}
\centerline{\includegraphics[width=0.35\textwidth,angle=-90]{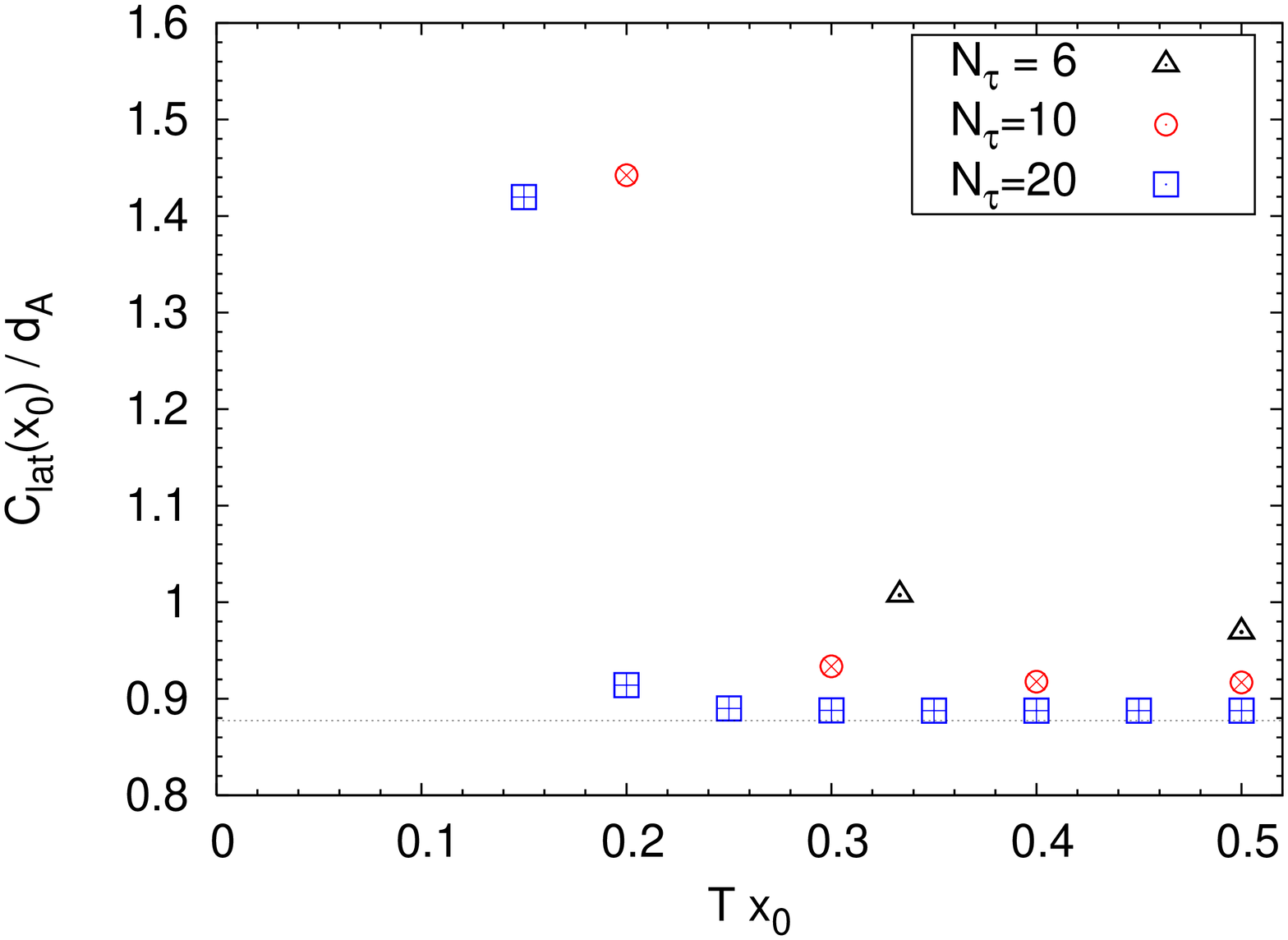}}
\caption{{\bf Top}: the treelevel $G_{00,00}(x_0,{\bf 0})$ correlator at finite
lattice spacing on the isotropic lattice in SU(3) gauge theory.  The
horizontal line is the continuum treelevel prediction, $4\pi^2/15$.
{\bf Bottom}: the treelevel $G_{03,03}(x_0,{\bf 0})$ correlator at
finite lattice spacing on the isotropic lattice.  The horizontal line
is the continuum trelevel prediction, $4\pi^2/45$~\cite{Meyer:2009vj}.}
\label{fig:t00andt03}
\end{figure}

The perturbative calculations of thermal spectral functions and
Euclidean correlators are reviewed in section (\ref{sec:pt}). Here we
describe the results of free field calculations performed in
lattice perturbation theory (see \cite{Capitani:2002mp} for a general
review of this subject).  The motivation is to quantify the size of
discretization errors for standard lattice actions and discretizations
of the local operators in a controlled framework, namely leading order
perturbation theory.  Thus these calculations allow one to make an
informed decision in choosing a discretization, and also allow one to
estimate what lattice spacing it will be necessary to reach in order
to achieve a certain accuracy. We will consider correlators in the
mixed representation where they depend on time $t$ and spatial
momentum $\bk$.  This representation has the advantage that the
correlator is `on-shell'.  The approach to the continuum of on-shell
correlators is considerably simpler than that of off-shell correlators
because the latter involve an integration over the contact
region~\cite{Symanzik:1983dc}.  For a given lattice spacing
$a=1/(T\Nt)$, there will thus be a range of time separations $t$ and
spatial momenta $\bk$ for which the discretization errors is below a
prescribed level, say $3\%$.

As an example, it is interesting to quantify how badly the breaking of
translation invariance on the lattice affects the conservation of
energy and momentum as measured by a discretization of $T_{00}$ and
$T_{0k}$. Indeed, the correlators $G_{00,00}(t,\bk=\boldsymbol{0})$ and
$G_{03,03}(t,\bk=\boldsymbol{0})$ are $t$ independent in the continuum,
but on the lattice they suffer of discretization errors which vanish
as $\sim a^2$ for a fixed $t$. 
\fig(\ref{fig:t00andt03}) illustrates this point. It also shows that 
the discretization errors are numerically large, particularly in the
energy correlator. A possible explanation for this large
discretization effect is that the flattening results from a delicate
cancellation of contributions which by power counting diverge as
$1/t^5$ for small $t$.

% \begin{figure}
% \centerline{\includegraphics[width=0.35\textwidth,angle=-90]{FIGS/t00-00piT-xi2.ps}}
% \centerline{\includegraphics[width=0.35\textwidth,angle=-90]{FIGS/t03-00piT-xi2.ps}}
% \caption{The ratio of lattice to continuum treelevel correlators for 
% $T_{00}$ (top) and $T_{03}$ (bottom) \cite{Meyer:2009vj}.}
% \label{fig:t00andt03}
% \end{figure}

\begin{figure}
\centerline{\includegraphics[width=0.35\textwidth,angle=-90]{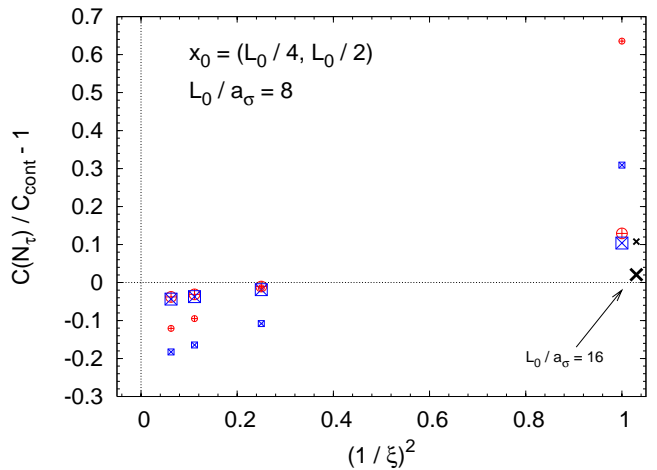}}
\caption{The cutoff effects on the tensor correlator for $x_0=\beta/4$ and $\beta/2$, 
corresponding to small and large symbols
respectively~\cite{Meyer:2009vj}. The spatial lattice spacing $\as$ is
held fixed, the temporal lattice spacing $\at$ is varied between a
quarter and one times $\as$.  The $\square$'s refer to $T_{12}$ and
the $\otimes$'s to $\half(T_{11}-T_{22})$.  In addition, the two
crosses at $\xi=1$ indicate the reduction of the cutoff effect on the
$\half(T_{11}-T_{22})$ correlators when increasing $\beta/\as$ from 8 to
16.}
\label{fig:t12cutoff}
\end{figure}

The resolution in the Euclidean time variable is obviously important
to constrain the spectral function. But the computational cost of
decreasing the lattice spacing grows very rapidly (see section
(\ref{sec:ccscaling}). The situation may potentially be improved by
choosing an anisotropic lattice, where the temporal lattice spacing is
smaller by a factor $\xi$ than the spatial lattice spacing. This
strategy has been followed in certain thermodynamic
studies~\cite{Namekawa:2001ih} and in studies of quarkonium in the
deconfined phase of SU(3) gauge theory~\cite{Jakovac:2006sf}. In
perturbation theory it is observed that the discretization errors on
thermodynamic potentials and on time-dependent correlators are more
sensisitive to the value of the temporal lattice spacing than on the
spatial one by a significant factor. At the same time the cost of
producing independent configurations only scales as $1/\at$, whereas
reducing the lattice spacing on the isotropic lattice would scale as
$1/a^4$.  This constitutes the main motivation to use an anisotropic
lattice. An example is given in \fig(\ref{fig:t12cutoff}), which
displays the cutoff effect affecting the shear channel correlator
$G_{13,13}(t,\bk=\boldsymbol{0})$ in the SU($\Nc$) theory at maximum
separation $t=\beta/2$. Clearly, reducing the temporal lattice spacing
overall reduces the discretization error, although it changes sign for
a particular value of the anisotropy near 2. The figure suggests that
reducing the temporal lattice spacing by a factor 2 is almost as good
as reducing the lattice spacing in each direction by the same amount.

The results described above are obtained at tree level, and
renormalization effects only set in at the next order.  A drawback of
the anisotropic action is that it requires a tuning of certain bare
parameters in addition to the gauge coupling and the quark masses.  In
the pure gauge theory, the only such additional parameter is the bare
anisotropy factor $\xi_0$ that enters the action. In effect it must be
tuned in order to maintain a constant renormalized anisotropy $\xi$ as
the spatial lattice spacing is varied. The calculation of
$\xi(\xi_0,g_0)$ has been carried out using various methods for the
Wilson action~\cite{Klassen:1998ua,Engels:1999tk}.  Including the
quarks, there are additional dimension-four operators whose
coefficients need to be tuned. Methods have been developed to achieve
this in the context of hadron spectroscopy~\cite{Morrin:2006tf}. In
addition the anisotropic lattice complicates the normalization of
local operators used in calculating thermodynamic properties or
correlation functions. A possible strategy is then to determine the
renormalization factors of a particular discretization based on the
renormalization of the anisotropy $\xi$; this discretization can
thereafter be used to calibrate other discretization.

The discretization effects in high-temperature meson correlators and
spectral functions at nonzero-momentum and fermion mass were
investigated by Aarts and Martinez-Resco in~\cite{Aarts:2005hg}.  In
this case the spectral function can be analyzed straightforwardly at
finite lattice spacing because the transfer matrix expression of the
meson correlators is simple~\cite{Luscher:1976ms}.  We give as an
example the Euclidean correlator of a quark bilinear
$\bar\psi\Gamma\psi$ for (unimproved) Wilson fermions:
\ba
G_H(t,\bp) &=& \frac{4\Nc}{L^3}\sum_{\bk} 
\tr\{ S(t,\bk) \Gamma S(-t,\bp+\bk)\Gamma\}~~~
\\
= \frac{4\Nc}{L^3}\sum_{\bk} &\Big[  &
 a_H^{(1)}S_4(t,\bk)S_4^\dagger(t,\bp+\bk)
\\
&& -a_H^{(2)}S_i(t,\bk)S_i^\dagger(t,\bp+\bk)
\nn
&& -a_H^{(3)}S_u(t,\bk)S_u^\dagger(t,\bp+\bk)
\Big]\,, 
\nonumber
\ea
where for $t\neq0$
\ba
S_4(t,\bk) &=& 
\frac{\sinh(E_{\bk}/\xi)}{2{\cal E}_{\bk}\cosh(E_{\bk}/2T)}
\cosh(E_{\bk}(t-{\txts\frac{\beta}{2}})) \,,
\\
S_i(t,\bk) &=&
\frac{1}{\xi}\, \frac{i\sin k_i}{2{\cal E}_{\bk} \cosh(E_{\bk}/2T)}\,
\sinh(E_{\bk}(t-{\txts\frac{\beta}{2}})) \,,
\nn
\!\!\!\!\!\!\!\! S_u(t,\bk) &=& - \frac{1-\cosh(E_{\bk}/\xi) + {\cal M}_{\bk}}
{2{\cal E}_{\bk} \cosh(E_{\bk}/2T)}
\sinh(E_{\bk}(t-{\txts\frac{\beta}{2}}))
\nonumber
\ea
and ($r$ is the `Wilson parameter'~\cite{Wilson:1974sk}, usually set to unity)
\ba
\xi\,{\cal M}_{\bk} &=& r \sum_{i=1}^3 (1-\cos k_i) + m\,,
\\
\xi{\cal K}_{\bk} &=& \sum_{i=1}^3 \gamma_i \sin k_i\,,
\\
\cosh(E_{\bk}/\xi) &=& 1+ \frac{{\cal K}_{\bk}^2 +{\cal M}_{\bk}^2} {2(1+{\cal M}_{\bk})}\,,
\\
{\cal E}_{\bk} &=& (1+{\cal M}_{\bk}) \sinh(E_{\bk}/\xi)\,.
\ea
Due to the UV cutoff, the spectral function on the lattice vanishes
for frequencies larger than the energy of the largest momentum in the
Brillouin zone, $\bp=\frac{\pi}{a}(1,1,1)$. This frequency is roughly
\be
 \omega_{\rm max} = \frac{2}{a}\log\Big(1+\frac{6+am}{\xi}\Big). 
\ee
In addition, the spectral function admits several cusps corresponding
to the edges of the Brillouin zone.  For smaller frequencies, the finite
lattice spacing affects predominantly the mismatch between the
continuum and lattice lightcone, which can be substantial~\cite{Aarts:2005hg}.

From a practical point of view, the conservative way to proceed is to
extrapolate the Euclidean correlator to the continuum, see for
instance \cite{Meyer:2007dy,Ding:2010ga}. Then the question of the
analytic continuation takes place entirely in the
continuum~\cite{Ding:2010ga}. This way of proceeding has the advantage
that analytic studies of the spectral function in the continuum, where
perturbative calculations can be pushed more easily to higher orders,
can be used directly.

\subsection{The Electric Conductivity\la{sec:econd}}

\begin{figure}
\centerline{\includegraphics[width=0.45\textwidth]{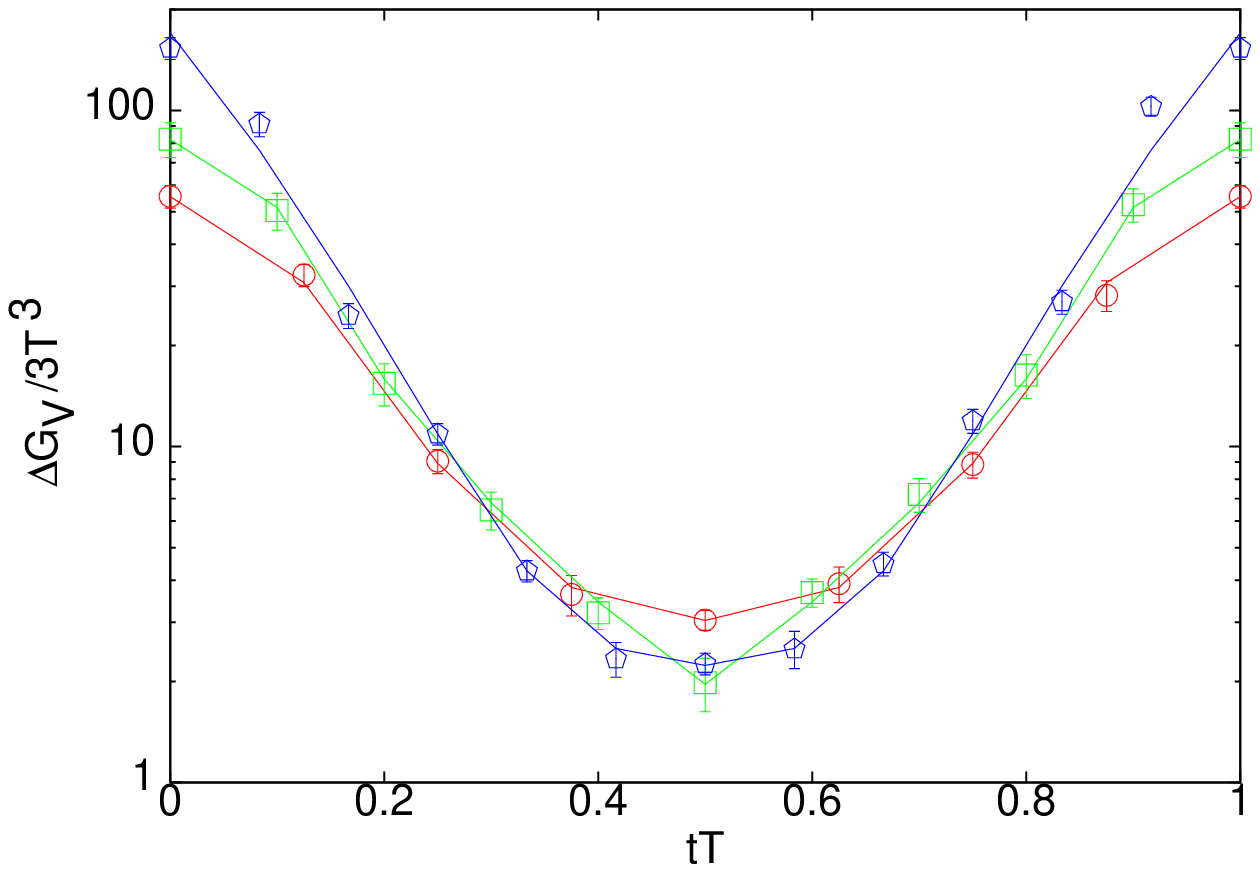}}
\centerline{\includegraphics[width=0.45\textwidth]{FIGS/LATCOR/G_ViVi_m1_zaa.eps}}
\caption{Vector current correlation functions calculated with Kogut-Susskind fermions
on quenched lattices at zero spatial momentum.
Top:  $T=2T_c$, with $\Nt=8$ (circles), 10 (squares) and 12 (pentagons)~\cite{Gupta:2003zh}.
Bottom: $T\approx0.6T_c$ with $\Nt=24$~\cite{Aarts:2006cq}.
On both plots the curves are the result of Bayesian fits.}
\label{fig:gupta}       
\end{figure}

% \begin{figure}
% \centerline{\includegraphics[width=0.45\textwidth]
% {FIGS/LATCOR/Rec_Momenta_0_0_0-1p5Tc_without-zeromode_vector.eps}}
% \caption{Vector charm current correlation functions using 
% O($a$) improved Wilson fermions on quenched isotropic lattices at $1.5T_c$.
% The ratio to the reconstructed correlator from $0.75T_c$ is taken~\cite{Ding:2009ie}.
% In `$V_{\rm sub}$', the contribution of the transport peak has been subtracted.}
% \label{fig:ding}
% \end{figure}

% \begin{figure}
% \centerline{\includegraphics[width=0.45\textwidth]{FIGS/LATCOR/vc_xi4.eps}}
% \caption{Charmonium correlators in the vector channel \cite{Jakovac:2006sf}
% on an anisotropic lattice, $\xi=4$, using the Fermilab lattice formulation.
% The enhancement over the reconstructed correlator
% is significant and is likely due to the contribution of a transport peak.}
% \la{fig:petr}
% \end{figure}

\begin{figure}
\centerline{\includegraphics[width=0.45\textwidth]
{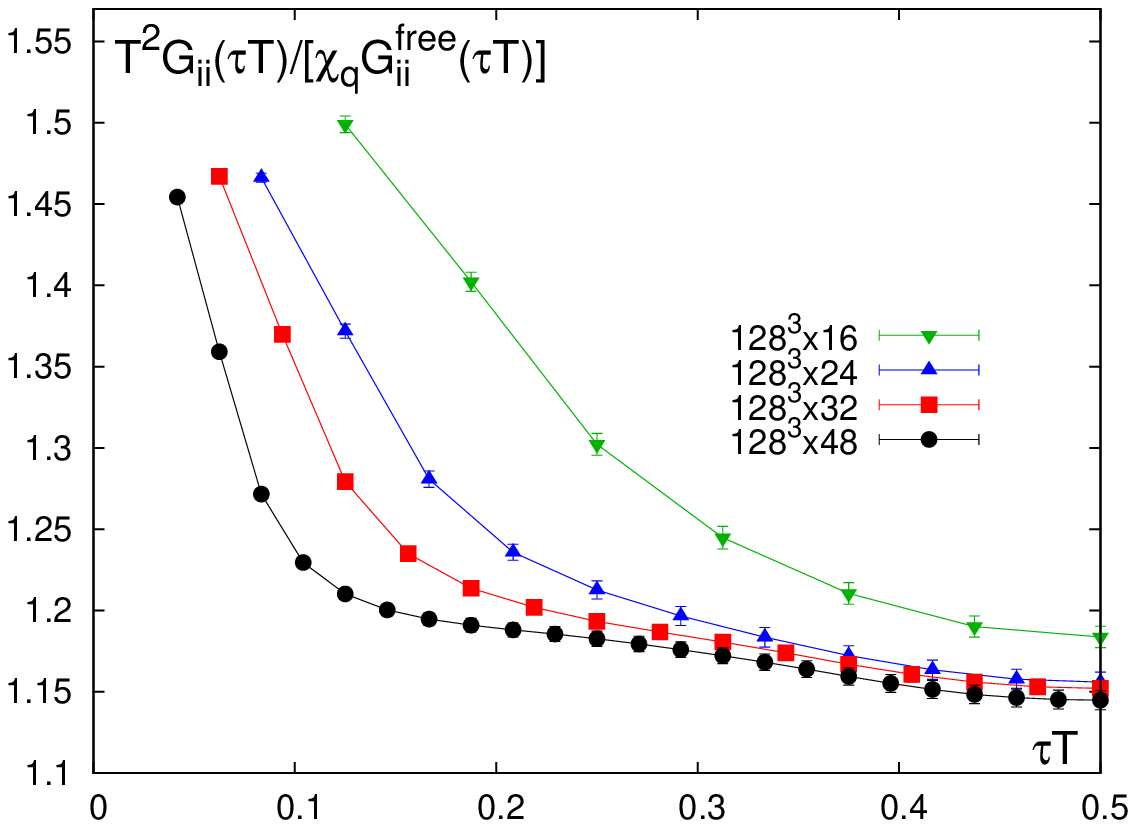}}
\centerline{\includegraphics[width=0.45\textwidth]
{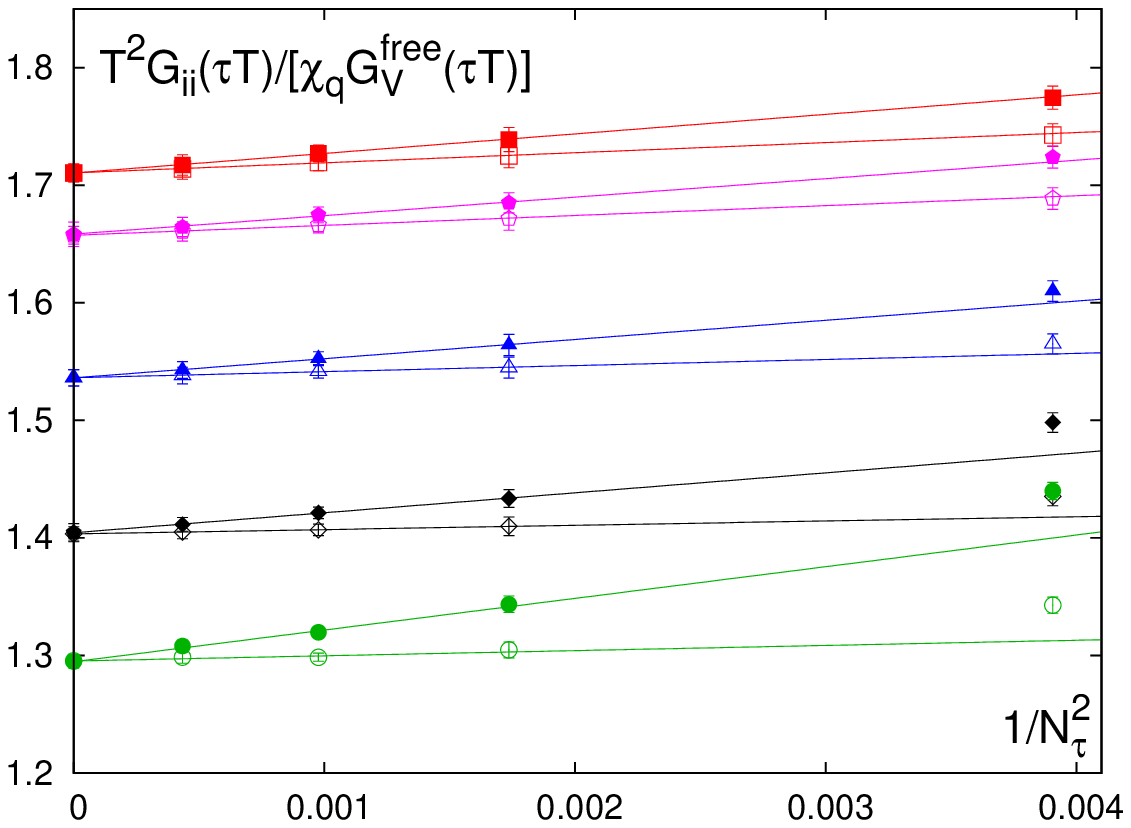}}
\caption{Top: Vector current correlation function calculated with O($a$) improved 
Wilson fermions on quenched lattices at zero spatial momentum~\cite{Ding:2010ga} and $T\simeq 1.45T_c$.
Bottom: from the same publication, the vector correlation function, $G_{ii} (\tau, T )$, normalized by
the quark number susceptibility and the free vector correlation function calculated in the continuum (full
symbols) and on lattices with temporal extent $\Nt$ (open symbols). The results of five values of
Euclidean time are displayed, $\tau T$ = 0.1875, 0.25, 0.3125, 0.375 and 0.5 (top to bottom), on lattices with temporal
extent $\Nt$ = 16, 24, 32 and 48.
}
\label{fig:francis}
\end{figure}

As reviewed in section (\ref{sec:elec_cond}), the electric
conductivity is phenomenologically important in that it determines the
soft photon emission by the quark-gluon plasma, and it can be
extracted from the vector current correlator via a Kubo formula.  The
early calculations of finite-temperature vector correlation functions
were carried out with Kogut-Susskind fermions. The correlators from
two such calculations, by S.~Gupta~\cite{Gupta:2003zh} (up to
$\Nt=12$) and Aarts et al~\cite{Aarts:2006cq} (up to $\Nt=24$), are
displayed in \fig(\ref{fig:gupta}). As is seen most clearly from the
bottom plot, there are effectively two channels contributing,
corresponding respectively to the vector current
$\bar\psi\gamma_i\psi$ and the quark bilinear
$\bar\psi\gamma_0\gamma_5\gamma_i\psi$; the contribution of the latter
comes with a + sign on odd time-slice and a - sign on the even
ones~\cite{Aarts:2006cq}. Due to this effect, the analysis proceeds
separately on the even and on the odd sites, effectively resulting in
a loss of resolution in the time variable by a factor 2. For this
reason most recent calculations have been based on Wilson-type
fermions.

A very recently published calculation using O($a$) improved Wilson
fermions is displayed in \fig(\ref{fig:francis}), with temporal
extents up to $\Nt=48$~\cite{Ding:2009ie}.  The Euclidean correlator
of the spatial components of the vector current at $1.45T_c$ has been
normalized by its value in the non-interacting limit, as well as by
the quark number susceptibility. The latter normalization factor has
the benefit of cancelling out the normalization of the vector current.
On the bottom panel, this data is extrapolated to the continuum. One
notices that normalizing by the free correlator computed at the
corresponding value of $\Nt$ reduces the slope of the extrapolation,
without completely eliminating it. The figure thus shows that a
continuum extrapolation is mandatory for a precision analysis of
Euclidean correlators.  The achieved accuracy is below $1\%$. The
continuum extrapolated data provides a solid basis for a study of the
corresponding spectral function, which we will discuss in section
(\ref{sec:MEM}).

\begin{figure}
\centerline{\includegraphics[width=0.45\textwidth]
{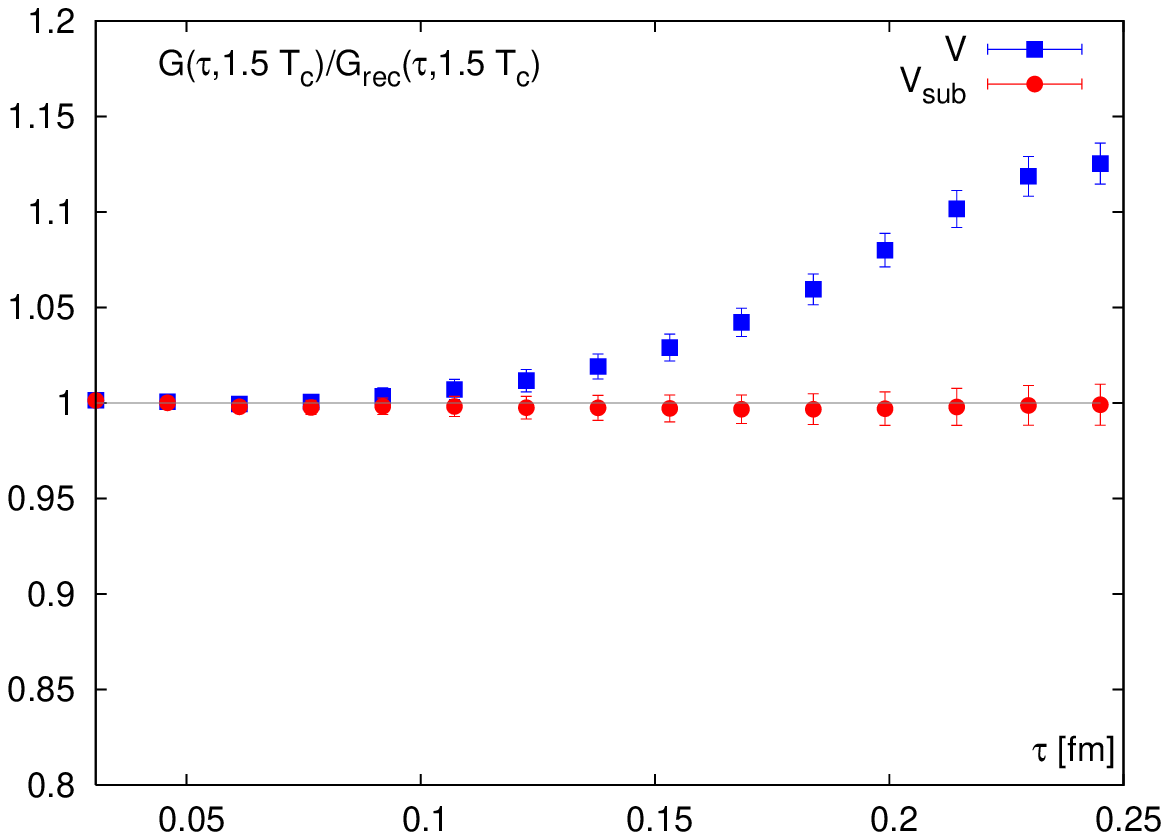}}
\centerline{\includegraphics[width=0.45\textwidth]{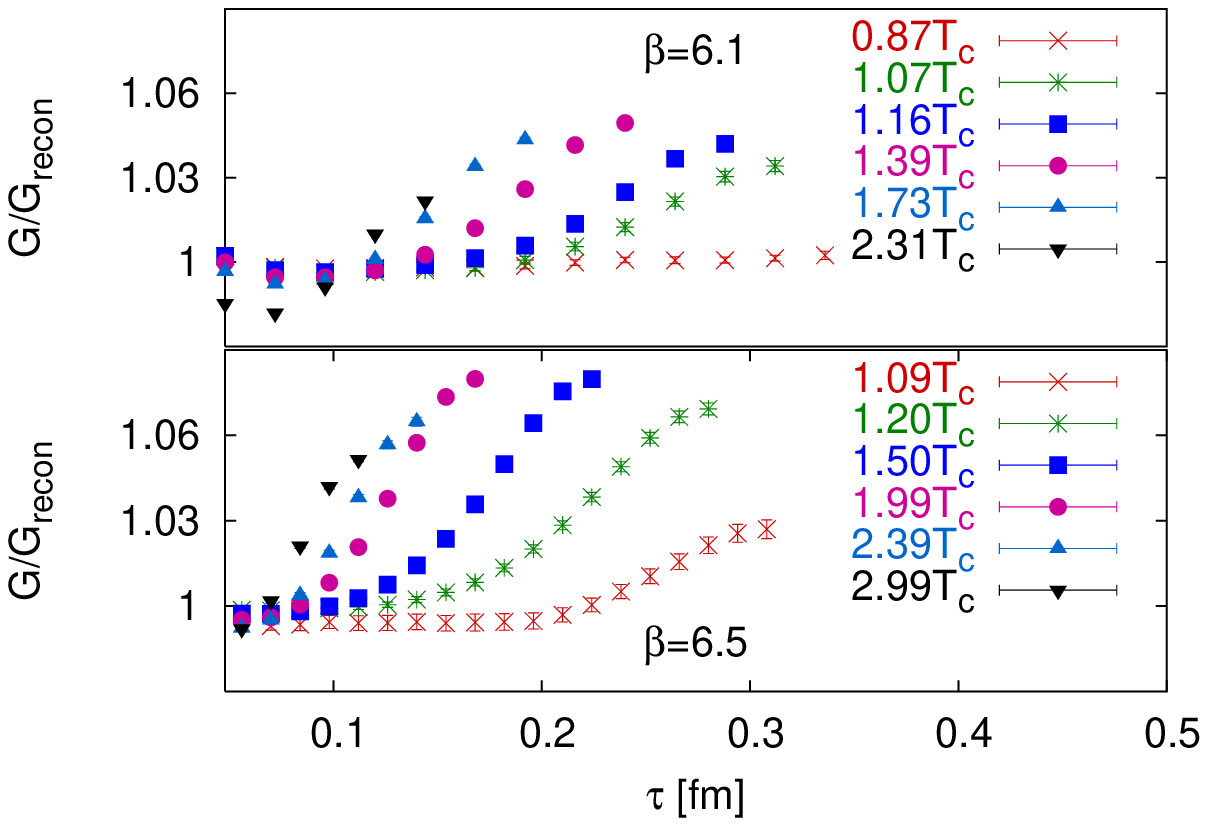}}
\caption{ Vector charm current correlation functions in quenched QCD.  The ratio to
  the reconstructed correlator from the confined phase is taken.  \underline{Top}: O($a$)
  improved Wilson fermions on isotropic lattices at
  $1.5T_c$~\cite{Ding:2009ie}. The correlator in the denominator is reconstructed from $0.75T_c$.  
  In `$V_{\rm sub}$', the contribution
  of the transport peak has been subtracted.  \underline{Bottom}: anisotropic
  Fermilab lattice formulation ($\xi=4$) at several
  temperatures (\cite{Jakovac:2006sf}, fig. from \cite{Bazavov:2009us}).  
  The correlator in the denominator is reconstructed from a very low temperature.
  The enhancement over the
  reconstructed correlator is significant and is likely due to the
  contribution of a transport peak.  }
\label{fig:ding-jako}
\end{figure}

The Euclidean correlator of the charm current has also been
investigated extensively, with the primary motivation to study the
`melting' of charmonium states in the quark-gluon plasma.  This effect
is expected on general grounds at sufficiently high
temperature~\cite{Matsui:1986dk}, and is described in detail by
resummed perturbative calculations, see in
particular~\cite{Laine:2008cf,Burnier:2007qm}. From the point of view
of these studies, the presence of a transport peak of width $\sim
T^2/M$, see \eq(\ref{eq:rho_L}), represents an undesirable
contribution~\cite{Umeda:2007hy,Mocsy:2005qw}, which can approximately
be subtracted by taking the difference of the correlator at two
Euclidean-time points\cite{Datta:2008dq,Ding:2009ie}. The functional
form of the transport peak was calculated in~\cite{Petreczky:2005nh}
based on the Langevin equation, and the result is reviewed in section
(\ref{sec:boltz}). If the charm quark is heavy enough to be well
described by the static approximation, then the parameters of the
transport peak can be obtained from a correlator calculated in
heavy-quark effective theory (see the next section).

Figure (\ref{fig:ding-jako}) displays the ratio of the Euclidean
correlator at $1.5T_c$ to the reconstructed correlator taken at
$0.75T_c$. The latter was obtained by first solving for
$\rho(\omega,0.75T_c)$, and then applying the definition
(\ref{eq:Grec1-main}). The top panel presents data obtained on an
isotropic lattice~\cite{Ding:2009ie}, while the lower one uses an
anisotropic lattice with $\as/\at=4$~\cite{Jakovac:2006sf}. These
calculations are performed in the quenched approximation using
`clover' improved Wilson fermions. Similar studies had been carried
out earlier \cite{Datta:2003ww}. An important result of these analyses
is that once the transport peak contribution is removed from the 
high-temperature correlator, the subtracted correlator at $1.5T_c$
is indistinguishable from the correlator at $0.75T_c$ to the present 
accuracy (top panel of \fig \ref{fig:ding-jako}).

\begin{figure}
\centerline{\includegraphics[width=0.5\textwidth]{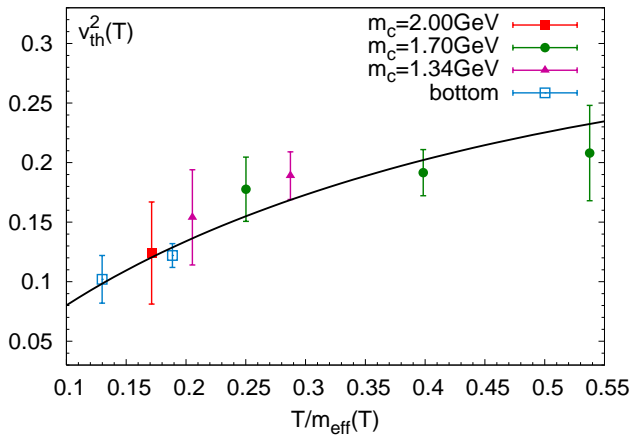}}
\caption{Thermal velocity of the heavy quarks as a function of temperature,
as extracted from the area under the transport peak in the vector current correlator~\cite{Datta:2008dq}.}
\label{fig:vel}
\end{figure}

At vanishing spatial momentum, the transport peak appearing in the
density-density correlator is an exact delta function, and its weight
is the static susceptibility $\chi_s$.  The transport peak appearing
in the current-current correlator on the other hand acquires a width
of order $T^2/M$. Its area is $\chi_s \<v_{p}^2\>$ (see section
\ref{sec:boltz}).  Therefore the thermal velocity of the heavy quarks
can be extracted from the ratio.  The main task consists in separating
the low-frequency from the high-frequency contribution in the current
correlator. An interesting analysis was carried out
in~\cite{Datta:2008dq} assuming that for $\omega>2M$, the spectral
density is unchanged from $T=0$, which is consistent with the lattice
data.  Subtracting this high-frequency contribution from the Euclidean
current correlator, what remains is an essentially $t$-independent
function. Its ratio to the static susceptibility yields an estimate of
the thermal velocity, displayed in \fig(\ref{fig:vel}). These results
are compared to a prediction based on kinetic theory.  Equating
expression (\ref{eq:chi_HQ}) with the susceptibility measured on the
lattice defines an effective (temperature-dependent) quark
mass~\cite{Datta:2008dq}; inserting this effective mass into the
equipartition prediction $\frac{1}{3}\<v^2\>= \frac{T}{M}$ leads to
the black curve in \fig(\ref{fig:vel}). Note that this curve is
obtained without using the current correlator, and clearly the
agreement with the data is excellent. This is evidence that the
kinetic description works. The velocity appears to be
solely a function of $\frac{T}{M}$, rather than $T$ and $M$
separately.

The transport peak has a width given by $\eta$, the drag coefficient
(see \eq(\ref{eq:rho_L})) which is of order $1/M$. Because of the
smoothness of the kernel relating the spectral density to the
Euclidean correlator, it is difficult to separately determine the
width and height of the peak in the longitudinal current correlator
from Euclidean data. This is crucial, since the height corresponds to
the diffusion constant $D$.  However, one can use heavy-quark
effective theory to determine the momentum diffusion coefficient
$\kappa$, in terms of which $D$ and $\eta$ can be expressed as
$\eta=\kappa/(2MT)$ and $D=2T^2/\kappa$ in the heavy quark limit. This
is the subject of the next section.

\subsection{Heavy Quark Diffusion\la{sec:hq}}

% \begin{figure}
% \centerline{\includegraphics[angle=-90,width=0.45\textwidth]{FIGS/LATCOR/corrE-7p483hyp.ps}}
% \caption{The Euclidean force-force correlator from simulations with the Wilson action at
 %$\beta=7.483$~\cite{Meyer:2010tt}. The black curve
% indicates the NLO result of Burnier et al~\cite{Burnier:2010rp}.  This
% result has been used to determine the normalization of the E-field by
% matching the lattice data to it at $tT=\frac{1}{4}$ and $3.1T_c$.}
% \label{fig:1012a}
% \end{figure}

\begin{figure}
\centerline{\includegraphics[angle=-90,width=0.5\textwidth]{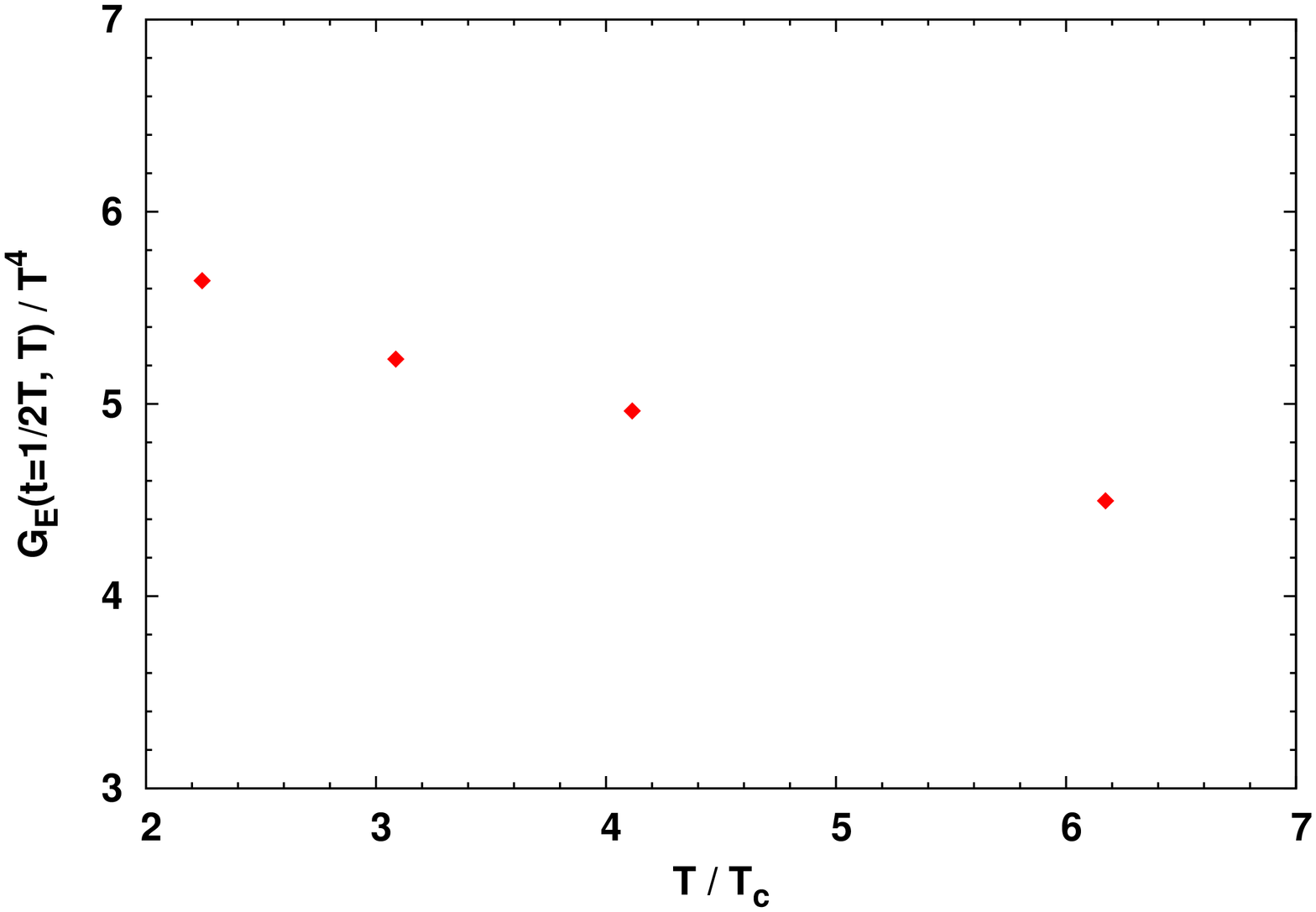}}
\centerline{\includegraphics[angle=-90,width=0.5\textwidth]{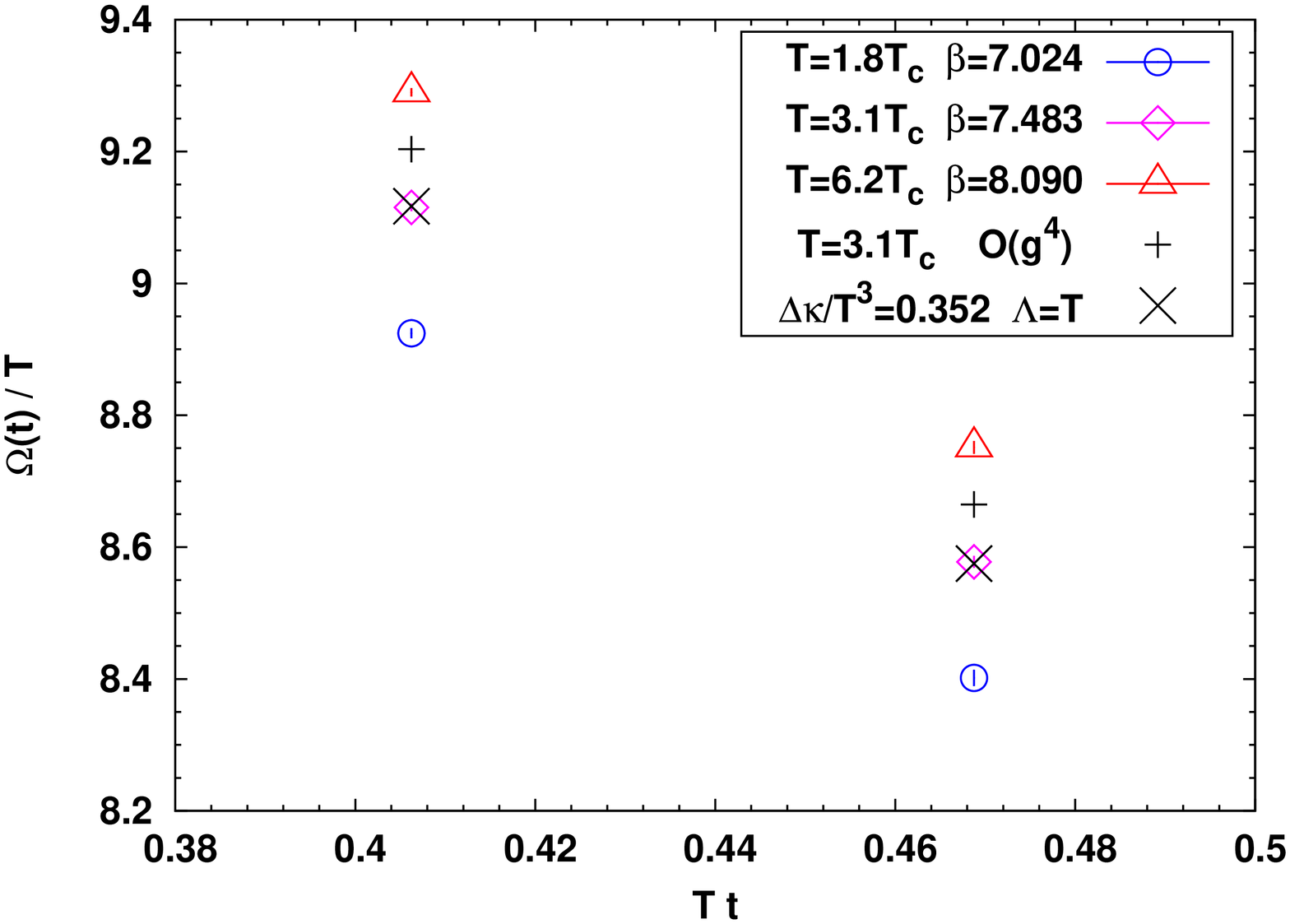}}
\caption{\underline{Top}: the Euclidean force-force correlator at $t=1/2T$ from simulations
with the Wilson action at $\beta=7.483$ and
$\Nt=8,12,16,22$~\cite{Meyer:2010tt}. The absolute normalization has
been determined by matching the data at $tT=\frac{1}{4}$ and
$T=3.1T_c$ with the NLO result of Burnier et al~\cite{Burnier:2010rp}.
\underline{Bottom}: the quantity $\Omega(t)$, defined in \eq(\ref{eq:rat}). The
`$+$' denote the $O(g^4)$ prediction at $3T_c$~\cite{Burnier:2010rp}.
The `$\times$' denote the function $\Omega(t)$ that results from
adding the low-frequency contribution $\Delta\rho(\omega)=\frac{1}{\pi}\Delta\kappa
\tanh(\omega/2T)\theta(\Lambda-|\omega|)$ to the O($g^4$) spectral
function, with parameters $\Delta\kappa$ and $\Lambda$ given in the
caption.}
\label{fig:1012b}
\end{figure}

In the current correlator of a heavy quark flavor, the tail of the
transport peak contains the information on the momentum diffusion
coefficient $\kappa$, see \eq(\ref{eq:kappa_M}).  Here we review a
first calculation of the HQET correlator (\ref{eq:G_HQET}) in the
SU(3) gauge theory~\cite{Meyer:2010tt}, from which $\kappa$ can be
extracted from the Kubo-type formula (\ref{eq:kappa_HQET}). The
classical interpretation of the correlator is that it measures the
(Lorentz) force correlator along the worldline of a heavy quark.

On the lattice the chromo-electric field is not protected from
renormalization (unlike in the continuum). Since this normalizing
factor has not yet been calculated, the next-to-leading order
perturbative calculation of Burnier et al~\cite{Burnier:2010rp} was
used to normalize the lattice correlator. The condition that the
lattice correlator agree with the perturbative correlator at
relatively short distances ($tT=\frac{1}{4}$) and $T=3.1T_c$ was
used. The resulting values of the Euclidean correlator at the maximal
separation of $t=\frac{\beta}{2}$ are displayed in
\fig(\ref{fig:1012b}). The temperature was varied by changing $\Nt$
from 8 to 22 at fixed $6/g_0^2=7.483$. The temperature variation
between $2T_c$ and $6T_c$ is on the order of $20\%$. The sign of the
variation is the same as predicted by the perturbative expression: the
magnitude of the force-force correlator in the gluon plasma increases
as the temperature is decreased.

A quantity that measures the relative fall-off of the Euclidean
correlator is the quantity $\Omega(t)\geq 0$ defined by
\be
\frac{G_E(t-a/2)}{G_E(t+a/2)} =
\frac{\cosh\big[\Omega(t)({\beta}/{2}-(t-a/2))\big]}
     {\cosh\big[\Omega(t)({\beta}/{2}-(t+a/2))\big]}
\la{eq:rat}
\ee
It has a continuum limit, in which $\Omega
\tanh\Omega(\beta/2-t)=-\frac{d}{dt}\log G_E(t)$. One can interpret $\Omega(t)$
as the location of a delta function in the spectral function which by
itself reproduces the local fall-off of the Euclidean correlator.  The
function $\Omega(t)$ is displayed in \fig(\ref{fig:1012b}, bottom
panel) for three different temperatures, where the temperature is
varied this time by varying the bare coupling $g_0^2$ at fixed
$\Nt=16$.  Because the statistical samples of the
numerator and the denominator in \eq(\ref{eq:rat}) are highly
correlated, the numerical results for these ratios have uncertainties
at the few-permille level. Figure \ref{fig:1012b} also displays the
perturbative prediction at $3.1T_c$~\cite{Burnier:2010rp}; it is based
directly on \eq(\ref{eq:rat}) rather than on the continuum version of
this equation, to allow for a more direct comparison with the lattice
data. The lattice $\Omega(t)$ differs from the perturbative one only
by a few percent, namely the latter falls off slightly more steeply.
Since the difference is so small, it could partly be due to
discretization effects. To reduce their influence, only
the largest values of $t$ were considered. One may ask nonetheless, 
how large a difference in the transport coefficient $\kappa$ could this
discrepancy possibly correspond to. An estimate is obtained by adding
a low-frequency correction to the perturbative spectral function,
\be
\Delta\rho(\omega)=                                                                     
\frac{1}{\pi} \cdot \Delta\kappa\, \tanh(\omega/2T)\;\theta(\Lambda-|\omega|).
\la{eq:Drho}
\ee
Other functional forms than (\ref{eq:Drho}), such as a Breit-Wigner
curve, would perhaps be more realistic, but would not change the
conclusions in any significant way.  Adding such a term to the
spectral function has the effect of making the Euclidean correlator
flatter, and indeed, by adjusting $\Delta\kappa$, one can obtain good
agreement for the largest two $t$ values between the perturbative
prediction modified by \eq(\ref{eq:Drho}) and the lattice data.  At
$T=3.1T_c$, for $\Lambda=T$ and $\Delta\kappa/T^3=
0.352(38)$, agreement is obtained with the lattice data at the two
largest $t$ values.  This represents a substantial enhancement of
$\kappa$ over the leading-order perturbative value.
An equally good agreement is obtained if one chooses
$\Lambda=2T$, which leads to $\Delta\kappa/T^3\approx0.204(22)$.
While it is too early to draw phenomenological conclusions, this
increase appears to be not quite sufficient to explain the
experimentally observed elliptic flow of heavy
quarks~\cite{Abelev:2006db,Adare:2006nq,CasalderreySolana:2006rq}.

One of the applications of this type of calculation in the HQET
framework is that it can provide the parameters of the transport peak
in bottomium, and perhaps even charmonium correlators. We refer the
reader to section (\ref{sec:boltz}) for a review of this
connection~\cite{CaronHuot:2009uh}.  This could be instrumental in the
analysis of the `melting' of quarkonia states.

\subsection{Shear and Bulk Viscosities\la{sec:EMT}}

% \begin{figure}
% \centerline{\includegraphics[angle=-90,width=0.35\textwidth]{FIGS/LATCOR/Cshear-log.ps}}
% \centerline{\includegraphics[angle=-90,width=0.35\textwidth]{FIGS/LATCOR/Cshear-tlimp.ps}}
% \caption{Euclidean correlation function $G_E(x_0)$ of shear stress in the SU(3) gauge theory
% at zero spatial momentum. Filled symbols correspond to $1.65T_c$, 
% open symbols to $1.24T_c$. In the bottom plot, the correlators have been normalized to 
% their treelevel values.}
% \label{fig:0704}
% \end{figure}

\begin{figure}
\centerline{\includegraphics[angle=-90,width=0.45\textwidth]{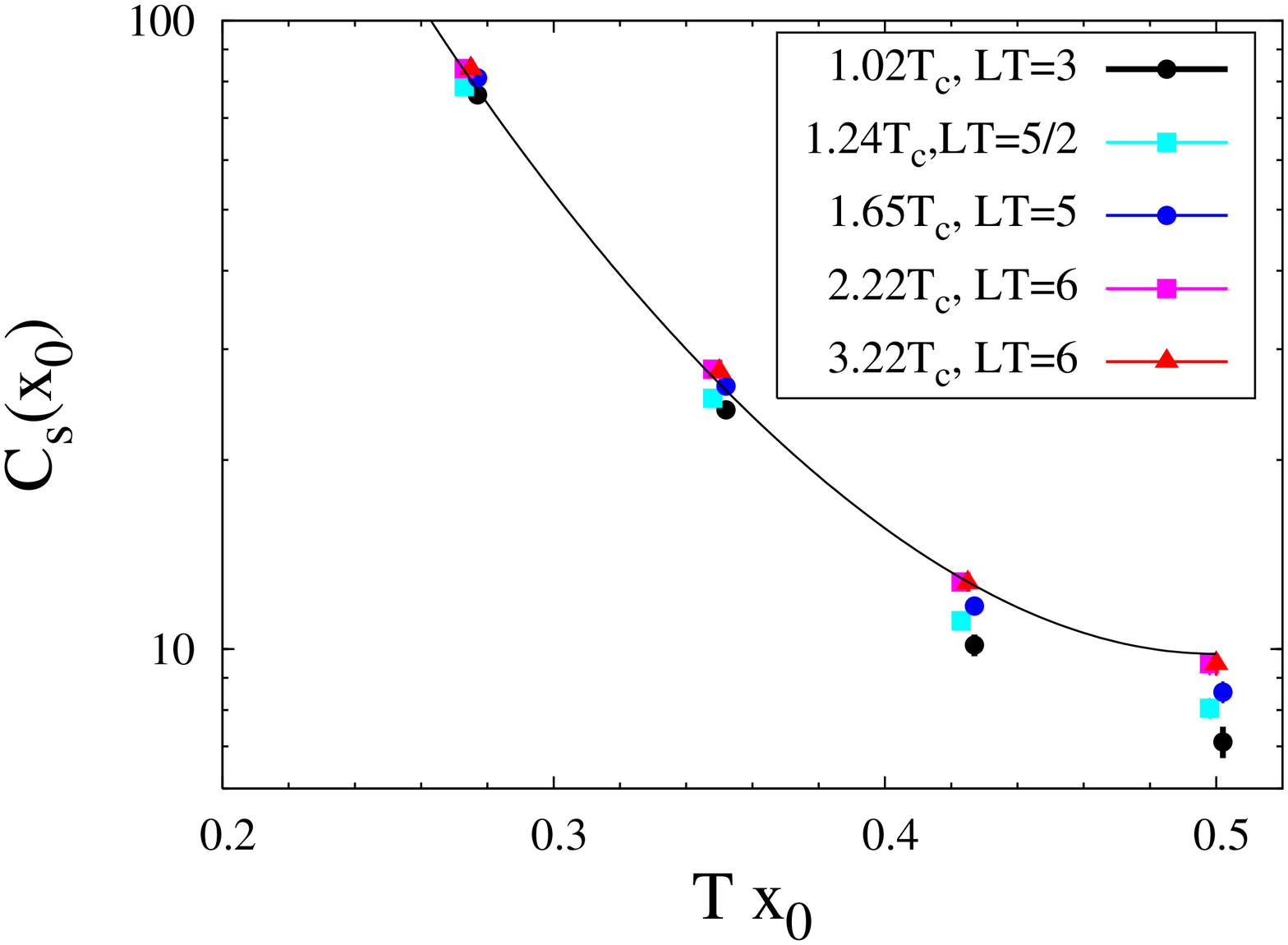}}
\centerline{\includegraphics[angle=-90,width=0.45\textwidth]{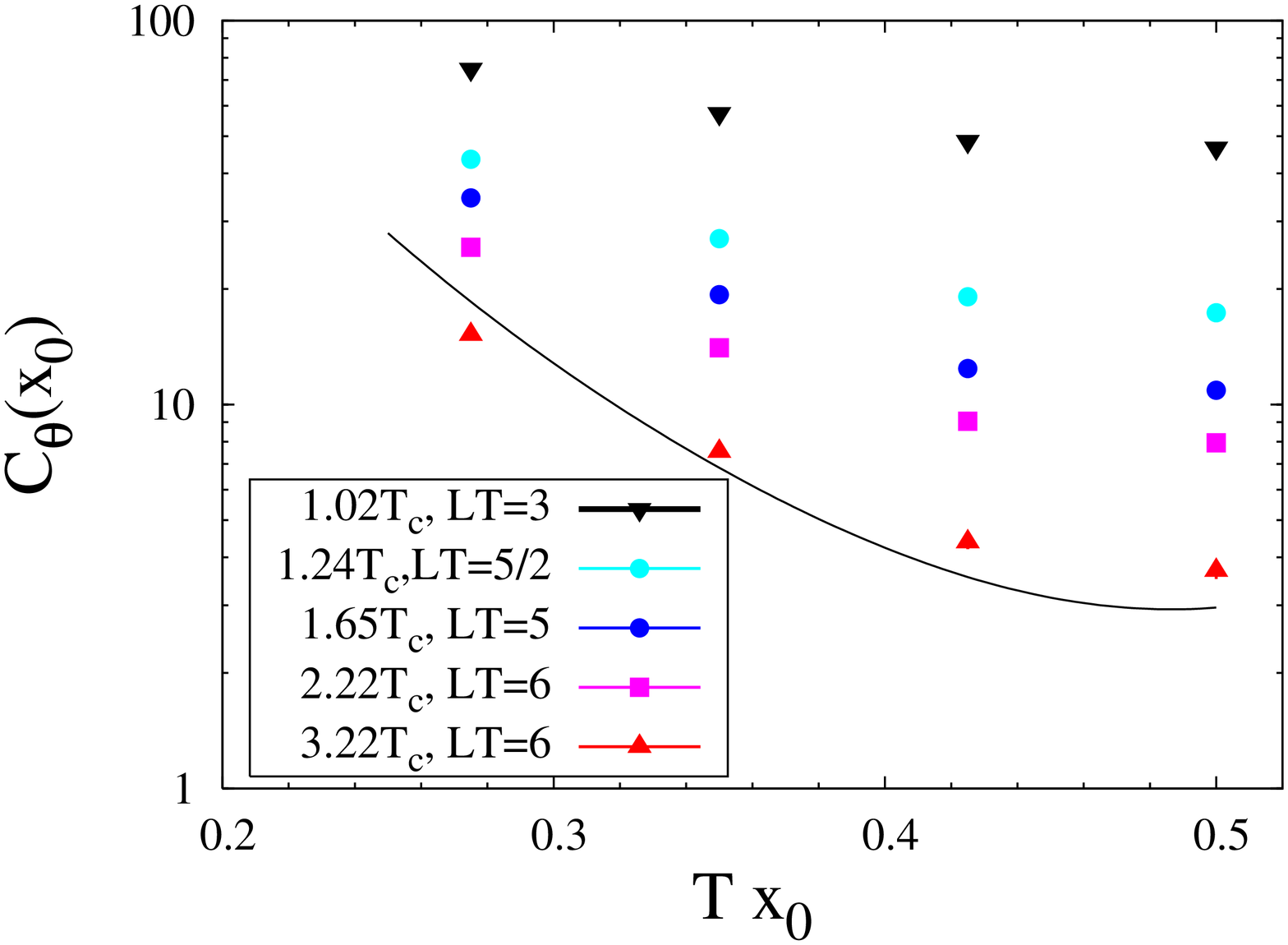}}
\caption{Euclidean correlation function $G_E(x_0)$ of $T_{12}$ (top) and $T_{\mu\mu}$
(bottom) in the SU(3) gauge theory, at zero spatial momentum and several temperatures in 
the deconfined phase with $\Nt=8$~\cite{Meyer:2007dy,Meyer:2007ic}. 
The solid curves represent the treelevel correlators.}
\label{fig:0805}
\end{figure}

The calculation of the shear viscosity is of the greatest importance
for heavy ion phenomenology. But for technical reasons the associated
spectral function is also probably the hardest to constrain by using
lattice simulations. Therefore results have only been obtained in the
SU(3) gauge theory (see also the SU(2) results investigating the behavior
of correlators near the second-order phase transition~\cite{Huebner:2008as}).

The Euclidean correlators in the shear and in the bulk channels are
displayed in \fig(\ref{fig:0805}). They are given as function of
Euclidean time $t$ and are evaluated at vanishing spatial momentum,
$\bk=0$.  The lattices used in these calculations are isotropic and
have $\Nt=8$.  The treelevel correlator is displayed as a solid black
line.  Clearly the departure from this curve is small. This is due to
the large contribution from the ultraviolet frequencies, which results
in a $1/t^5$ behavior at short distances and dominates the Euclidean
correlator for all $t$. The data was obtained with a two-level
algorithm~\cite{Meyer:2002cd,Meyer:2003hy}, which exploits the
locality of the action and the operator $T_{\mu\nu}$, both of which
are formed of color-traces of 1x1 plaquettes. The algorithm averages
out the short-wavelength fluctuations of the two operators separately
before evaluating their correlation. In this way, the `statistics'
accumulated for each of them multiplies with the statistics
accumulated for the other. These stochastic sub-averages do not lead
to a bias in the final expectation value~\cite{Luscher:2001up}. As a
consequence of using this algorithm, the relative statistical
uncertainty on the data points is roughly independent of $t$.

\begin{figure}
\centerline{\includegraphics[angle=-90,width=0.45\textwidth]{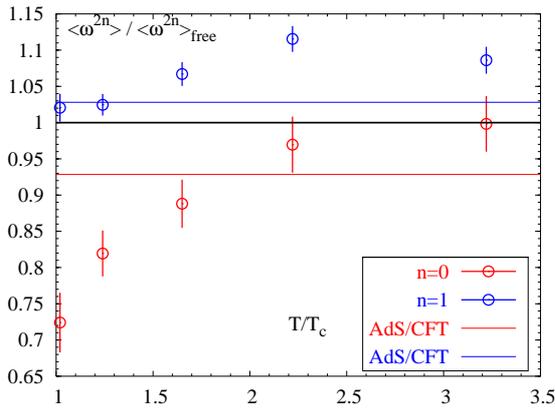}}
\caption{The lowest two moments $n=0$ and 1 defined in \eq\ref{eq:moments}, normalized
to their treelevel values~\cite{Meyer:2008dq}. 
For comparison, the same quantities computed in the strongly coupled
${\cal N}=4$ SYM theory are displayed~\cite{Teaney:2006nc}.} 
\label{fig:0805b}
\end{figure}

In order to see the departure from the free-field behavior, one has to
inspect the correlator in greater detail.  The moments
(\ref{eq:moments}) are directly computable on the lattice.  The first
two moments of the shear correlator, which amount to the value and the
curvature of the Euclidean correlator at $t=\frac{\beta}{2}$, are
displayed in \fig(\ref{fig:0805b}) as a function of temperature. They
were computed on $\Nt=8$ lattices and are normalized to their
treelevel values, which can be obtained straightforwardly by
integrating the spectral functions (\ref{eq:rho_q.eq.0}). While below
$T=2T_c$ one observes a clear departure of the lowest moment from the
non-interacting limit, a higher accuracy (both statistical and
systematic) would be needed to see a statistically significant
departure from the non-interacting limit at the higher temperatures.
The next moment is overall closer to the non-interacting limit, as one
would expect on the basis that it weights the higher frequencies more
strongly than the lowest moment.  Before interpreting the departure of
less than $10\%$ from the free value, the discretization errors would
need to be brought under a control comparable to the statistical
errors. Note, the strong statistical correlations between Euclidean
data points result in a strong reduction in the statistical error for
the second derivative.  The figure also displays the ratio of the
${\cal N}=4$ SYM moments, calculated at infinite and zero
coupling~\cite{Teaney:2006nc,Kovtun:2006pf}. This confirms how
numerically insensitive the Euclidean correlator is to interactions
among the constituents of the plasma. It also shows that the size and
magnitude of the deviations from the free approximation is the same in
both theories. The numerical insensitivity of the Euclidean correlator
to the transport properties was pointed out in~\cite{Aarts:2002cc} (weak
coupling regime) and in~\cite{Teaney:2006nc} (strong coupling regime).

\begin{figure}
\centerline{\includegraphics[angle=-90,width=0.5\textwidth]{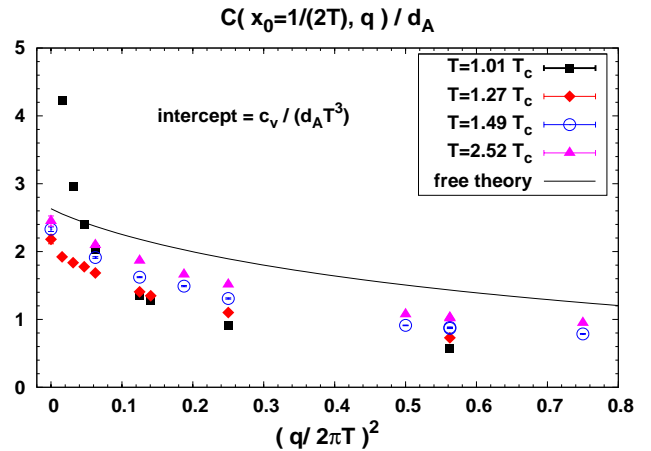}}
\caption{The Euclidean energy density correlator for fixed $t=\frac{\beta}{2}$,
as a function of the spatial momentum $\bq$~\cite{Meyer:2009jp}. In these calculations,
the lattice size is $12\times48^3$ at the two lower temperatures and 
$10\times 20^3$ at the two higher temperatures. The lattice is anisotropic, $\as/\at=2$.
At $1.01T_c$ the intercept is as high as $13.6(1.9)$.} 
\label{fig:0907}
\end{figure}

The conservation of the EMT, $\partial_\mu T_{\mu\nu}=0$, implies in
particular, for ${\bq}=q\hat e_3$ and with Euclidean sign conventions,
\ba 
\omega^4\,\rho_{{00},{00}}(\omega,{\bq}) &=&
-\omega^2\,{q}^2\,\rho_{{03},{03}}(\omega,{\bq}) = q^4
\rho_{33,33}(\omega,{\bq}), 
\nn
-\omega^2\,\rho_{{01},{01}}(\omega,{\bq}) &=&
q^2\,\rho_{13,13}(\omega,{\bq})\,.  
\la{eq:0202e1} 
\ea 
In order to be more sensitive to the infrared part of the shear and
sound channels, it is useful to combine the information from all the
Euclidean correlators that belong to these
channels~\cite{Meyer:2008gt}.  Indeed the spectral function for the
energy density correlator goes like $\bq^4$ at high frequencies
instead of $\omega^4$. Similarly, the momentum density spectral
function goes like $\omega^2\bq^2$ at high frequencies.

The energy-density correlator at non-zero spatial momentum $\bq$ is
particularly interesting. Its main contribution comes from the sound
peak, \eq(\ref{eq:ImGRsound}). The area under this peak is to leading
order the specific heat, $c_v$. At second order in (conformal)
hydrodynamics, the position of the peak also depends on the relaxation
time of the medium (\eq\ref{eq:vsq}). Figure (\ref{fig:0907}) displays
the Euclidean correlator at the fixed time $t=\frac{\beta}{2}$ and as
a function of $\bq$. The correlator falls off fairly slowly, by about
a factor 2 between $\bq=0$ and $|\bq| = \pi T$. This is in qualitative
agreement with the free field expectation. The main difference between
the free field prediction, displayed as a black curve, and the lattice
data is the value of the specific heat which gives the intercept in
\fig(\ref{fig:0907}). In section (\ref{sec:combi}), we describe an
attempt to fit simultaneously several sound channel correlators as a
function of both $t$ and $\bq$.

Very close to the phase transition, a much steeper fall as a function
of $\bq$ is observed. The phase transition is weakly first order in
the SU(3) gauge theory. However the specific heat is very large just
above $T_c$~\cite{Boyd:1996bx}, and this is confirmed by
\fig(\ref{fig:0907}). However for perturbations of wave-vector
$|\bq|\simeq \half \pi T$, the correlator is of the same typical
magnitude as at the higher temperatures. This sharp fall-off of the
energy-density correlator between $|\bq|=0$ and $\half \pi T$ means
that the area under the sound peak falls off rapidly in this
wave-vector interval. This could be due to a corresponding fall-off in
the static susceptibility of $T_{00}$ as a function of the wavelength
of the perturbation. In any case, the Euclidean correlator is
incompatible with the leading-order form of the spectral function
(\ref{eq:ImGRsound}). Possibly, the fluctuations of the order
parameter (the Polyakov loop) have to be treated as slow modes in
hydrodynamics (see for instance~\cite{Kunihiro:2009mm}).  We will
return to the closely related question of the bulk channel near $T_c$
in section (\ref{sec:combi}).

\subsection{Computational Cost Scaling\la{sec:ccscaling}}

\begin{figure}
\centerline{\includegraphics[angle=-90,width=0.5\textwidth]{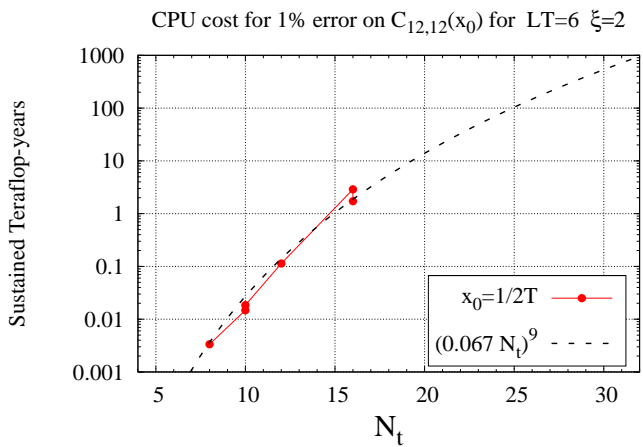}}
\caption{Computational cost scaling for the shear channel correlator~\cite{Meyer:2009jp}.} 
\label{fig:0907cost}
\end{figure}

An important question for the field is how computationally expensive it is 
to improve our knowledge of the spectral functions. A first step is to 
look at the computational cost scaling of the Euclidean correlators.

Let us consider the correlator of $T_{12}$ in the SU($N$) gauge theory, 
out of which the shear viscosity can in principle be extracted.
As long as we are in the deconfined phase, the correlator behaves 
as in a scale-invariant theory, up to small deviations. In other words, 
once the overall dimensionality is scaled out by an appropriate power of 
the temperature $T$, the correlator is to a good approximation a function 
of the variable $tT$. At the level of the lattice simulation, it is of 
course a function of $\Nt,\Ns$ and the bare coupling $g_0$. The statement 
of  scale-invariance translates into the lattice correlator being 
approximately independent of $g_0$, as long as the system is in the
high-temperature phase. The observable, normalized to be finite
both in the continuum and infinite-volume limit, reads 
\be\la{eq:cost_obs}
\lim_{L\to\infty} \lim_{a\to0} \frac{\beta^5}{L^3} 
\Big< {\txts\int} \ud\bx \,T_{12}(t,\bx)  
      {\txts \int} \ud\by T_{12}(0,\by)\Big\>_{\rm conn}.
\ee
Using a straightforward algorithm to compute this observable,
the root of the variance is approximately
\be
\frac{\beta^5}{L^3}\, L^3 
\Big\< {\txts\int} \ud\by \,T_{12}(0,\by)  T_{12}(0,\by) \Big\>_{\rm conn} \propto \Nt^5.
\ee
Therefore the number of independent `measurements' needed scales as
$N_s^0 N_t^{10}$.  This is a very poor scaling indeed, and it means that reducing
the lattice spacing by a factor two at a fixed temperature requires about 
three orders of magnitude as much statistics. Note also that both the observable
(\ref{eq:cost_obs}) and the root of the variance are independent of the volume
(in the large volume regime).

The scaling is however dependent on the way the gauge fields are
sampled.  The fields can be sampled in two levels, and we denote by
$[\dots]$ an average done at the lower level of the
algorithm~\cite{Luscher:2001up,Meyer:2003hy}.  Specifically, the
estimator in a two-level algorithm reads
\be
\frac{\beta^5}{L^3} \Big\langle \big[ {\txts\int} \ud\by\, T_{12}(t,\by) \big] 
        \big[ {\txts\int} \ud\bx\, T_{12}(0,\bx) \big]  \Big\rangle_{\rm conn}.
\ee
The root of its variance is approximately
\be
\sqrt{2} \frac{\beta^5}{L^3}\, L^3 
\Big\langle\big[ {\txts \int} \ud\bx T_{12}(0,\bx) \big]^2\Big\rangle_{\rm conn}
\ee
which scales in the same way as the observable. Therefore the number
of measurements at the higher level of the algorithm is independent of
$\Nt$ and $\Ns$.  On the other hand, the number of measurements at the
lower level of the algorithm must grow like $N_t^5$ in order to yield
a good estimate of the sub-average $\big[ \int \ud\bx \, T_{12}(0,\bx) \big]$.
Therefore the scaling of the number of measurements, all levels confounded, 
is $N_s^0N_t^5$.

Based on these power-counting estimates, the two-level algorithm
yields a reduction of the number of measurements needed for a given
statistical error by five powers of $\Nt$.  In both cases, the number
of measurements must be multiplied by the volume, $N_t N_s^3$, to
obtain the computational cost scaling. At the very least then,
ignoring any critical slowing down, the computational cost scales as 
$N_t^6N_s^3$ with the two-level algorithm.

The prefactors of a scaling law are obviously important, too.
\fig(\ref{fig:0907cost}) displays the actually performed number of
flops for a given statistical error in SU(3) gauge theory
calculations~\cite{Meyer:2009jp}.  These calculations were performed
on anisotropic lattices $\xi=2$, $t$ is set equal to
$\frac{\beta}{2}$, and the parameter $LT=\Ns/\Nt=6$ is kept constant
along the curve. Therefore one expects a scaling with $N_t^9$, which
is consistent with the empirically observed variance of the observable,
although the range of $\Nt$ is not long enough to exclude other scaling
laws.

A recent calculation of the isovector vector current correlator at
$1.5T_c$ in quenched QCD~\cite{Ding:2010ga} shows that, at equal
statistics, the error bars of the correlator at a fixed time
separation only weakly depends on the lattice spacing. Since the
configurations used were separated by a large number of update sweeps,
one can assume that this reflects the behavior of the variance with
the lattice spacing.  It is clearly very different from the behavior
in a flavor-singlet channel discussed above.  This can be understood
on the basis of the fact that the variance can be interpreted in
partially quenched QCD. The cutoff dependence of the variance is then
dominated by the operators of lowest dimension entering the
operator-product expansion of the four-point function. The relevant
operator product is $j_{\rm iso}(y) \tilde j_{\rm iso}(0)$, where
$j_{\rm iso}$ is the isovector current and $\tilde j_{\rm iso}$ is a
`replica' thereof involving two unrelated flavors $\tilde u$ and
$\tilde d$.  Because of the non-trivial flavor quantum numbers, no
operator of dimension less than 6 can contribute to the OPE, and
therefore the variance is UV-finite by power-counting. When the
correlator is calculated with the `point-to-all' technique, neither
the observable nor the variance have a volume dependence in leading
order.  The only source of increase in computational requirements as
the lattice spacing is lowered is therefore the increase in the
condition number of the Dirac operator, roughly $\propto \Nt$ in the
deconfined phase, and the obvious $N_s^3\Nt$ increase in cost for
handling a larger lattice.

\subsection{A variational method?\la{sec:future}}

One lesson of linear response theory is that the operator $B$
used to perturb the Hamiltonian need not be the same operator as the
operator $A$, whose relaxation one wants to study.  It is sufficient
for it to have the same quantum numbers as $A$.  One can even study
the correlation of a whole set of operators $B$ with $A$. Let us take
$A\doteq n(t,\bk)=\int \ud\bx\,e^{i\bk\cdot\bx} n(x)$ for
illustration. Each of the correlators $G^{AB}(t)$ must decay with the
same exponent $\exp{-D\bk^2 t}$, only the coefficient is given by the 
static susceptibility $\chi_s^{AB}\equiv G_R^{AB}(i\epsilon)$, which depends on $B$:
\be
G^{AB}(t) \stackrel{t\to\infty}{\sim}
 \chi_s^{AB} \Gamma \, e^{-\Gamma t},\qquad \Gamma(\bk) = D\bk^2.
\la{eq:Gdecays}
\ee
The susceptibility $\chi_s^{AB}$ expresses how well the operator $B$
is able to induce an excitation of the particle number density with
wavelength $2\pi/k$.  If one were able to find an operator $B$ such
that $G^{AB}(t)$ is a pure decaying exponential \emph{even at short
times}, then one could determine separately the static susceptibility
$\chi_s^{AB}$ and the decay rate $\Gamma = D\bk^2$. What functional
form of the Euclidean correlator one should try to achieve by choosing
an appropriate $B$ operator is given by
\eq(\ref{eq:GEG}). Since the exponential (\ref{eq:Gdecays}) is the
slowest decaying mode, the search for $B$ amounts to finding the
operator $B$ whose Euclidean correlation with $A$ decays as slowly as
possible. This suggests a variational procedure to find the optimal
$B$ operator. Since this procedure can be repeated for every $\bk$,
one can check that the decay rate $\Gamma$ is quadratic in $\bk$, thus
confirming that one is in the regime described by hydrodynamics. We
note that a variational approach to the calculation of charmonium
spectral function has been proposed recently (in a different
form)~\cite{Ohno:2010xx}.

% \newpage

\section{Euclidean Correlators and the Analytic Continuation Problem\la{sec:EC}}
% Completion end of October.

In this section, we discuss the problem of determining the spectral
function $\rho(\omega)$ from Euclidean correlators.  This question
arises either when performing analytic calculations in Euclidean space
before continuing them to real frequencies; or when Monte-Carlo
calculations are done in Euclidean space, and one wants to determine
the main features of the spectral function. Unless explicitly
indicated, we will restrict ourselves to the two-point function of a
Hermitian operator. Therefore the spectral function is a real, odd
function of frequency, and it is given by the imaginary part of the
retarded correlator.

As a historical note, this problem was first studied in the context of
lattice QCD by Karsch and Wyld in the
mid-eighties~\cite{Karsch:1986cq}. In non-relativistic systems, the
problem had already been formulated and different numerical methods
had been proposed to perform the analytic
continuation~\cite{serene,thirumalai,scalapino}.  A regularization of the
inverse problem based on smoothness was introduced in~\cite{jarrell}.
The idea to impose a sum rule constraint was also used
in~\cite{jarrell}.  In the early nineties, the maximum entropy method
was introduced~\cite{Gubernatis:1991zz} and reviewed in
\cite{Gubernatis:1996} in the context of the many-body problem, as well
as in~\cite{Koonin:1996xj} in the context of nuclear physics.  The
recent Ref.~\cite{PhysRevB.82.165125} compares different methods to
extract the optical conductivity. There are still new methods that are
being proposed~\cite{PhysRevB.62.6317}, see notably the stochastic
method introduced in~\cite{krivenko}.

Let $\tilde G_E(\omega_\ell)\equiv G_E^{(\ell)}$ be the
frequency-space Euclidean correlator.  It is initially calculated for
the Matsubara frequencies $\omega_\ell=2\pi T\ell$.  However, since 
it is analytic, it can be continued into the complex plane. In
particular, based on \eq(\ref{eq:l.neq.0}), the retarded correlator
can be obtained by performing the substitution 
\be G_R(\omega) =
\tilde G_E(\omega_\ell \to -i[\omega+i\epsilon]), 
\la{eq:analyconti}
\ee 
and then using
\eq(\ref{eq:rho-ImGR}) one arrives at the spectral function, 
\be
\rho(\omega) = \frac{1}{\pi} \im G_R(\omega).  
\ee 
Examples of spectral function calculations performed in this way can
be found
in~\cite{Laine:2006ns,Asaka:2006rw,CaronHuot:2009uh,Burnier:2010rp}.
The uniqueness of the solution is guaranteed by the Carlson theorem.
This theorem of complex analysis implies that the only function $f$
that is analytic in the upper half plane, is bounded at infinity by
$e^{|cz|}$ with $|c|<\pi$, and vanishes at $z=in$ for $n\geq 0$ is
$f=0$. The condition $|c|<\pi$ is obviously necessary, because
$\sinh\pi z$ fulfills the other conditions.  The analytic
continuation, as a mathematical problem, has been studied in detail
in~\cite{Cuniberti:2001hm}.

\subsection{Formulation of the problem in configuration and in frequency space\la{sec:xw}}

Here we will be mostly concerned with the numerical version of the
problem. The first question one may want to address is in what form to
formulate the problem. Both on the Euclidean side and on the Minkowski
side, one has a choice of working in frequency space or in coordinate
space.  If one chooses frequency space on both sides, the naive
procedure (based on \eq\ref{eq:analyconti}) consists in fitting an
analytic function to the retarded correlator $G_R$ at the discrete
Matsubara frequencies along the imaginary axis, and then evaluate the
function for frequencies just above the real axis. One can also work
entirely in coordinate space, in which case the kernel relating the
Minkowski-time correlator and the Euclidean correlator is given in
\eq(\ref{eq:GEG}).  One then makes an ansatz for the Minkowski-time
correlator and fits it to the Euclidean data. The formulation of the
problem that has been used almost exclusively is the formulation based
on the coordinate-space Euclidean correlator $G_E(t)$ and the
frequency-space Minkowski correlator, which are related by the
relation (see \eq(\ref{eq:ClatRho}), 
\be 
G_E(t) = \int_0^\infty
\ud\omega\, \rho(\omega) \, \frac{\cosh
  \omega({\beta}/{2}-t)}{\sinh{\omega\beta/2}}.  \la{eq:ClatRho2} 
\ee

We are primarily interested in the low-frequency part of the spectral
function. This leads to significant shortcuts. In the fully
frequency-space based approach, the odd derivatives of the spectral
function at the origin can be obtained from the derivatives of the
retarded correlator along the imaginary axis, see
\eq(\ref{eq:dGR}). Since the transport properties can be extracted in
this way via Kubo formulae, there is no need to literally evaluate the
retarded correlator for an argument along the real axis.  In the fully
coordinate-space based approach, the low-frequency behavior of the
spectral function is encoded in the late-time exponential fall-off of
the Minkowski-time correlator, which allows one to extract for
instance a diffusion coefficient ($G(t,k)\sim e^{-Dk^2 t}$) without
having to explicitly perform the passage to frequency space.
Finally, in the formulation (\ref{eq:ClatRho2}), the transport properties
are read off the spectral function using the Kubo formulae.

Working in frequency space, one thus has to interpolate between the
Matsubara frequencies to obtain an estimate of the slope of the
retarded correlator at the origin, yielding the transport coefficient.
If one uses the coordinate-space Euclidean correlator as a starting
point, one has to solve an integral equation numerically, either
\eq(\ref{eq:ClatRho2}) or \eq(\ref{eq:GEG}). In all three cases, the
task represents a numerically ill-posed problem. In the interpolation
case, one may see this by noticing that in practice, only frequencies
up to a certain magnitude are numerically available. In that
situation, a polynomial which vanishes at $\omega=i\ell \cdot 2\pi T$
for $\ell=0,\dots, \ell_{\rm max}$ can always be added with an
arbitrary coefficient to a given tentative solution of the problem
without spoiling the agreement with the available data.  The slope at
the origin is then modified by an arbitrary amount. Of course, such a
high-degree polynomial can be excluded if one takes into account the
information that the retarded correlator grows at most with a certain
power (in an asymptotically free theory, or in theories satisfying
non-renormalization theorems such as ${\cal N}=4$ SYM, this power is
given by dimensional analysis).

\subsection{A numerically ill-posed problem\la{sec:ill-posed}}

To illustrate the ill-posed nature of the problem more quantitatively,
let us consider the problem at zero temperature, when it reduces
to the well-studied problem of inverting the Laplace transform,
\be
G_E(t) = \int_0^\infty \ud\omega \, e^{-\omega t} \rho(\omega).
\ee
Bertero, Boccacci and Pike~\cite{BerteroI} proved the following 
property. The input `data' $G^{\rm data}_E(t) = G_E(t) + \delta G_E(t)$
is necessarily imperfect, but is assumed to satisfy the quality criterion
\be
 \int_0^\infty \ud t\; |\delta G_E(t)|^2 \leq \epsilon^2.
\ee
Let $\delta\rho$ be the function of frequency of which $\delta G_E(t)$ is the Laplace transform.
Then the size of the set $\{\delta\rho\}$, in the sense of the $L^2(0,\infty)$ norm, is infinite.
Suppose however we introduce a limit on the `oscillations' of 
$\delta\rho(\omega)$ in the form
\be
\int_0^\infty \ud\omega \,\omega |\delta\rho'(\omega)|^2 \leq E^2.
\ee
Let $S$ be the set of $\delta\rho$ satisfying this bound.
Then a lower bound on the $L^2$-size of $S$ can be given for small $\epsilon/E$,
\be
{\rm sup}_{\delta\rho\in S} ||\delta\rho||\gtrsim \frac{\pi E}{2|\log(\epsilon/E)|}.
\ee
From here one clearly sees that (a) letting $E$ be arbitrarily large results
in an infinite $L^2$-uncertainty on $\delta\rho$ and (b) 
for a given $E$, the $L^2$-uncertainty only decreases logarithmically
in $\epsilon$. When $G_E^{\rm data}(t)$ comes from a Monte-Carlo simulation,
improving the accuracy is thus exponentially computationally expensive in the desired 
$L^2$-accuracy of the spectral function.

While this fact may seem discouraging, the authors of~\cite{BerteroI}
also prove that the situation improves drastically if the spectral
function is known to satisfy certain analyticity properties.  Suppose
that $\rho$ is analytic in the sector $\omega e^{i\phi}$,
$\forall |\phi|<\alpha$ around the real axis, and that $\rho$ 
is square-integrable along the direction $\phi$. Then they provide the estimate
\be
{\rm sup}_{\delta\rho\in S} ||\delta\rho|| \propto
 \epsilon^\beta,\qquad \beta= \frac{2\alpha}{\pi+2\alpha}.
\ee
In words, the uncertainty decreases much faster in this case.  As a
very favorable example, in the ${\cal N}=4$ SYM vector channel
spectral function (\ref{eq:N4vc}), $\alpha=\frac{\pi}{4}$.

\subsection{Linear methods\la{sec:lin}}

%
% see /home/meyerh/CTPHOPPER/ctphopper-home/PLAQUETTES/INV_PROBL/NEW-METHOD/COMPA-WITH-BACKGILB 
%
% % \cite{PhysRevLett.75.517,Pike}

We start by describing a class of linear methods to solve \eq(\ref{eq:ClatRho2}).
A Fredholm equation of this type can trivially be modified by multiplying the kernel
$K(t,\omega)= \frac{\cosh\omega(\beta/2-t)}{\sinh\omega \beta/2}$ by a
function of $t$ times a function of $\omega$,
\ba
\tilde G_E(t) &=& \int_0^\infty \ud\omega\, \tilde\rho(\omega) \tilde K(t,\omega),
\la{eq:ClatRho3}
\\
\tilde G_E(t) &=& \phi(t) G_E(t),\quad  \tilde\rho(\omega) = \rho(\omega)/m(\omega),
\\
\tilde K(t,\omega) &=& \phi(t) m(\omega) K(t,\omega).
\ea
Since \eq(\ref{eq:ClatRho3}) has the same form
as \eq(\ref{eq:ClatRho2}), we will assume in the following that this
`preconditioning' has been done and that the kernel is regular everywhere.
The twiddle over $G_E$, $\rho$ and $K$ will be dropped. 

We assume that the input data in the inversion problem is a set of $N$
points of the Euclidean correlator, $\{G_i\equiv G_E(t_i)\}_{i=1}^N$,
along with their covariance matrix 
\be
S_{ii'}=\< \delta G_i \delta G_{i'} \>.
\ee
For any linear method, the estimator of the spectral
function can be written in the form
\be
\widehat\rho(\omega) = \sum_{i=1}^N q_i(\omega) G_i,
\la{eq:lin_rhohat}
\ee
where the $q_i(\omega)$ form a set of functions that depends on
the specific method. Ignoring for now the statistical
fluctuations affecting the values $G_i$, we can substitute for it
expression (\ref{eq:ClatRho3}) into \eq(\ref{eq:lin_rhohat}) and
obtain the relation between the estimator $\widehat\rho(\omega)$ and
$\rho(\omega)$,
\ba
\widehat{\rho}(\omega) &=& 
\int_0^\infty \ud\omega'\,\widehat\delta(\omega,\omega') \rho(\omega'),
\la{eq:rhohatrho}
\\
\delta(\omega,\omega') &=& \sum_{i=1}^N q_i(\omega) K(t_i,\omega').
\la{eq:resfct}
\ea
Clearly, in view of \eq(\ref{eq:rhohatrho}) the goal is to have
the \emph{resolution function} $\widehat\delta(\omega,\omega')$ as
close as possible to a delta function (by a measure which is relevant
to the problem).  The finiteness of $N$ results in a loss of
`sharpness' in relation (\ref{eq:rhohatrho}) between the estimator and
the genuine spectral function.  The estimator $\widehat\rho$ can thus
be thought of as a `fudged' version of the true spectral function. As
the number of points $N$ increases, the resolution function becomes
more sharply peaked, as illustrated below.

As the simplest approach, one may choose a set of $N$ linearly independent
functions $u_\ell(\omega)$, and write $\widehat\rho$ as a linear
combination of them, with coefficients to be determined,
\be
\widehat\rho(\omega) = \sum_{\ell=1}^N a_\ell u_\ell(\omega).
\la{eq:rhohat}
\ee
We also define 
\be
G_\ell(t) \equiv \int_0^\infty \ud\omega \,
K(t,\omega) \, u_\ell(\omega)\,.
\la{eq:G_ell}
\ee
The coefficients $a_\ell$ in \eq(\ref{eq:rhohat}) can be worked out
explicitly. The condition that $\widehat\rho$ reproduces the 
Euclidean correlator amounts to 
\be
\sum_{\ell=1}^N G_{i\ell} a_\ell = G_i,
\quad i=1,\dots,N
\la{eq:lin_sys}
\ee
where we have set $G_{i\ell}\equiv G_\ell(t_i)$.
A standard way to proceed is to invert the matrix $G_{i\ell}$ using
the singular-value decomposition,
\be
G = U w V^t,
\la{eq:Gsvd}
\ee
where $U$ and $V$ satisfy
$U^tU=VV^t=V^tV=\boldsymbol{1}$, and $w={\rm diag}(w_1,\dots,w_N)$.
We assume the $w_\ell$ to be ordered, $w_1>w_2>\dots$.
One then obtains the solution of the linear problem,
\ba
q_i(\omega) &=& \sum_{\ell=1}^N \sum_{\ell'=1}^N U_{i\ell'} 
\frac{1}{w_{\ell'}} V_{\ell\ell'}   u_\ell(\omega),
\quad i=1\dots N,
\la{eq:qi1}
% \\  z_\ell &=& w_\ell^{-1}.
\ea
\eq(\ref{eq:qi1}, \ref{eq:lin_rhohat}) provide an estimator $\widehat\rho$ 
of the spectral function which reproduces exactly the input data
points $G_i$ when inserted into the integral equation
(\ref{eq:ClatRho3}).

\subsubsection{Regularization}
So far we have ignored the finite accuracy with which the Euclidean
correlator is known.  The next question to address is therefore how
accurately the Euclidean correlator $G_E(t)$ must be known in order to
determine the coefficients $a_\ell$ with good accuracy.  If $S_{ii'}$
is the covariance matrix of the Euclidean correlator points, then the
covariance matrix of $\widehat\rho(\omega)$ and
$\widehat\rho(\omega')$ reads
\be
\sigma(\omega,\omega') = \sum_{i,i'=1}^N q_i(\omega)S_{ii'} q_{i'}(\omega),
\la{eq:sig_ww}
\ee
with $q_i(\omega)$ given by \eq(\ref{eq:qi1}).  The problem that
arises is that the $w_\ell$ fall off very rapidly as the
mode number $\ell$ increases.  In practice, the direct application
of \eq(\ref{eq:qi1}, \ref{eq:lin_rhohat}) then typically leads to a
wildly oscillating function $\widehat\rho$ with $100\%$ uncertainty.

It is the coefficients of modes of higher $\ell$ that are hardest to
pin down. Therefore the simplest cure to this instability problem is
to restrict $\ell$ to run from 1 to $M<N$ in \eq(\ref{eq:rhohat}).
Since there are now fewer unknowns than input data, one performs
an ordinary, linear fit to a lattice correlator, where the $\chi^2$
reads
\ba
\chi^2[a] &=& \sum_{i,i'=1}^N 
(G_i-\widehat G_i) (S^{-1})_{ii'} (G_{i'}-\widehat G_{i'}),
\la{eq:chi2}
\\
\widehat G_i &\equiv& 
\int_0^\infty \ud \omega\, K(t_i,\omega) \,\widehat\rho(\omega).
\la{eq:Gihat}
\ea
However this straighforward way of proceeding is sometimes unsatisfactory,
because the transition (as $M$ is increased) between having an
unacceptably large $\chi^2$ value and having a poorly conditioned
system (leading to large uncertainties in $\widehat\rho$) is quite
rapid. Therefore it is preferable to have a continuous parameter
which allows one to balance the stability of the procedure against
a faithful description of the Euclidean data. This is achieved by 
minimizing a function obtained by adding a regulating term to the $\chi^2$
(\ref{eq:chi2}), multiplied by a `Lagrange multiplier' $\lambda$:
\be
{\cal F}  = \chi^2 + \lambda R.
\ee
The minimization of ${\cal F}$ with respect to the fit parameters $a_\ell$ can
now be interpreted as the minimization of the $\chi^2$ under the
condition that $R$ take a given value. For $\lambda = 1/\mu$, it is
also equivalent to the problem of minimizing $R+\mu \chi^2$, which
means minimizing $R$ at a fixed value of the $\chi^2$.
There is a significant amount of freedom in choosing $R$.

If one has an \emph{a priori} idea of the functional form of the
spectral function, then $R$ can be chosen to measure the departure of
$\widehat\rho$ from that functional form. Typically the expected
functional form is \emph{smooth}, so that the regulating term $R$ tends to
make the estimator $\widehat\rho$ less wildly oscillating.  This kind
of regulator is used also in non-linear methods, such as the Maximum
Entropy Method (see \eq\ref{eq:SJ2}).  A typical example would be a `kinetic energy' term
$
R = \int_0^\infty \ud\omega\, \big(\frac{\ud\widehat\rho}{\ud\omega}\big)^2g(\omega),
$
where $g(\omega)>0$ is a weight function.

A different kind of regulator directly aims at reducing
the \emph{statistical variance} of $\widehat\rho$, for instance 
\be
R= \int_0^\infty \!\!\ud\omega\, \int_0^\infty \!\!\ud\omega'\, 
D(\omega,\omega') \sigma(\omega,\omega')
\la{eq:Reg}
\ee
for some symmetric weighting function $D(\omega,\omega')$.  If the
goal is to minimize the variance of $\widehat\rho$ at the origin, one
chooses $D(\omega,\omega')=\delta(\omega)\delta(\omega')$.  This sort
of regularization is employed e.g.~in the Backus-Gilbert method
(see \cite{Press:2007zz} and Refs.~therein), and has also been
proposed in the context of the analytic continuation of Euclidean
correlators in~\cite{krivenko}.

\subsubsection{Choosing an appropriate basis of functions}
% {A specific choice of basis and regulator}

For concreteness we give a specific realization of the general
procedure outlined above. Both in $t$-space and in $\omega$ space, one
should first specify a scalar product, which is given by a measure.
In $t$-space, the inverse covariance matrix $(S^{-1})_{ii'}$ provides
a natural measure, see for example expression (\ref{eq:chi2}). In
frequency space, we define a scalar product via a symmetric kernel
$D(\omega,\omega')$ as in \eq(\ref{eq:Reg}).  We will see that this
kernel ultimately allows one to adjust what part of the spectral
function one would like to reduce the statistical uncertainty on, and
also what shape one wants the resolution function to have.

We define the symmetric frequency-space kernel
\be
H(\omega,\omega') = \sum_{i,i'=1}^N (S^{-1})_{i,i'}
 K(t_i,\omega) K(t_{i'},\omega')\,.
\la{eq:H}
\ee
There are functions with positive eigenvalues
that satisfy the following generalized eigenvalue problem,
\ba\la{eq:u_ell}
&&\int_0^\infty \ud\omega' \, 
H(\omega,\omega')u_\ell(\omega') 
\\ && \qquad = w_\ell^2 
\int_0^\infty \ud\omega'\, D(\omega,\omega') u_\ell(\omega')\,,
\nonumber
\ea
and their orthogonality and completeness properties take the form
\ba
\!\!\! \int_0^\infty \ud\omega\, u_\ell(\omega) D(\omega,\omega') u_{\ell'}(\omega') = \delta_{\ell \ell'}\,,
\la{eq:orthogonality}
\\
\!\!\! \sum_{\ell=1}^{\infty} u_\ell(\omega) \int_0^\infty \ud\omega'' \,
D(\omega',\omega'') u_{\ell}(\omega'') = \delta(\omega-\omega').
\la{eq:completeness}
\ea
We order the eigenfunctions so as to make the sequence
$\omega_1>\omega_2\dots$ monotonically decreasing.  The motivation for
choosing this basis is that the functions with largest $w_\ell$
contribute most to the correlator, as seen in \eq(\ref{eq:qi}) below.
Therefore it is natural to determine the components of
$\widehat\rho(\omega)=\sum_{\ell=1}^M a_\ell u_\ell(\omega)$ in the linear
subspace that the correlator is overall most sensitive too.

\begin{figure}
\centerline{\includegraphics[width=0.35\textwidth,angle=-90]{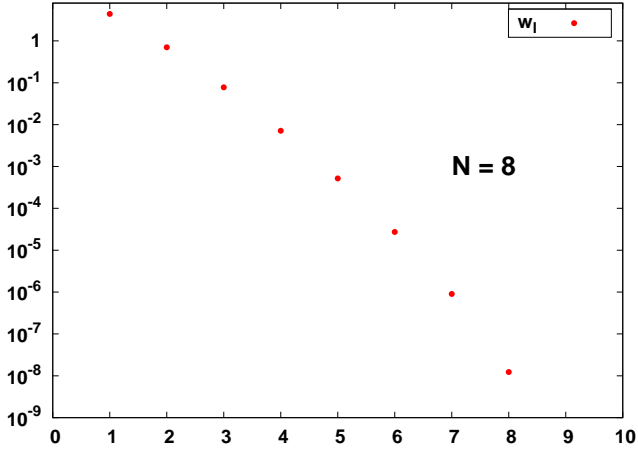}}
\caption{Non-vanishing eigenvalues $w_\ell$ corresponding to operator
 $H(\omega,\omega')$ (\eq\ref{eq:Hn}
with $K(t,\omega)= \frac{\cosh\omega(\frac{\beta}{2}-t)}{\cosh\omega\beta/2}$)
for $N=8$ and $Tt_i=\half-\frac{i}{20}$, $i=0\dots N-1$.}
\label{fig:w_ell}
\end{figure}

With $G_{i\ell}\equiv G_\ell(t_i)$, one easily checks that 
\be
w_\ell^{-1} w_{\ell'}^{-1} \sum_{i,i'=1}^N G_{i\ell} (S^{-1})_{ii'} G_{i'\ell'}
=\delta_{\ell\ell'}.
\la{eq:key}
\ee
In other words, if $S=L_sL_s^{t}$ is the Cholesky decomposition of $S$, the $N\times M$ 
matrix
\be
U_{i\ell} \equiv \sum_{i'=1}^{N} (L_s^{-1})_{ii'} G_{i'\ell} \,\frac{1}{w_\ell}
\la{eq:Vil}
\ee
is orthogonal, $U^t U  = \boldsymbol{1}$.  
We now restrict ourselves to $M\doteq N$.
The solution to \eq(\ref{eq:lin_sys}) then yields a spectral function of 
the form \eq(\ref{eq:lin_rhohat}), with
\be
q_i(\omega) = \sum_{i'=1}^N   (S^{-1})_{ii'}
\sum_{\ell=1}^N \frac{1}{w_\ell^2} \cdot G_{i'\ell}\, u_\ell(\omega).
\la{eq:qi}
\ee
The resolution function (\ref{eq:resfct}) is given here by 
\be
\widehat\delta(\omega,\omega')= \sum_{\ell=1}^N u_\ell(\omega)
\int_0^\infty \ud\omega'' D(\omega',\omega'') u_\ell(\omega'').
\la{eq:resolfct}
\ee

If $S_{ii'}$ is the
covariance matrix of the Euclidean correlator points, then the
covariance matrix of the coefficients $a_\ell$ is given by
\be
\sigma_{\ell,\ell'} = \frac{1}{w_\ell w_{\ell'}} \delta_{\ell\ell'}.
\la{eq:sig_ll}
\ee
Since the eigenvalues $w_\ell$ appear in the denominator
of \eq(\ref{eq:sig_ll}), the contribution of mode $\ell$ is guaranteed
to be well determined (in terms of statistical uncertainty) if $w_\ell$
is not too small.
Finally, one can convert the covariance of the coefficients $a_\ell$
into the covariance between two values of $\widehat\rho$, 
weighted by the kernel $D$,
\be
\la{eq:tot_sig}
\int_0^\infty \!\!\ud\omega\, \int_0^\infty \!\!\ud\omega'\,
D(\omega,\omega') \sigma(\omega,\omega')
= \sum_{\ell=1}^N \frac{1}{w_\ell^2}.
\ee

To interpret \eq(\ref{eq:resolfct}) and (\ref{eq:tot_sig}), it is
easiest to set
$D(\omega,\omega')=f(\omega)\delta(\omega-\omega')$. Then these
formulae become
\ba
\widehat\delta(\omega,\omega')  &=& \sum_{\ell=1}^N 
 u_\ell(\omega) f(\omega') u_\ell(\omega'),
\\
\int_0^\infty \ud\omega \,f(\omega) \sigma^2(\omega,\omega) 
&=& \sum_{\ell=1}^N \frac{1}{w_\ell^2}.
\ea
Thus one sees that if $f$ is chosen to be peaked at some frequency $\omega_0$, 
the resolution function $\delta(\omega_0,\omega')$ will be more concentrated
around that point. But this comes at the cost of the whole variance
$\sum_{\ell=1}^N \frac{1}{w_\ell^2}$ being concentrated around $\omega_0$.
Suppose that the spectral function is expected to have a narrow peak
at $\omega_0$. To balance the two effects against eachother, one can maximize
the figure-of-merit $S/\sqrt{\sigma^2+B}$, where $S$ is the area under the
peak, $\sigma\doteq \sigma(\omega_0,\omega_0)$ and $B$ is the area under the 
`smooth' part of the spectral function. The task is reminiscent of choosing
a bin size in experimental particle physics when trying to extract a signal $S$
from a background $B$.

To regularize the problem, an option is to minimize $\chi^2+\lambda R$, 
with $R$ given by \eq(\ref{eq:Reg},\ref{eq:sig_ww}), but with 
\be
q_i(\omega) = \sum_{i'=1}^N   (S^{-1})_{ii'}
\sum_{\ell=1}^N z_\ell \cdot G_{i'\ell}\, u_\ell(\omega),
\la{eq:qi2}
\ee
and the $z_\ell$ are treated as the fit parameters.

\subsubsection{Illustration and an example from the literature}

For illustration, a possible set of orthogonal functions 
is formed by the $N$ non-zero modes of the kernel
\be
H(\omega,\omega') = \sum_{i=1}^N K(t_i,\omega) K(t_i,\omega').
\la{eq:Hn}
\ee
Upon discretizing the frequency variable, $\omega_j= j\delta\omega$,
these modes can be obtained conveniently from the singular-value decomposition
of the matrix $K_{ij}\equiv \sqrt{\delta\omega} K(t_i,\omega_j)$, $K^t =  Uw V^t$.

The eigenvalues $w_\ell$ are displayed in \fig(\ref{fig:w_ell}) for
the case $m(\omega)=\tanh(\omega/2T)$. Note that these eigenvalues are
now solely a property of the kernel $K(t,\omega)$. Unfortunately, they
fall off very rapidly as $\ell$ increases.  This means that already
the fourth or fifth normalized eigenfunction $u_\ell(\omega)$ make
only a small contribution to the Euclidean correlator
(recall \eq(\ref{eq:key})), in other words the correlator has little
sensitivity to it.

\begin{figure}
\centerline{\includegraphics[width=0.37\textwidth,angle=-90]{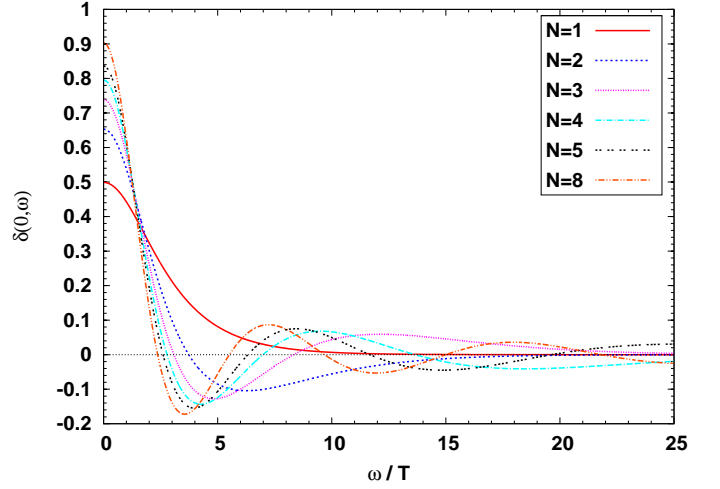}}
\caption{Resolution function 
$\delta(\omega',\omega)=\sum_{\ell=1}^N u_\ell(\omega) u_{\ell}(\omega')$ 
at the origin ($\omega'=0$), for different values of $N$. The 
$u_\ell(\omega)$ are eigenfunctions of $H(\omega,\omega')$ (\eq\ref{eq:Hn}
with $K(t,\omega)= \frac{\cosh\omega(\frac{\beta}{2}-t)}{\cosh\omega\beta/2}$).}
\label{fig:resfct}
\end{figure}

The resolution function centered around the origin is displayed
in \fig(\ref{fig:resfct}).  It is clear that it becomes more strongly
peaked as $N$ increases, but that it takes quite a large $N$ for it to
become narrow on the scale of the temperature $T$.

\begin{figure}
\centerline{\includegraphics[width=0.37\textwidth,angle=-90]{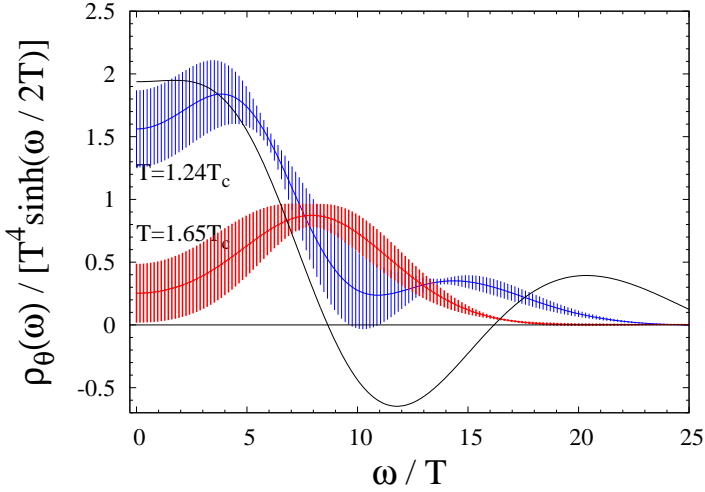}}
\caption{The spectral function $\widehat\rho(\omega,T)$ in the bulk 
channel of SU(3) gauge theory reconstructed by a linear method 
from $N_t=12$ lattices at two different temperatures~\cite{Meyer:2007dy}. 
The bulk viscosity is estimated by $\zeta/T^3=(\pi/18)\times {\rm intercept}$.
The oscillating curve is the (rescaled) resolution function $\widehat\delta(0,\omega)$.}
\label{fig:bulk}
\end{figure}

In~\cite{Meyer:2007dy}, the bulk channel was studied, more precisely
the $\bp=0$ correlation function of $\theta$. The spectral function 
was parametrized as 
\be
\rho(\omega) = m(\omega)[1+\sum_{\ell=1}^N a_\ell u_\ell(\omega)]
\ee
with 
\be
m(\omega) = \frac{A\omega^4}{\tanh(\frac{\beta}{4}\omega) \tanh^2(c\beta\omega)}.
\la{eq:mw2}
\ee
The constant $A$ is given by leading order perturbation theory, 
\eq(\ref{eq:rho_q.eq.0}.  The
sensitivity of the final result to the parameter $c$ was probed by
varying it by a factor 2 around the value $\frac{1}{4}$.  The
$u_\ell(\omega)$ were the eigenmodes of the operator (\ref{eq:Hn}),
with $K(t,\omega) =
m(\omega)\frac{\cosh\omega(\frac{\beta}{2}-t)}{\sinh\frac{\beta}{2}\omega}$
and $m(\omega)$ given by \eq(\ref{eq:mw2}).  Since $N$ was only 4, the
determination of the coefficients $a_\ell$ did not require invoking a
regulating term. The resulting spectral function, as well as the
resolution function associated with this reconstruction, are displayed
in \fig(\ref{fig:bulk}). Due to the large UV tail of $m(\omega)$, the
resolution function $\widehat\delta(0,\omega)$ is rather broad in
units of temperature. This clearly signals that only the gross
features of $\widehat\rho(\omega)$ are reliable. The
actual spectral function may well contain more rapidly varying features
which are washed out in this analysis, due to the paucity of the data.

% For $M>N$, the system is underdetermined; when $M< N$, the system is
% overdetermined. An additional ingredient is thus needed to make the
% problem well-posed.  For instance, one may parametrize the spectral
% function as a $\rho(\omega) = m(\omega)[1+ r(\omega)]$, and subtract
% the contribution of the $m(\omega)$ term on both sides
% of \eq(\ref{eq:ClatRho3}).  The idea is now that if $m(\omega)$ is
% already a reasonable approximation to the spectral function, the
% residual $r(\omega)$ should be a `small' correction.  For this reason
% a standard way to proceed is to use the singular-value decomposition
% of the $N\times M$ matrix $G_{i\ell}$,

\subsubsection{The method of Cuniberti, de Micheli and Viano}

% \hyphenation{fre-quen-cy}

%%%%%%%%%%%%%%%%%%%%%%%%% FIGURE %%%%%%%%%%%%%%%%%%%%%%%%%%%%%%%%%%%%%%%%%
%
\begin{figure}[t]

% \hspace*{1.0cm}%
\centerline{
\begin{minipage}[c]{6.5cm}
\begin{picture}(160,160)(-80,-80)
\SetScale{1.0}  
\SetWidth{0.5}
\Line(7,0)(7,75)%
\Line(-7,0)(-7,75)%
\CArc(0,0)(7,180,360)%
\ArrowLine(7,65)(7,75)%
\ArrowLine(-7,75)(-7,65)%
\BBox(-97,67)(-63,83)%
\SetWidth{1.0}
\LongArrow(-80,0)(80,0)%
\LongArrow(0,-80)(0,80)%
\ZigZag(0,0)(0,70){2}{16}%
\ZigZag(-50,0)(-50,70){2}{16}%
\ZigZag(50,0)(50,70){2}{16}%
\Text(75,75)[c]{$\tau$}
\Line(70,70)(80,70)%
\Line(70,70)(70,80)%
\Text(-80,75)[c]{$\cal{G}^+(\tau)$}%
\Text(12,70)[l]{$t$}%
\Text(11,55)[l]{$J^+(t)$}%
\Text(-55,-10)[c]{$-\beta$}%
\Text(50,-10)[c]{$\beta$}%
\end{picture}
\end{minipage}}%
\vspace*{1.0cm}%
\centerline{
\begin{minipage}[c]{6.5cm}
\begin{picture}(160,160)(-80,-80)
\SetScale{1.0}  
\SetWidth{0.5}
%\GBox(-80,-80)(0,80){0.8}
\CBox(-80,-80)(0,80){Cyan}{Cyan}
\Line(7,75)(7,-75)%
\ArrowLine(7,-65)(7,-75)%
\Line(32,-3)(38,3)%
\Line(32,3)(38,-3)%
\Line(67,-3)(73,3)%
\Line(67,3)(73,-3)%
\BBox(-97,67)(-63,83)%
\SetWidth{1.0}
\LongArrow(-80,0)(80,0)%
\LongArrow(0,-80)(0,80)%
\Text(75,76.5)[c]{$\zeta$}
\Line(70,70)(80,70)%
\Line(70,70)(70,80)%
\Text(-80,75)[c]{$\tilde{J}^+(\zeta)$}%
\Text(12,-70)[l]{$\omega$}%
\Text(11,-55)[l]{$\tilde{R}(\omega)$}%
\Text(35,-10)[c]{$2\pi T$}%
\Text(70,-10)[c]{$4\pi T$}%
\end{picture}
\end{minipage}}%
\caption[a]{
The complex planes, basic functions, and analytic structures 
(figure from~\cite{Burnier:2011jq}).
} 
\la{fig:planes}
\end{figure}
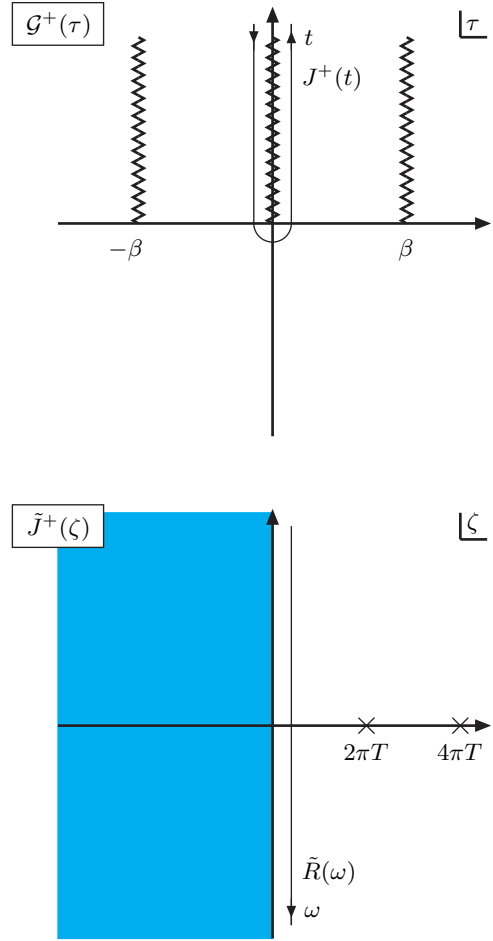

The analytic structure of Minkowski and Euclidean correlators was studied 
thoroughly in~\cite{Cuniberti:2001hm}.
The basic quantities defined there are the one-sided sum 
\be
 {\mathcal G}^+(\tau,\cdot) \equiv 
 T \sum_{\omega_n \ge 0} \tilde {\mathcal G}
 (\omega_n,\cdot) e^{-i \omega_n \tau} , 
\ee
which is analytic for 
$\im\tau < 0$ but has cuts in the upper half-plane; 
its discontinuity across the cut starting at the origin, 
\be
 J^+(t,\cdot) \equiv i \bigl[ {\mathcal G}^+(\epsilon + i t,\cdot) - 
 {\mathcal G}^+(-\epsilon + i t,\cdot) \bigr], 
\ee
$t>0$, $\epsilon = 0^+$, 
which equals the retarded real-time correlator, 
${\mathcal R}(t,\cdot)$; 
as well as its Laplace transform 
\be
 \tilde J^+(\zeta,\cdot) \equiv
 \int_0^\infty \! {\rm d}t \, e^{-\zeta\, t} J^+(t,\cdot), 
\ee
which is analytic for $\re \zeta > 0$. 
For $\zeta = \omega_n$, $\tilde J^+(\zeta,\cdot)$
reduces to the Fourier components $\tilde {\mathcal G}(\omega_n,\cdot)$,
and therefore constitutes the desired analytic continuation 
to a complex half-plane. 
The value of $\tilde J^+(\zeta,\cdot)$
along the axis $\zeta = \epsilon -i \omega $, 
$\omega \in \mathbb{R}$, 
yields the Fourier transform of the retarded correlator, 
$\tilde{\mathcal R}(\omega,\cdot)$, 
whose imaginary part in turn equals the spectral function. 
The basic analytic structure is illustrated in  \fig\ref{fig:planes}.

Cuniberti et al.~\cite{Cuniberti:2001hm} also proposed a practical
procedure to analytically continue the Euclidean correlator.  It was
recently reviewed and tested in a realistic example by Burnier et
al.~\cite{Burnier:2011jq}. Here we follow the notation and treatment
of the latter publication. From our point of view the most important
result of Cuniberti et al. is that the retarded correlator $G(t)$ can
be obtained as a series in Laguerre polynomials (normalized so that
$L_\ell(0)=1$),
\ba
\!\!\! G(t) &=& \exp\big[-\exp(-2\pi Tt) \big] \,
\sum_{\ell=0}^\infty a_\ell \,  L_\ell(2e^{-2\pi Tt}),
\la{eq:G_Laguerre}
\\
 a_\ell &\equiv &
 2 (-1)^\ell \sum_{n=0}^\infty
 \frac{(-1)^n}{n!} G_E^{(n+1)} \,  _2 F_1(-\ell,n+1;1;2)
%  \;, \quad \ell = 0, 1, 2, \ldots 
\;.~~ \la{eq:al_ideal}
\ea
Here ${}_2F_1$ is the hypergeometric function; Burnier et al. provide
a recurrence relation to compute the combination appearing in
\eq(\ref{eq:al_ideal}). From here the spectral function can be
obtained by Fourier transformation of $G(t)$, \eq(\ref{eq:rhodef}),
exploiting if applicable the symmetry properties of $G(t)$
\eq(\ref{eq:Gisodd},\ref{eq:Gisreal}). However, if the goal is to
determine transport properties, they can also be read off directly
from the $t\to\infty$ behavior of $G(t)$ (see e.g. \eq\ref{eq:Gxixi-L}
and \ref{eq:rhoxixi-L}). Finally we note that formula (\ref{eq:G_Laguerre})
is valid for $G_E(t)$ continuous at $t=0$~mod~$\beta$. In relativistic 
theories this property will in general only be fulfilled after 
a subtraction of the hardest UV contributions.

Since the required input into formula (\ref{eq:G_Laguerre}) is the
Euclidean correlator in frequency-space, we may assume that the latter
has been calculated on the lattice for the $N$ first Matsubara
frequencies.  The practical recipe that the result
(\ref{eq:al_ideal}) suggests is to truncate the sum
(\ref{eq:G_Laguerre}) at some $\ell=\ell_{\rm max}$~\cite{Burnier:2011jq}. 
Various criteria can be used to motivate the choice of $\ell_{\rm  max}$. 
In the original proposal, based on mathematical arguments the suggestion was to look for a 
plateau in $\sum_{\ell=1}^{\ell_{\rm max}} |a_\ell|^2$ as a function of $\ell_{\rm max}$.
An alternative, more physical condition put forward in~\cite{Burnier:2011jq} is 
to require that $G(t)\stackrel{t\to\infty}{\to}0$~\cite{Arnold:1997gh}, 
which is equivalent to $\sum_{\ell=1}^N a_\ell = 0$. We remark that through
the dependence of $\ell_{\rm max}$ on the data, one introduces a non-linearity
of the spectral function on the input data.

In the case that the channel under consideration admits a diffusion pole, 
the asymptotic behavior of $G(t)$ is (disregarding non-linear effects that
generically lead to power-law tails~\cite{Kovtun:2003vj}),
\be
G(t) \sim \chi D\bk^2 \,e^{-D\bk^2 t}.
\ee
It is interesting to see how this behavior arises from a formula such
as (\ref{eq:G_Laguerre}). At large times, $e^{-2\pi Tt}$ will be very
close to zero, and therefore the Laguerre polynomials are evaluated
near the origin in this regime.  Since the leading term, $\sum_\ell
a_\ell$, must vanish for the retarded correlator to vanish at
infinity, the exponential fall-off must arises from the derivative of
the Laguerre polynomials at the origin.  It is thus clear that the
expression with a finite number of terms will always fall off as
$e^{-2\pi Tt}$ for sufficiently large $t$. However, this asymptotic behavior 
can of course be different when all the terms are included. Specifically, 
we note the formula
\be
x^\beta = \Gamma(\beta+1) \sum_{i=0}^\infty (-1)^i 
\left(\begin{array}{c} \beta \\ i \end{array} \right)  L_i(x),
\ee
which for $x= 2e^{-2\pi T t}$ and $\beta=D\bk^2/2\pi T$ allows one 
to express the diffusion constant $D$ directly in terms of the large-$\ell$
behavior of $a_\ell$.

Burnier et al. performed a practical test of the method on a
UV-subtracted Euclidean correlator. With $N$ equally spaced data
points in Euclidean time, they tested the accuracy of the
reconstructed spectral function, assuming a relative accuracy of
$10^{-3}$ or better on the Euclidean mock data. In this case, for the
kind of test spectral function used, the accuracy achieved on
$\rho(\omega)$ is better than $50\%$.  Another lesson is that doubling
$N$ from 24 to 48 only moderately improves the result, and the
accuracy must be drastically improved to benefit from the additional
points. We expect the latter to be a general qualitative conclusion.

\subsection{The Maximal Entropy Method\la{sec:MEM}}

The Maximal Entropy Method (MEM) was first applied to the
determination of spectral functions in lattice QCD
in~\cite{Nakahara:1999vy}. The method is described in detail
in~\cite{Asakawa:2000tr} (see also the older
review~\cite{Gubernatis:1996}).  Some technical improvements have been
introduced more recently (\cite{Aarts:2007wj,Engels:2009tv}).  It is
noteworthy that the method is also currently being used to study the
spectral function in non-relativistic
systems~\cite{PhysRevB.81.155107}.

From a pragmatic point of view, this method is also based on the
minimization of $\chi^2 +\lambda R$, where $R$ acts as a regulator.
But the difference with linear methods is that neither the $\chi^2$ nor
$R$ are any longer quadratic in $\widehat\rho$.  Using a more standard
notation the quantity that is maximized is thus not simply
$e^{-\frac{1}{2}\chi^2}$, but rather
\ba \la{eq:SJ}
\!\!\! P&\propto & \exp\left\{-\frac{1}{2}\chi^2+\alpha S\right\},
\\
\la{eq:SJ2}
\!\!\! S[g]&=& \int_0^\infty\!\! \ud\omega \, \left[ 
\widehat\rho(\omega)-m(\omega)-\widehat\rho(\omega)\log\frac{\widehat\rho(\omega)}{m(\omega)}
\right].
\ea
For Gaussian distributed data, minimizing the $\chi^2$ corresponds to 
the condition of maximum likelihood of the parameters, given the observed data.
The functional $S$ is called the Shannon-Jaynes entropy, and measures the departure
of the reconstructed $\rho(\omega)$ from the prior function $m(\omega)$.
The relative weight given to the `prior knowledge', encoded by $S$, and the 
observed data, encoded by the $\chi^2$, is fixed by the parameter $\alpha$.
We shall return to the question of how to set its value shortly.

Similar to the linear methods, the singularity of the kernel should be
removed by multiplying it by a function which is proportional to
$\omega$ at small frequencies, and compensating by dividing
$\rho(\omega)$ by this function~\cite{Aarts:2007wj,Engels:2009tv}.

The spectral function $\rho(\omega)$ is assumed to be positive.
It is parametrized as follows,
\be
\la{eq:mem-ansatz}
\widehat \rho(\omega) = m(\omega) \, \exp f(\omega),
\ee
where $m(\omega)$ is the same `prior' function as in \eq(\ref{eq:SJ2}).
The exponent $f$ is parametrized using a similar basis of functions as 
described in (\ref{sec:lin}),
\ba
f(\omega) &=& \sum_{\ell=1}^M a_\ell u_\ell(\omega)
\ea
where $u_\ell(\omega)$ are the $M$ eigenfunctions of highest
eigenvalue of the eigenvalue problem (\ref{eq:u_ell}), with $M\leq N$.
In the recent literature, the basis was obtained by performing the
singular-value decomposition of the matrix $K_{ij} \equiv
K(t_i,\omega_j)$, $K^t=Uw V^t$.  This corresponds to choosing
$H(\omega,\omega') = \sum_{i=1}^N K(t_i,\omega) K(t_i,\omega')$ and
$D(\omega,\omega')=\delta(\omega-\omega')$ in \eq(\ref{eq:u_ell}). The
solution to the minimization of $\half \chi^2-\alpha S$ is then obtained
iteratively, for example with the Marquardt-Levenberg
algorithm~\cite{Press:2007zz}.

\begin{figure}
\centerline{\includegraphics[width=0.47\textwidth]{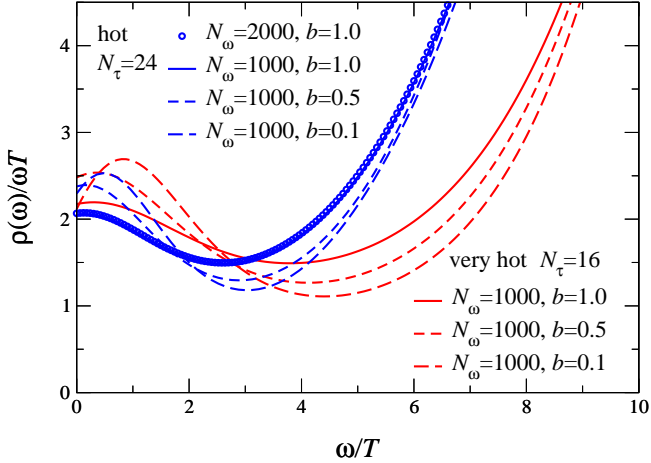}}
\caption{The spectral function $\widehat\rho(\omega,T)$ in the vector
channel of quenched QCD obtained by a MEM analysis~\cite{Aarts:2007wj}.
The normalization of the spectral function differs by a factor $2\pi$ from ours;
the conductivity $\sigma$ is $\frac{T}{6}\times$ the intercept.
The prior used for $\rho/\omega$ was $m(\omega) \propto (b N_t + \frac{\omega}{T})$.}
\label{fig:gert}
\end{figure}

A reasonable value of the parameter $\alpha$ is one for which the
$\chi^2$ per degree of freedom is of order unity. More refined ways of
choosing $\alpha$, or more generally averaging over it, can be found
in~\cite{Asakawa:2000tr}.  To calculate the error on the spectral
function, a simple and robust approach is to use the jacknife
method. Splitting the data into batches has the advantage of giving
some information on the spread of $\widehat\rho$, and has been found
to yield somewhat more stable results~\cite{PhysRevB.81.155107}.  One
may also quote the error not on a particular value of
$\widehat\rho(\omega)$, but rather on its average over a chosen
frequency interval~\cite{Asakawa:2000tr}, which has a stabilizing
effect.  An example of a spectral function in the vector channel of
quenched QCD obtained from a MEM analysis is reported
in \fig(\ref{fig:gert}).  See the recent reference \cite{Ding:2010ga}
for further examples.

The choice of non-quadratic regulator (\ref{eq:SJ2}) has an
information-theoretic justification. An argument for this form of $S$
is given in \cite{Asakawa:2000tr}.  The argument is based on the
Bayesian idea that one has an \emph{a priori} expectation for the
outcome of $\widehat\rho$, which is encoded in the function
$m(\omega)$.  The idea is then to construct the most probable spectral
function by maximizing the conditional probability
$P[\widehat\rho|D\,H]$, where $D$ denotes the data and $H$ the prior
information.

Generally speaking, non-linear regulators have an advantage over
linear regulators when the spectral density exhibits sharp peaks over
a slowly varying `background'~\cite{Press:2007zz}. This is quite easy
to understand in the case of MEM, since it is the exponent
in \eq(\ref{eq:mem-ansatz}) that is expanded in a basis of functions,
and therefore a modest increase of this linear combination can lead to
a strong increase in $\widehat\rho(\omega)$ itself.  For this reason,
the maximum entropy method is probably the best currently available
method to analyze quarkonium correlators, where rapid changes of the
spectral function are expected around threshold.

\subsection{Frequency-space method\la{sec:lin-omega}}

Here we describe an approach to determining transport coefficients
proposed in~\cite{Meyer:2010ii} starting from the Euclidean correlator
in frequency space.  The idea is based on \eq(\ref{eq:dGR}), which
relates derivatives of the retarded correlator in the imaginary
frequency direction to derivatives of the spectral function along the
real axis; and (\ref{eq:l.neq.0}), which relates the retarded
correlator for frequencies along the positive imaginary axis to the
Euclidean correlator.  By the Carlson theorem, the exact knowledge of
the Euclidean correlator at the Matsubara frequencies, where it can be
calculated by Euclidean path integral methods, determines its analytic
continuation to all frequencies, since physical correlation functions
can grow at most with a power-law at large frequencies. At the point
$\omega$ it then coincides with the retarded correlator evaluated at
$i\omega$, so that
\ba
\frac{d\tilde G_E(\omega)}{d\omega} &\stackrel{\omega=0}{=}& - \pi \frac{d\rho}{d\omega},
\la{eq:GEderiv}
\\
\frac{d^3\tilde G_E(\omega)}{d\omega^3} &\stackrel{\omega=0}{=}& \pi \frac{d^3\rho}{d\omega^3},\quad {\rm etc.}
\ea
This allows one, in principle, to determine the transport properties, since the 
right-hand side of \eq(\ref{eq:GEderiv}) is precisely the expression appearing 
in the Kubo formulae.
To illustrate this point, it is useful to note that the contribution of a 
diffusion pole $\rho=\chi_s \frac{D\bk^2}{\omega^2+(D\bk^2)^2}$ to the Euclidean 
correlator takes the form
\be
\tilde G_E(\omega,\bk) = \frac{\chi_s D\bk^2}{|\omega|+D\bk^2}.
\la{eq:GEdiffusion}
\ee
Thus one sees that a non-zero value of $\rho(\omega)/\omega$ near the origin
translates into a non-analyticity of the Euclidean correlator in $\omega^2$ 
around the origin. 

For a function known at $\omega_\ell=2\pi T\ell\equiv
\omega_M\cdot\ell$, one can write a polynomial of degree $n$ going
through the $n+1$ first points $\{G_E^{(\ell)}\}_{\ell=0}^n$.  It has
a slope at the origin given by
\be
\la{eq:GEprime}
\tilde G_E'(\omega=0)_{\rm eff} \equiv  -\frac{n!}{\omega_M}
\sum_{\ell=1}^n \frac{(-1)^{\ell}}{\ell!(n-\ell)!} 
\frac{G_E^{(\ell)}-\tilde G_E^{(0)}}{\ell}\,.
\ee
When expressed 
in terms of dispersion integrals as in 
\eq(\ref{eq:lincomb1}-\ref{eq:lincomb2}), 
the degree of convergence
in the UV of this expression 
improves with the order of the polynomial: with some 
combinatorics one can show that it is 
asymptotically $\rho(\omega)/\omega^{n+3}$ for $n$ even and 
$\rho(\omega)/\omega^{n+2}$ for $n$ odd.
For instance, the slope at the origin of the $n=2$ polynomial 
is given by
\ba\la{eq:slope2}
&& -\frac{1}{2\omega_M} (3\tilde G_E^{(0)} - 4G_E^{(1)} + G_E^{(2)})
\\
&=&
-6\omega_M^3\int_{-\infty}^\infty
 \ud\omega \,\frac{\rho(\omega)}{\omega}
\frac{1}{(\omega^2+\omega_M^2)\,(\omega^2+4\omega_M^2)}\,,
\nonumber
\ea
where the linear combination (\ref{eq:lincomb2}) appears.  Thus the
slope of the quadratic polynomial is negative-definite, just as the
actual slope of the retarded correlator $\frac{d}{d\omega}\tilde
G_R(i\omega)$. One might be tempted to use it as a Euclidean estimator
for the transport coefficient (for instance the diffusion coefficient
$D$).  However, if the spectral function admits a diffusion pole of
width $D\bk^2\ll \omega_M$, expression (\ref{eq:slope2}) amounts
to $-\frac{3}{2}\chi_s/\omega_M$, while actually
$\frac{d}{d\omega}\tilde G_R(i\omega)= - \frac{\chi_s}{D\bk^2}$.  It
is clear why that is: a polynomial interpolating between the Euclidean
points is only a good approximation to the retarded correlator if the
latter does not contain a scale $\omega_0=D\bk^2\ll \omega_M$ over which it
varies significantly. 

However, one knows \emph{a priori} that the number density correlator
must exhibit a diffusion pole. Therefore it is appropriate to `help' 
the determination of the slope by including a term of the form 
(\ref{eq:GEdiffusion}) in a fit to the Euclidean data set $\{G_E^{(\ell)}\}$.
The remainder can be assumed to analytic in $\omega^2$ near the origin.
We note that the idea of parametrizing the frequency-space Euclidean
correlator by a rational function, and then obtaining the spectral
function from its discontinuity across the real axis was one of the
first proposed~\cite{thirumalai}. Extracting a derivative along the imaginary
axis should be comparatively more stable.

No attempt has been made yet to use this method in practice, mainly
because the calculation of off-shell correlators such as
$G_E^{(\ell)}$ is technically more complicated than working in
coordinate space.  The attractiveness of this method is that one does
not need to reconstruct explicitly the whole spectral function, and
this is a special feature of the transport properties -- as opposed
to, for instance, determining the width of a quarkonium state.
Recently, the method has been tested to determine the optical
conductivity in a Hubbard model and found to work quite well in this
case (\cite{PhysRevB.82.165125}, see section \ref{sec:method-compa}).

\subsection{Combining Numerical Euclidean Data with Analytic Results \la{sec:combi}}

In attacking the inverse problem of determining the spectral function,
one can take a `purist' approach, and apply a ready-to-use formula
which is exact in the limit of perfect Euclidean data, such as
(\ref{eq:G_Laguerre}) or (\ref{eq:GEprime}).  At the other extreme, if
one has control over the functional form of the spectral function,
then a fit with a small number of free parameters is
appropriate. However in practice and for the foreseeable future the
first approach is bound to be of limited use due to imperfect data,
while in QCD it seems unlikely that one will have good enough control
over the functional form over the whole real axis of frequencies.
Therefore it will remain useful to exploit prior knowledge about the
spectral function, such as the parametric location of a diffusion or
sound pole, even when a `purist' approach is adopted.  Here we wish to
collect the kind of analytic information one may want to make use of.

Knowing even partially the functional form of the spectral function is of
course a great advantage. Information about the functional form is
provided by `straightforward' perturbation theory at frequencies
$\omega\gg T$.  Power corrections are predicted by the
operator-product expansion.  In QCD at zero chemical potential, the
leading power corrections at finite temperature are suppressed by four
powers of the frequency, because that is the dimension of the
lowest-dimensional gauge-invariant operators (the mass operator
$\bar\psi\psi$ comes accompanied by one power of the quark mass, since
it is a chirally non-invariant contribution).  
At low frequencies and momenta on the other hand, hydrodynamics
is expected to predict the pole structure of the correlators correctly,
see section (\ref{sec:kubo}). The main unknown is the domain 
filling the gap between these two asymptotic regimes. 
This interval is precisely the region that the Euclidean correlator
at the first few Matsubara frequencies is most sensitive too.

Sum rules can be used as a constraint on the spectral function, if the
RHS of the sum-rule is known. One example is the difference of vector
and axial-vector current
correlators~(\eq\ref{eq:sr-av1}-\ref{eq:sr-av3}).  Because the
difference of the vector and axial-vector spectral functions falls off
rapidly at large frequencies, the difference should be easier to
parametrize and be more sensitive to the low-frequency region.  The
shear and bulk channels also obey sum rules once the $T=0$ correlators
have been subtracted, \eq(\ref{eq:conti-sr}) and (\ref{eq:bsr1})
respectively.  In all these cases a subtraction is necessary in order
to make the spectral integral converge. This entails the loss of the
positivity of the function one is trying to reconstruct. This is
clearly a drawback, since the positivity property has a stabilizing
effect. It has to be decided on a case-by-case basis whether it is
preferable to deal with non-positivity or with a large ultraviolet
contribution to the signal. This is the major added difficulty one
faces in relativistic field theories as compared to non-relativistic
systems.

Exploiting the fact that different correlators at finite
$(\omega,\bk)$ are related by Ward identities can help to constrain
the unknown spectral function. This observation is particularly useful
in the sound channel, where the energy-density correlator is most
sensitive to the sound peak, while the longitudinal pressure
correlator is much more sensitive to the UV tail of the sound-channel
spectral function.

\begin{figure}
\centerline{\includegraphics[angle=-90,width=0.5\textwidth]{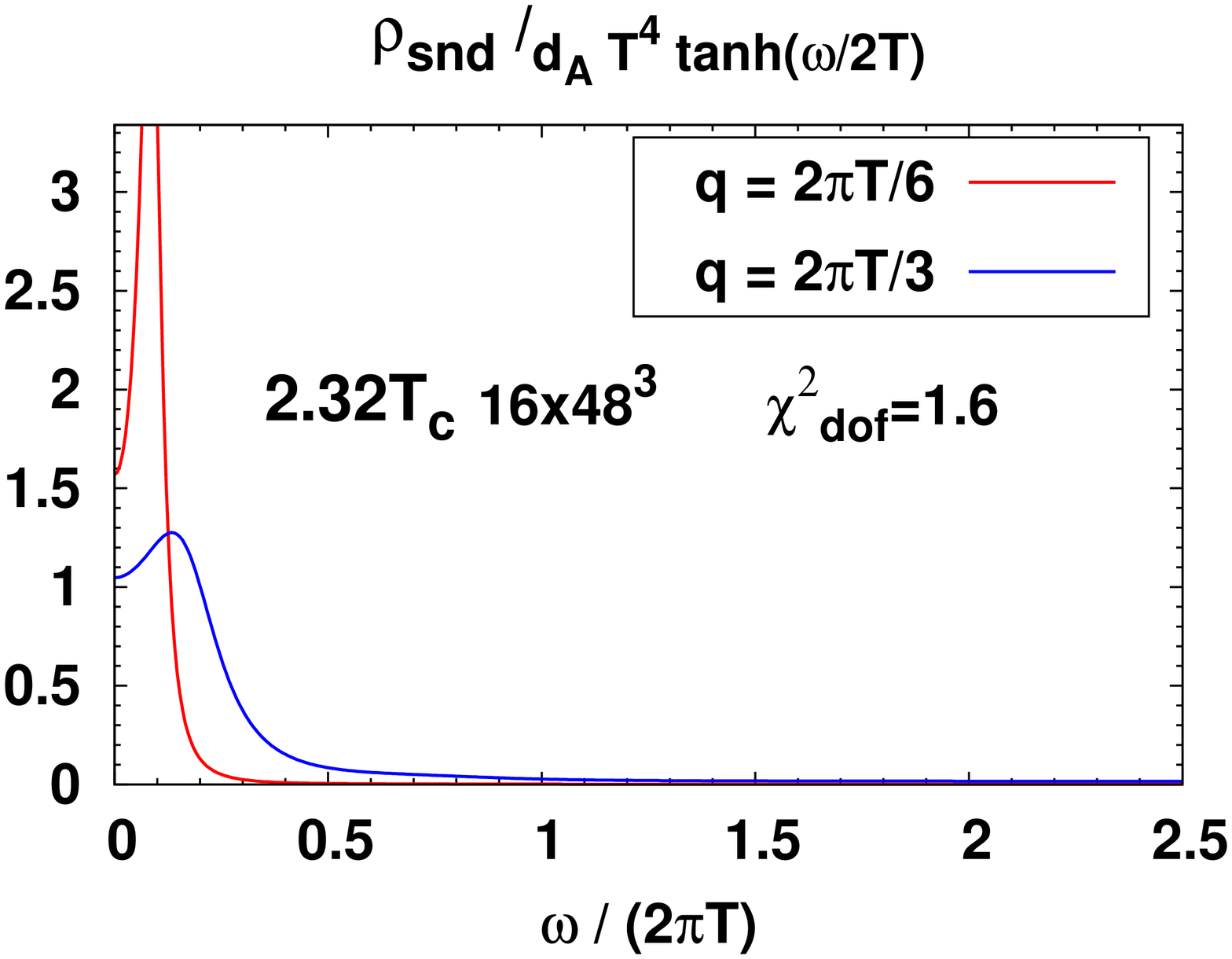}}
%\vspace{-0.4cm}
\centerline{\includegraphics[angle=-90,width=0.5\textwidth]{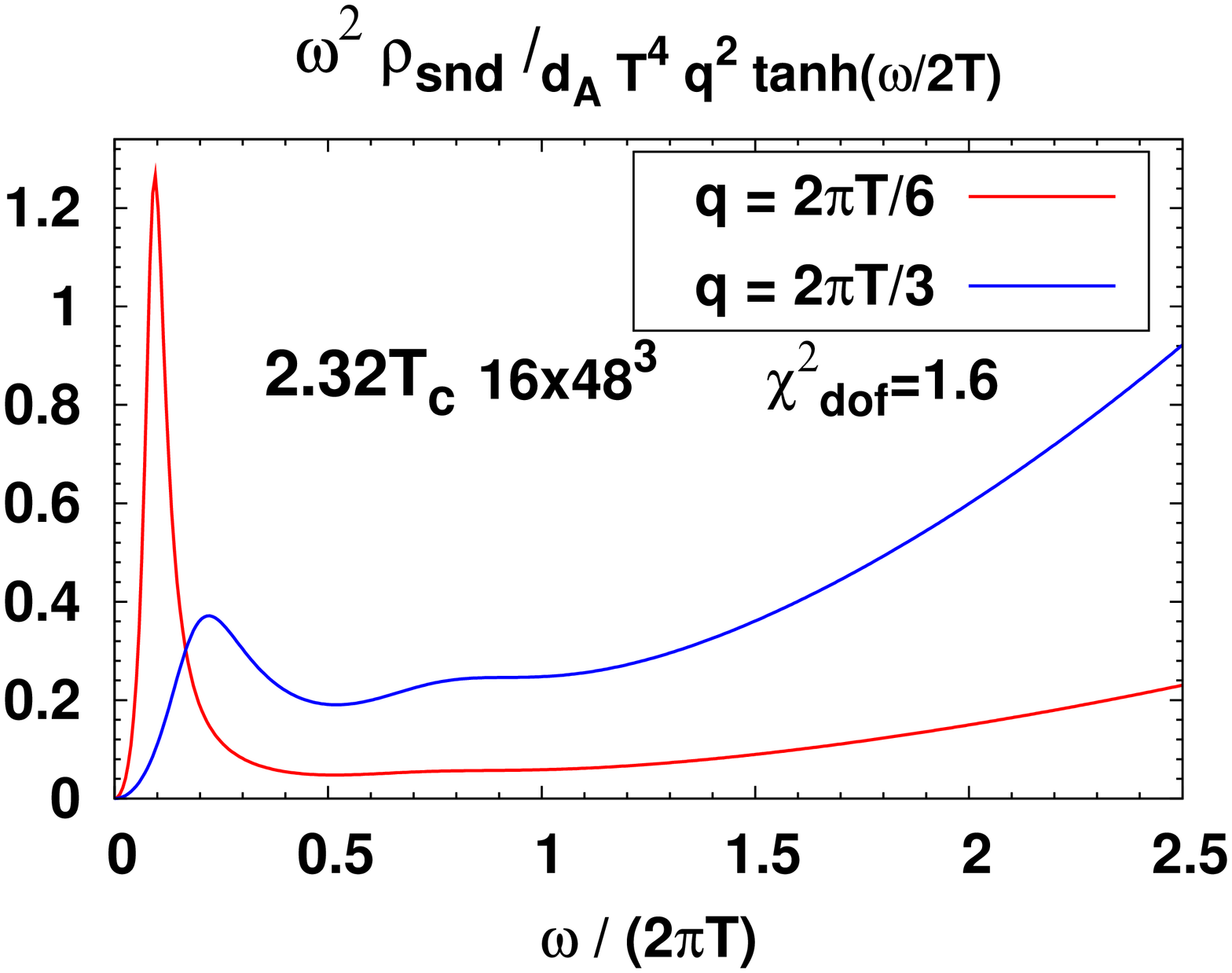}}
%\vspace{-0.4cm}
\centerline{\includegraphics[angle=-90,width=0.5\textwidth]{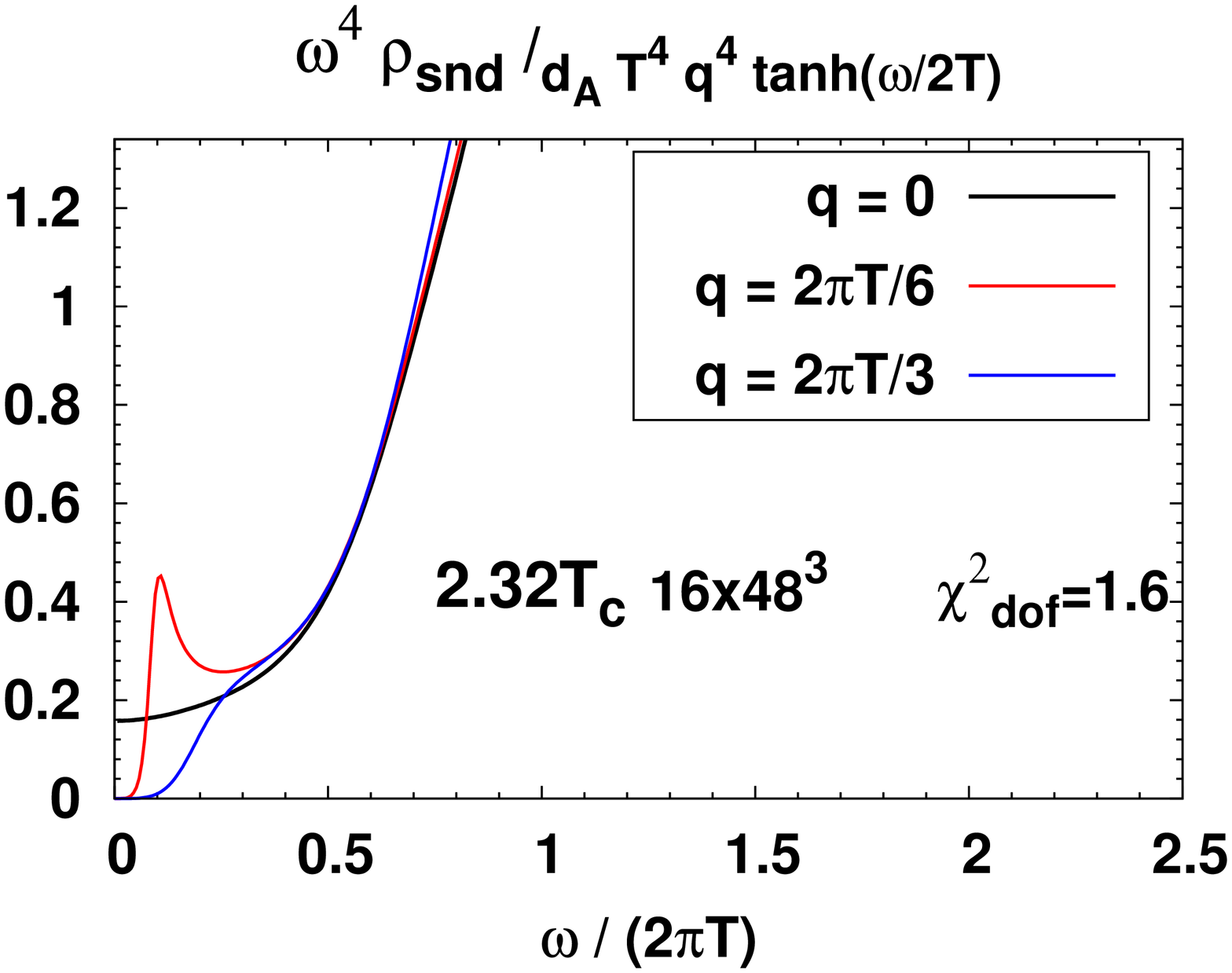}}
\caption{The spectral function obtained from the fit (\ref{eq:soundfit})~\cite{Meyer:2009jp}.}
\label{fig:0907sf}
\end{figure}

\begin{table}
\begin{tabular}{c@{~~~}c@{~~~}c@{~~~}c@{~~~}c}
% \toprule
\hline
                     &    $1.58T_c$  &   $2.32T_c$ & free gluons & $\lambda=\infty$ SYM \\
                     \hline
$(\eta+\frac{3}{4}\zeta)/s$ & 0.20(3)  & 0.26(3) & $\infty$ & $\frac{1}{4\pi}\approx0.080$ \\
$2\pi T \tau_\Pi$    & 3.1(3)& 3.2(3) & $\infty $& $2-\log2\approx1.31$ \\
$\frac{\eta+\frac{3}{4}\zeta}{Ts \tau_\Pi}$ & 0.40(5) & 0.51(5) &  0.17& 0.38 \\
% \bottomrule
\hline
\end{tabular}
\la{tab:0907fit}
\caption{Sound channel spectral function  at $2.3T_c$. In the Table,
lattice results are compared to free gluons~\cite{York:2008rr} and to the strongly coupled
SYM results~\cite{Policastro:2001yc,Baier:2007ix}. Stat. errors only are given.
We expect $\zeta$ to be negligible at these temperatures.}
\end{table}

Thanks to the known tensor structure of the vector and 
energy-momentum tensor correlators (see section \ref{sec:tensor}),
one can exploit a richer set of Euclidean data if one is willing to
analyze two channels simultaneously. For instance, 
letting $\theta_{\rm q}$ be the polar
angle of ${\bq}$, one can exploit also momenta that are not aligned
along a lattice axis via~\cite{Meyer:2009jp}
\ba \la{eq:tensorstruc}
G_E^{03,03}(t,{\bq})&=& \int_0^\infty d\omega
\frac{\cosh\omega (\half \beta -t)}{\sinh \half \beta\omega }\cdot
\\
&& \Big\{
 \rho_{\rm snd}(\omega,{\bq})\,\cos^2\theta_{\rm q}+
\rho_{\rm sh}(\omega,{\bq})\,\sin^2\theta_{\rm q} \Big\}.
\nonumber
\ea
Here $\rho_{\rm sh}$ and $\rho_{\rm snd}$ are the shear and sound
channel spectral functions, using the notation $\rho_{\rm
snd}(\omega,{\bq})\equiv-\rho_{{03},{03}}(\omega,{\bq})$ and
$\rho_{\rm sh}(\omega,{\bq})\equiv -\rho_{{01},{01}}(\omega,{\bq})$.
On the anisotropic lattice, the correct normalization of all
components of the energy-momentum tensor involves many independent
normalization factors.  Apart from reproducing the correct
thermodynamic properties, they must obey the Ward identities. This has
been used to determine some of the normalization constants in the
SU(3) gauge theory~\cite{Meyer:2009jp}.  In the latter reference, an
attempt was made to fit the sound-channel correlators with the ansatz
\be \la{eq:soundfit}
\rho_{\rm snd} = \rho_{\rm low} + \rho_{\rm med} + \rho_{\rm high},
\ee
where
\ba
\frac{\rho_{\rm low}(\omega,q,T)}{\tanh(\omega/2T)} &=& 
\frac{2{\widehat\Gamma_s}}{\pi}
\frac{(e +P)\,\omega^2q^2}{(\omega^2-c_s^2(q)q^2)^2+({\widehat\Gamma_s}\omega q^2)^2}
\frac{1+{\sigma_1} \omega^2}{1+{\sigma_2} \omega^2}\,
\la{eq:rho-low} \nn
\frac{\rho_{\rm med}(\omega,q,T)}{\tanh(\omega/2T)} &=& 
 \omega^2q^2\tanh^2(\frac{\omega}{2T}) 
\frac{{\ell \,\sigma}}{{\sigma}^2 + (\omega^2-q^2-{M}^2)^2}\,,
\la{eq:rho-med} \nn
\frac{\rho_{\rm high}(\omega,q,T)}{\tanh(\omega/2T)} &=& 
\omega^2q^2\tanh^2(\frac{\omega}{2T}) \frac{2d_A}{15(4\pi)^2} \,,
\la{eq:rho-high}
\ea
and $c_s(q^2)$ is given by \eq(\ref{eq:vsq}).  In the present specific
example 7 parameters are fitted to a total of 48 data points, obtained
at $T=2.3T_c$ on a $16\times48^3$ lattice with anisotropy $\as/\at=2$.
The parameters are
$\widehat\Gamma_s,\sigma_1,\sigma_2, \tau_\Pi, \ell, \sigma, M$.
Figure (\ref{fig:0907sf}) displays the reconstructed sound spectral
function based on lattice data at $2.3T_c$, and the results for the
fit parameters of interest are given in \tab(\ref{tab:0907fit}).  In
this fit, the contribution of the $\delta$-function to the $T_{33}$
correlator at $\bq=0$ according to (\ref{eq:rho3333}) was not properly
taken into account. This neglect leads to an overestimation of the
sound attenuation length $\Gamma_s=\frac{\frac{4}{3}\eta+\zeta}{e+p}$
(although the error bars on the $T_{33}$ correlator being
significantly larger than on the $T_{03}$ correlator, the effect may
be small).

The fit yields an acceptable correlated $\chi^2$ value per degree of
freedom, $\chi^2/{\rm d.o.f}=1.6$. An interesting point is that no
prominent features are needed to fit the Euclidean data apart from the
hydrodynamics structure at low frequencies and the perturbative UV
tails at high frequencies.  In this sense the result is reminiscent of
the strongly coupled SYM spectral functions displayed
in \fig(\ref{fig:sf-sound-AdSCFT}).  Based on the perturbative
calculation~\cite{Arnold:2006fz}, at $2.3T_c$ the bulk viscosity is
expected to be negligible compared to the shear viscosity. The values of
the shear viscosity are numerically quite small, and the `relaxation time' $\tau_\Pi$
is quite well constrained by the fit. This comes from the sensitivity 
of the Euclidean correlator to the precise location of the sound pole.
We note that the ratio $\frac{\Gamma_s}{\tau_\Pi}$ is strongly constrained
by the requirement of causality~\cite{Pu:2009fj}, namely 
\be
\frac{\Gamma_s}{\tau_\Pi} \leq 1-c_s^2.
\ee
The values of the parameters given in \tab(\ref{tab:0907fit}) are 
close to but consistent with this upper limit.

A different strategy was adopted in~\cite{Meyer:2010ii} to interpret
the bulk channel Euclidean correlator using a maximum amount of
analytic information. The temperatures studied were 1.02, 1.24 and $1.65T_c$
in the SU(3) gauge theory. The first step is to switch to the
Euclidean correlator of the operator ${\cal O}_\star\propto
\big(T_{\mu\mu}+(3c_s^2-1)T_{00}\big)$ defined in \eq(\ref{eq:Ostar}).
This removes the delta function at the origin from the infinite-volume
spectral function of $T_{\mu\mu}$. Secondly, the difference between
the finite-temperature and zero-temperature spectral functions was
taken, which corresponds to \eq(\ref{eq:Grec2}) in terms of the
Euclidean correlators.  While this removes the bulk of the UV
contribution, which represents an undesirable background when
determining the low-frequency part of the spectral function, it also
introduces the (delta-function) contributions of the stable glueballs
with negative weight.  A property characteristic of SU($N$) gauge
theories is that the mass gap $m$ is numerically large, as compared to
$T_c$. In particular, the subtracted spectral function
$\Delta\rho(\omega,T)\equiv \rho(\omega,T)-\rho(\omega,0)$ is still
positive for $\omega<m$. The lightest two scalar glueball (G) masses
are known quite well, and also the matrix element $f_G\equiv \<{\rm
vac}|{\cal O}_{\star}|G\>$ for the lightest glueball has been
calculated~\cite{Meyer:2008tr,Chen:2005mg}.  Therefore is would be
desirable to compensate for these contributions by forming
$\Delta\rho(\omega,T)+ f_G^2 \delta(\omega-m) +
f_{G'}^2 \delta(\omega-m')$.  This linear combination is then still
positive up to the two-glueball threshold $2m$, which is about 3GeV.
However, due to the imperfect knowledge of the glueball matrix
elements, the subtraction of the one-particle contributions was not
carried out explicitly.

The asymptotic UV behavior of the difference of spectral functions is
given by the operator-product expansion, \eq(\ref{eq:rho_laine}).
Numerically, the OPE was assumed to be reliable starting at
$\omega\approx 4.8$GeV. The formula (\ref{eq:rho_laine}) then predicts
$\Delta\rho$ to be very small beyond that frequency. This sets a
boundary value for the spectral function beyond which it essentially
vanishes.

\begin{figure}
\centerline{\includegraphics[angle=-90,width=0.5\textwidth]{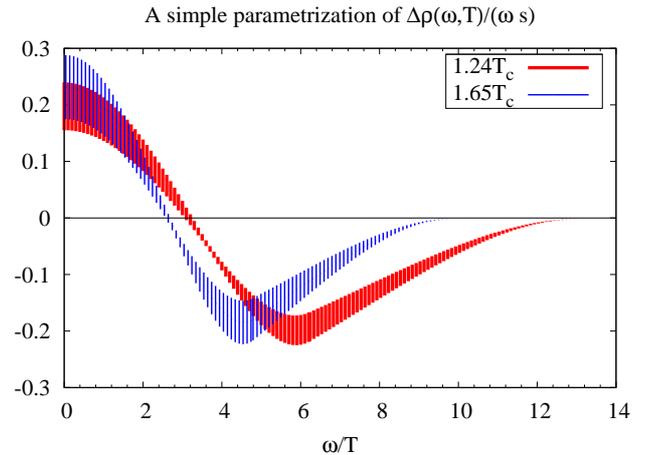}}
\caption{A smooth parametrization of the subtracted bulk-channel 
spectral function in SU(3) gauge theory, $\Delta\rho_\star(\omega,T)/(\omega\cdot s)$, 
compatible with the bulk sum rule and the lattice correlator at $t=\beta/2$\cite{Meyer:2010ii}. 
Recall that the ratio $\zeta/s$ is given by 
$\frac{\pi}{9}\times$ the intercept of this function.}
\label{fig:1002}
\end{figure}

The bulk sum rule in the form \eq(\ref{eq:bsr1}) provides one global
constraint on the subtracted spectral function in terms of
thermodynamic quantities,
\be
\frac{2}{e+p}\int_0^\infty \frac{d\omega}{\omega}
\,\Delta\rho_\star(\omega,T) = 3(1-3c_s^2) -4\frac{e-3p}{e+p}\,.
\la{eq:srA}
\ee
Secondly the value of the subtracted
Euclidean correlator at maximum time separation $t=\frac{\beta}{2}$
was used,
\ba\la{eq:Cmid}
\frac{T^4}{e+p}\, [G_E(t,T)-G_E^{\rm rec}(t,T) ]_{t=\beta/2}\,,
\\
= \frac{1}{e+p}\int_0^\infty \frac{d\omega}{T}\,
\frac{\Delta\rho_\star(\omega,T)}{\sinh \frac{\omega}{2T}}.
\nonumber
\ea
Expression (\ref{eq:Cmid}) turns out to be positive while expression
(\ref{eq:srA}) is negative.  Since both receive the same contribution
from low frequencies, but the higher frequencies 
contribute practically only to (\eq\ref{eq:srA}), one infers that the subtracted
spectral function must be positive at small frequencies and negative
at higher frequencies.  This can be quantified as follows. The
frequency axis was divided into three intervals, defined by the
separation points $\omega_1\approx m$ and $\omega_2=4.8$GeV.  In each
interval, the subtracted spectral function is assumed to to be a
quadratic polynomial in $\omega$, and the function is required to be
continuous and differentiable at the boundary points. Beyond
$\omega_2$, it is assumed to be given by the operator-product
expansion prediction.  Based on the two pieces of information
(\ref{eq:srA}) and (\ref{eq:Cmid}), the parametrization of the
subtracted spectral function obtained in
\cite{Meyer:2010ii} is displayed in \fig(\ref{fig:1002}).

This calculation thus provides evidence for the depletion of the
spectral density in the region of those glueballs that are stable or
lie slightly above the two-particle threshold, and secondly for the
appearance of a positive spectral weight for $\omega<m$.  The analysis
proceeded by maximizing the use of available analytic information on
the spectral function.

\subsection{Comparison of different methods\la{sec:method-compa}}

\begin{figure}
\centerline{\includegraphics[angle=-90,width=0.45\textwidth]{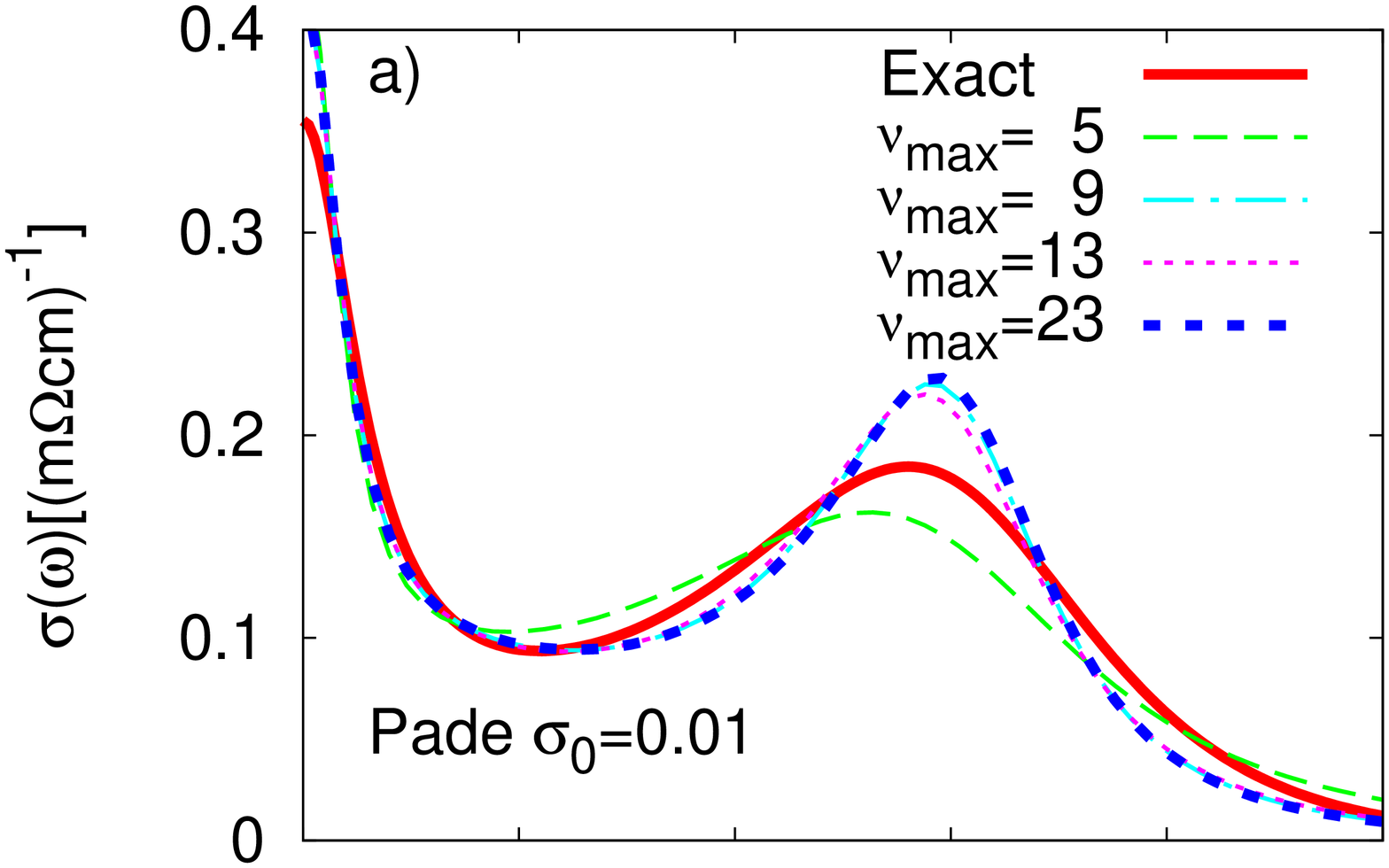}}
\vspace{-0.4cm}
\centerline{\includegraphics[angle=-90,width=0.45\textwidth]{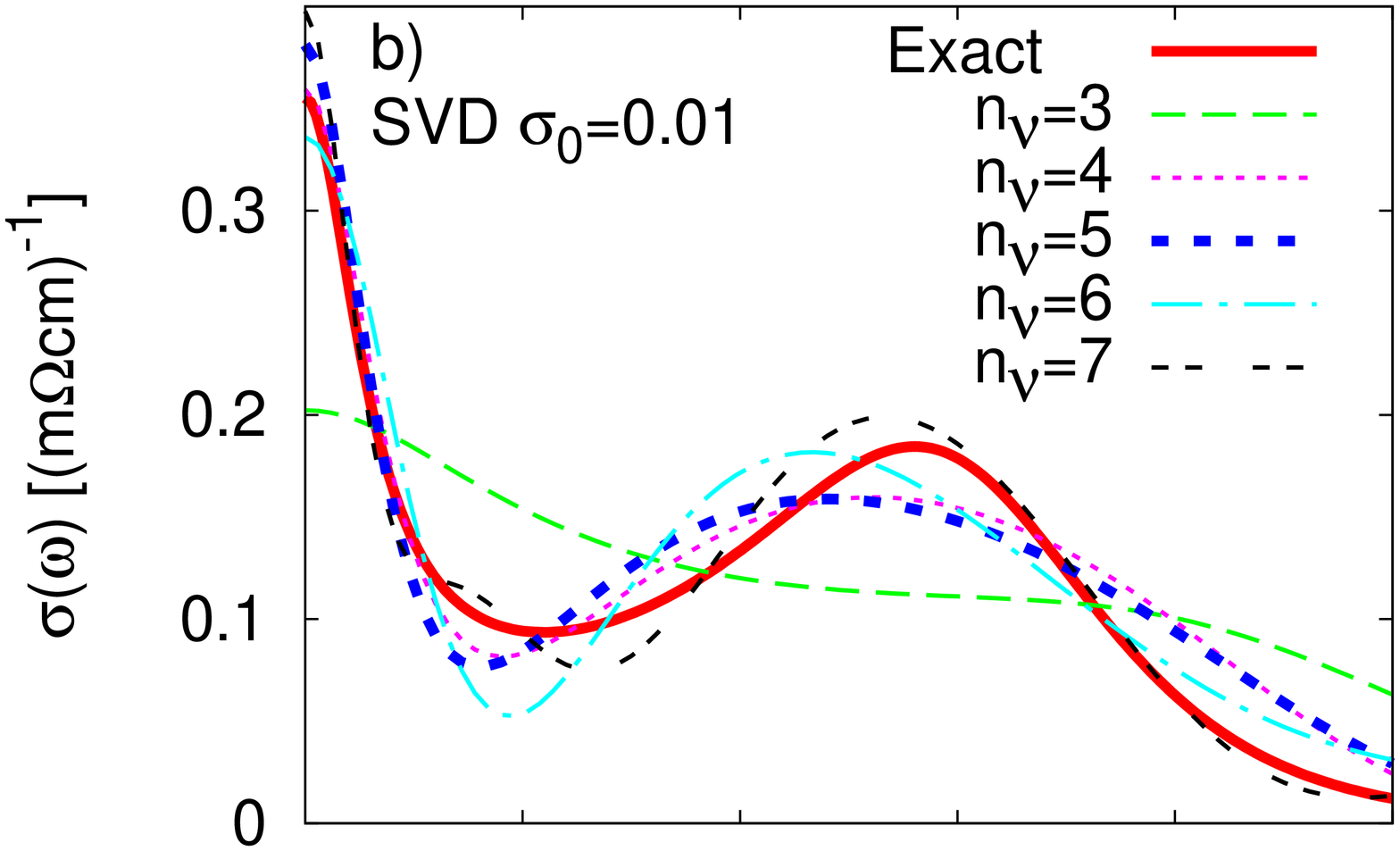}}
\vspace{-0.4cm}
\centerline{\includegraphics[angle=-90,width=0.45\textwidth]{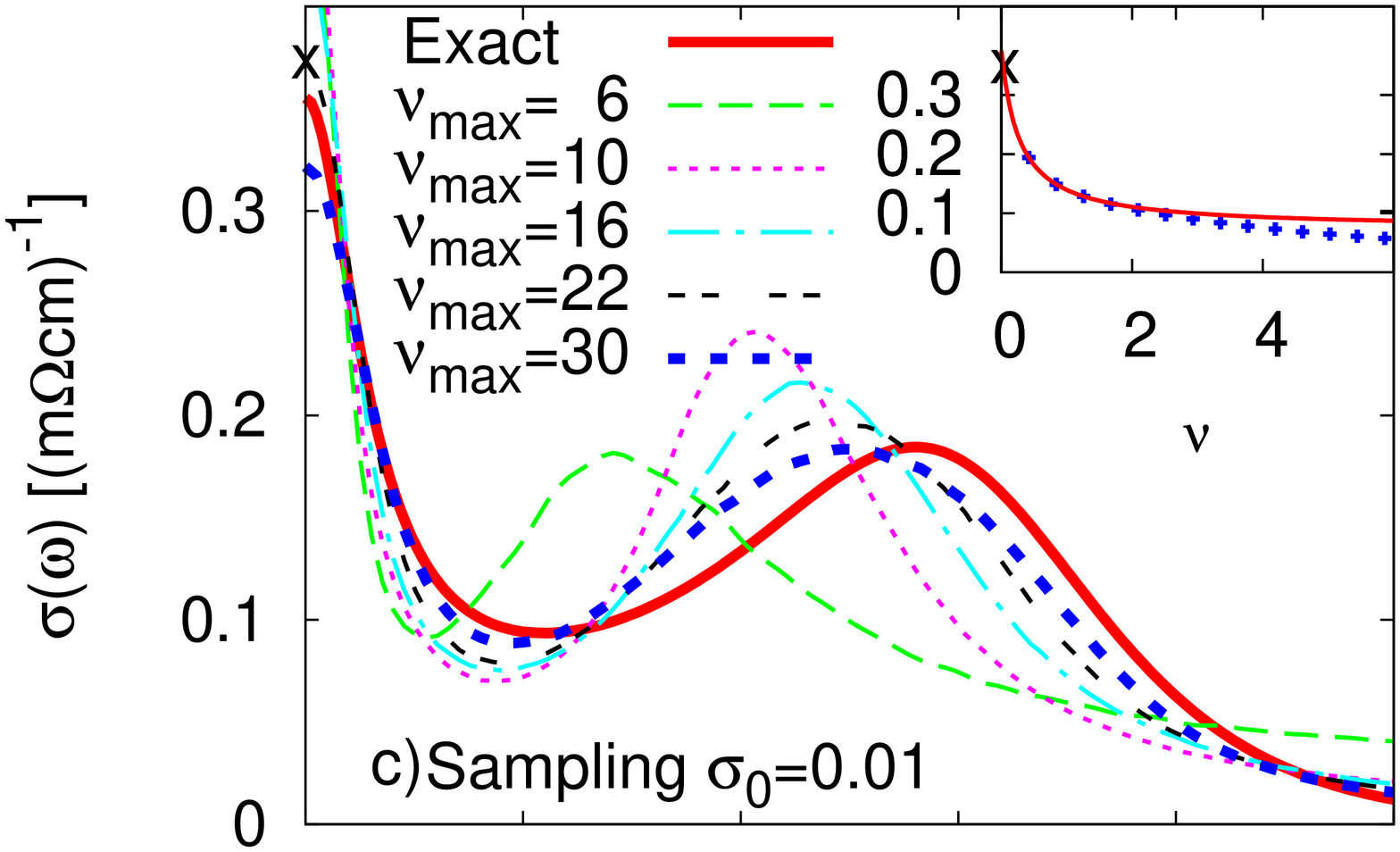}}
\centerline{\includegraphics[angle=-90,width=0.45\textwidth]{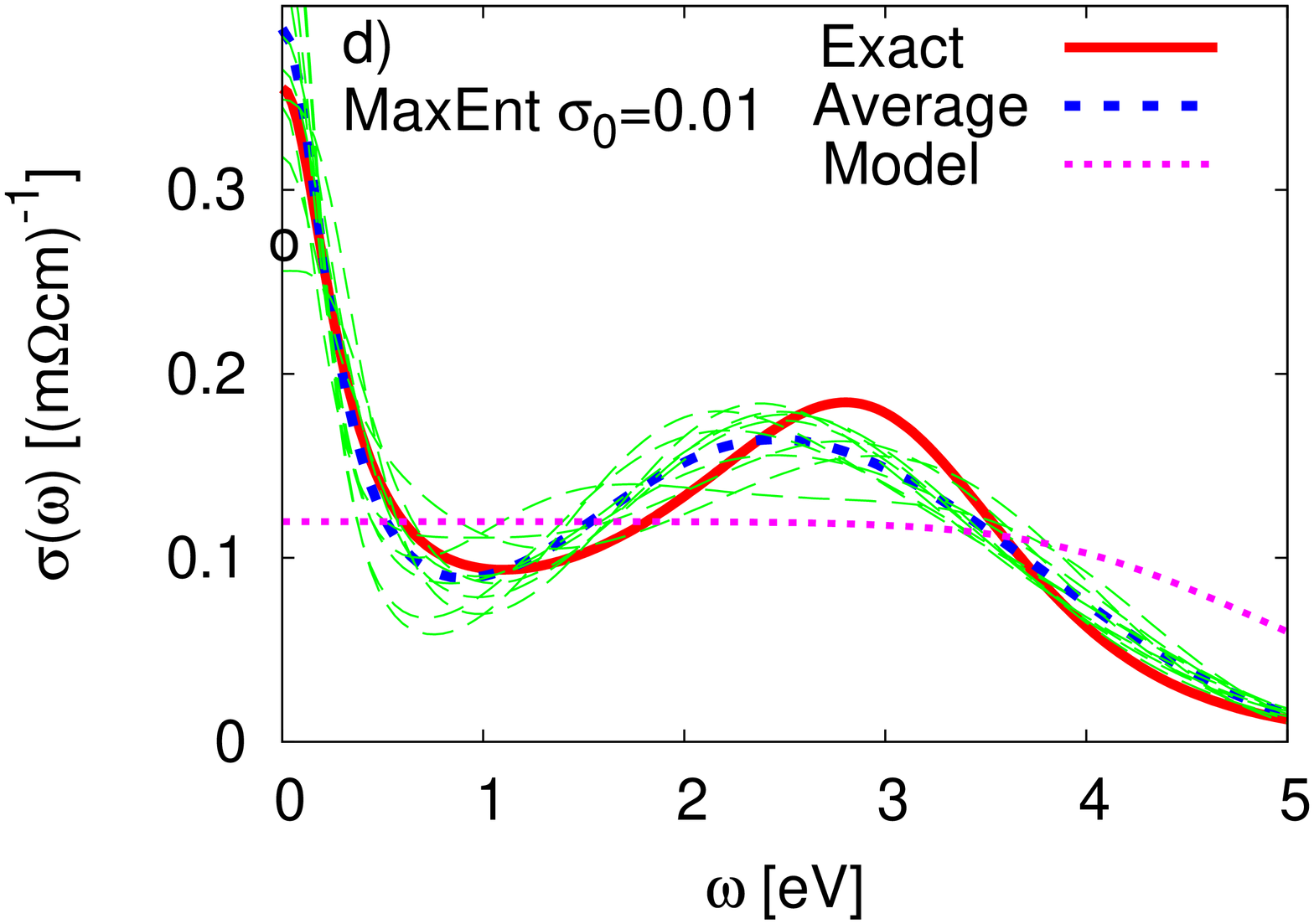}}
\caption{Comparison of different methods to reconstruct a known spectral function from
the frequency-space Euclidean correlator. See the text for a
description of the methods.  Study and figure
from~\cite{PhysRevB.82.165125}. The temperature is 0.067eV.}
\label{fig:test}
\end{figure}

It is of interest to know how well different inverse problem methods
compare, even if the outcome of such a comparison depends to some
extent on the character of the spectral function one is trying to
reconstruct from a Euclidean correlator.  Such a comparison was
recently done by Gunnarsson, Haverkort and
Sangiovanni\cite{PhysRevB.82.165125}. They started from the spectral function
\ba
%% sigma = pi rho / omega
\la{eq:rho-test}
&& \sigma(\omega) \equiv \frac{\pi \rho(\omega)}{\omega} = 
\Big[ \frac{W_1}{1+(\omega/\Gamma_1)^2} 
+ \frac{W_2}{1+((\omega-\epsilon)/\Gamma_2)^2} 
\nn
&&\quad  + \frac{W_2}{1+((\omega+\epsilon)/\Gamma_2)^2} \Big]
\frac{1}{1+(\omega/\Gamma_3)^6},
\ea
a superposition of Lorentzians weighted by an overal factor.
The values of the parameters were chosen to be
\ba
\la{eq:G1}
 \frac{\Gamma_1}{2\pi  T} = 0.71,
\qquad
\frac{\Gamma_2}{2\pi T} = 2.9,
\qquad
\frac{\Gamma_3}{2\pi T} = 9.5,
\\
\frac{\epsilon}{2\pi T} = 7.1,\qquad 
W_1 = 0.3,\qquad W_2=0.2\;.
\ea 
in this example. To put this in perspective, Moore and
Saremi~\cite{Moore:2008ws} obtained the spectral function in the shear
channel of SU(3) gauge theory and found a transport peak of width
roughly $\Gamma\approx 0.05 g^4 T$.  For an intermediate 't Hooft
coupling value of $\lambda_H=10-12$ thought to be relevant at a
temperature of a few $T_c$~\cite{Iqbal:2009xz}, this corresponds to
$\frac{\Gamma_1}{2\pi T}=$0.09 .. 0.12.  Thus the width (\ref{eq:G1})
should be considered to be quite large and for such a smooth spectral
function (on the scale of $2\pi T$) one would expect the inverse
problem to be a fairly controlled procedure.  We remark that there are
also channels where the QCD spectral functions do not have kinetic
theory peaks, for instance the tensor channel of the energy-momentum
tensor or the transverse channel of the vector current. So a
subtracted spectral function in a strongly interacting gauge theory
may in fact bear some resemblance with this form, displayed as a solid 
red curve in \fig(\ref{fig:test}).

After obtaining the frequency-space Euclidean correlator via
(\ref{eq:KK-main}) and giving them uncorrelated Gaussian noise
of relative standard deviation $\sigma_0$,
four different methods were used to reconstruct the
spectral function.  The four panels of \fig(\ref{fig:test}) display
these results.  We briefly summarize what these four methods are, and
refer the reader to the original article~\cite{PhysRevB.82.165125} for
the details.  
\begin{enumerate}
\item The Pad\'e approximation is one of the oldest methods
used in this problem~\cite{serene}. The procedure consists in fitting
a rational function ansatz to the Euclidean data (i.e. at imaginary
frequencies), and then evaluating the imaginary part of the function
along the real axis.
\item the second method is very similar to the method described in section
(\ref{sec:lin}), except that it is based on Euclidean-frequency data.
The spectral function is expanded in a set of functions which are
eigenfunctions of $K^\dagger K$, if $K$ is the kernel of the
transform~(\ref{eq:KK-main}) modified by a weighting factor to take advantage 
of the finite (effective) support of the spectral function.  
The eigenvalues of $K^\dagger K$ become
small rapidly and a cutoff $n_\nu$ on their number is introduced.
\item the sampling method: this is similar to the Maximum Entropy Method,
where however instead of using a default model the spectral function 
is given by an average weighted by $P[\sigma|G_E]$.
\item the Maximum Entropy Method with a flat prior distribution of 
the parameter $\alpha$ and a featureless prior function $m(\omega)$
displayed in \fig(\ref{fig:test}d). We note that the data is 
divided into batches and the result of MEM in each of them is averaged over.
\end{enumerate}
See \fig(\ref{fig:test}) for the outcome of this test. The sensitivity of
the results to variations in the parameters was also investigated by
doubling $\Gamma_1$ and by reducing $\sigma_0$ by an order of
magnitude.  The authors conclude that the Pad\'e approximation is
somewhat less reliable than the three others, while no clear winner 
was found among the SVD, MEM, and MEM sampling methods. For the 
spectral function studied, all three provide a decent approximation
in the range of parameters they investigated. Using several methods
has the advantage of giving some indication of which features are
genuine and which are artefacts of a particular method.

In addition, the method of extrapolating the Euclidean frequency-space
correlator to $\omega=0$ described in section (\ref{sec:lin-omega})
provides a rather good estimate of the conductivity, $\sigma(0)$, in
this example. This is illustrated in the inset of figure
(\ref{fig:test}c). It is a good method if one is only interested 
in the transport part of the spectral function.

% \newpage 
\section{Summary and Outlook\label{sec:concl}}

We started this review by presenting the motivation to study the
transport coefficients of the quark-gluon plasma in heavy-ion
collisions and in cosmology. We then summarized the theoretical
framework which underlies the determination of spectral functions from
Euclidean-space Monte-Carlo simulations. We reviewed the analytic
methods that have been used to study the spectral functions in various
kinematic regimes, such as the hydrodynamic expansion and kinetic
theory at low frequencies, and perturbation theory and the
operator-product expansion at high frequencies. The
recent lattice QCD calculations of Euclidean correlators in the
vector, shear and sound channels were reviewed in section~(\ref{sec:latt}).

In section~(\ref{sec:EC}), we summarized the main methods used to
infer information about spectral functions from Euclidean correlation
functions. Whatever conclusion one draws about the spectral function
must be shown to be a property of QCD as opposed to a combination of
QCD and model information.  In most cases that were described in this
review, the functional form of the spectral function is known both far
in the ultraviolet $\omega\gg T$ (from perturbation theory) and deep
in the infrared, $\omega\ll \tau_R^{-1}$ (from hydrodynamics), where
$\tau_R$ is the longest relaxation time in the system.  However
practically no solid information is available for the frequencies
between these two extremes for temperatures relevant to heavy-ion
collisions.  In addition the onset of the validity of perturbation
theory and of hydrodynamics is not known a priori, adding to the
uncertainty of the ansatz one should use.

We highlighted the differences between linear and non-linear
methods to solve \eq(\ref{eq:ClatRho2}). The main advantages of the
former is that it allows one to predict straightforwardly the maximum
resolution one may possibly achieve in frequency space for a given
kind of input data, \emph{before the data is available}, and secondly
that the solution is found directly by inverting a matrix rather than
by an iterative procedure. An attractive feature of the maximum
entropy method (MEM) is that it is based on a generalization of the
maximum likelihood principle to the situation where one has prior
knowledge on the spectral function. In addition, it conveniently
enforces the positivity of the spectral function by writing its
estimator as the exponential of a real function.  The functional form
of the ansatz means that one can more readily describe a spectral
function that contains both sharp `peaks' and smooth `backgrounds',
but the difficulty of determining both the width and height of narrow
peaks remains.

In view of the inherent difficulty of any analytic continuation, it is
also worth trying to develop methods that circumvent this step, such
as finite-volume and variational techniques.  We have only briefly
described some aspects of these research directions in Sections
(\ref{sec:finivol}) and (\ref{sec:future}). For instance, one should
study what signature a diffusion process occurring in a finite volume 
leaves on current correlators.

Nonetheless, what have we learnt about QCD spectral functions from the
corresponding Euclidean correlators? Fit ans\"atze motivated by
hydrodynamics at low frequencies and by perturbation theory at high
frequencies are able to describe the Euclidean correlators, see for
instance \fig(\ref{fig:0907sf}) where the estimated relaxation time
was $\tau_\Pi\simeq 3.1/2\pi T$ at $1.58T_c$.  Recently, the isovector
vector current correlator (extrapolated to the continuum) was
successfully fitted by an ansatz composed of the free-field
contribution, plus a Breit-Wigner peak centered at the origin. The
relaxation time corresponding to the width of the peak is well
constrained by the fit and about $5.6/2\pi T$ at $1.45T_c$. These
values certainly suggest that the relaxation times are not long on the
thermal time scale, as kinetic theory parametrically predicts
($1/g^4T$). On the other hand, they are not as small as their value in
the ${\cal N}=4$ SYM in the strong coupling limit, where
$\tau_\Pi \approx 1.3/2\pi T$. The fact that the relaxation time is
longer for the vector current is natural in perturbation theory, since
the color charge carried by the quarks is smaller than the charge
carried by the gluons~\footnote{Comparing the widths of the transport peaks in 
\fig(\ref{fig:chinn}) and (\ref{fig:rho-diff}), 
the ratio of relaxation times is approximately given by the 
ratio of the Casimir operators in their irreducible representations, 
$C_F/C_A$, which amounts to 4/9 for the SU(3) color group.}.
While the ordering of scales characteristic of
a weakly coupled plasma is not respected at temperatures relevant to
heavy-ion collisions (we knew this already from static quantities),
some remnant of the kinetic theory transport peak would thus appear to
be present, at least in the vector channel
(see \fig\ref{fig:gert}). These observations should be regarded as
preliminary; one's confidence in them would increase if $\tau_R T$
was seen to become progressively larger as the temperature is
raised.  By the methods used, it cannot be excluded at present that
the longest relaxation times in the system are in reality much longer
than these values, and that the spectral function has a more
complicated structure than the fit ans\"atze allow. The quoted results
merely provide the `simplest' explanation for the lattice data.

%  (the ratio of the Casimir operators in their
% irreducible representations is $4/9$ for the gauge group SU(3)). 

One of the goals of this review is to indicate the prospects for
decisive progress in the field using lattice QCD methods.  This
exercise splits up into two separate questions. The first is, what
accuracy is necessary on the Euclidean correlator in order to impact
our knowledge of the spectral functions. And secondly, how much
computing time is necessary to obtain Euclidean correlators with that
precision. We can illustrate this set of questions with the case of
the force-force correlator that determines how fast a heavy quark
diffuses in the quark-gluon plasma (see
section \ref{sec:hq}). Addressing the first question, we want to
determine the form of the spectral function at small frequencies, and
it is important that one is really interested in a subtracted spectral
function where the ultraviolet $\sim\omega^3$ tail has been removed.
An important aspect here is that even in the weak-coupling limit, no
narrow features are predicted to occur in the spectral function.  The
analysis of~\cite{Burnier:2011jq}, where the narrowest feature of the
(resummed perturbative) spectral function has width $\approx \pi T$,
then shows that one needs a relative error at least as small as
$10^{-3}$ on the UV-subtracted Euclidean correlator and $N_t=24$ to
reconstruct semi-quantitatively the spectral function.  In this
estimate, no physically motivated ansatz is made, it is a truncation
of an exact mathematical formula.

What amount of computing time is then necessary to determine the
Euclidean correlator to this high degree of accuracy? Recently, the
force-force correlator was calculated in the SU(3) gauge theory on
lattices with $N_t$ between ranging from 8 to
22~\cite{Meyer:2010tt}. An $N_t=16$ simulation around $6T_c$ cost
about 2Gflop years (see \fig\ref{fig:1012b}).  The cost increases
rapidly with $N_t$: to achieve the same accuracy at $N_t=22$ as at
$N_t=16$, an increase in CPU-time by a factor 20--25 is
necessary. From this data, we estimate that it would take about half a
month on a machine that sustains 100Tflops to achieve the relative
accuracy of $10^{-3}$ on the shape of the UV-subtracted correlator at
$N_t=22$.  In addition to the statistical errors, it is important to
control the systematic errors at the same level. Thus one needs to
check for finite-volume effects, and (probably more importantly in
this particular case) for discretization errors. This requires varying
the lattice spacing within a significant range. We note however that
the overall normalization of the correlator does not need to be known
to better than a few percent: it is the `shape' of the correlator that
must be controlled to very high accuracy in order to constrain the
spectral function.

The example above shows that when detailed analytic information is
available, and that relatively efficient numerical techniques can be
used, the problem of determining the spectral function and in
particular the transport coefficient say at the $30\%$ level can be
tackled with currently existing computing resources, albeit at the
high end. It appears likely that the calculations will continue to be
performed in the pure Yang-Mills theory for some time, since they are
much faster and the crucial dynamics in any case appears to be
associated with the gluons.  Thus with a dedicated effort it is likely
that in the next five to ten years, decisive progress will be made in
this direction.  Fortunately this is also the time scale on which
heavy-ion collision experiments are approved for, in particular the
CBM experiment at FAIR.  The only assumption (backed up by resummed
perturbation theory) made in this analysis on the spectral function is
that it does not vary strongly on a frequency intervals of about $\pi
T$.  Even before such a model-independent reconstruction of the
spectral function becomes possible, one may be able to draw
interesting lessons already when the Monte-Carlo data quality is
lower, for instance by making weak-coupling or strong-coupling
(AdS/CFT) inspired fits.

The isovector vector channel for light quarks is another case where we
expect significant progress to be made, because the cost does not
increase as fast with $N_t$, and the spectral function increases more
mildly, as $\omega^2$, at large frequencies. On the other hand, the
spectral function has a narrow transport peak at small frequencies in
the weak coupling regime (parametrically, the width is $g^4T$ and
numerically, $0.4T$ to $0.6T$ for intermediate coupling
values~\cite{Moore:2008ws}), and determining separately the width and
height of the transport peak is extremely difficult in that regime.
The difficulty may be overcome with extremely accurate data, and
possibly exploiting small but non-vanishing momenta by using twisted
boundary conditions~\cite{Sachrajda:2004mi,Bedaque:2004kc}.

Given the observation of elliptic flow at
RHIC~\cite{Adcox:2004mh,Adams:2005dq,Back:2004je,Arsene:2004fa} and
most recently at the LHC~\cite{Aamodt:2010pa} the shear viscosity is
undoubtedly the most important transport coefficient for heavy-ion
phenomenology.  The shear and sound channels are presumably the
hardest to tackle by Euclidean methods, because they both exhibit a
parametrically narrow transport peak at weak coupling, they have a
very hard UV contribution ($\omega^4$ in the spectral function), and
the variance is strongly UV-divergent. The multi-level
algorithm~\cite{Meyer:2003hy} only partly alleviates the latter
problem by reducing the power with which the variance diverges.  So
far these major difficulties have only partly been overcome by fitting a
particular ansatz (motivated by hydrodynamics and perturbation theory)
simultaneously to several correlators related by Ward identities (see
section~\ref{sec:combi}).

The methods discussed here may be of interest to physicists in other
fields where quantum Monte-Carlo simulations are used.  In particular
we have in mind cold Fermi gases, with which very controlled
experiments can be made. When tuned to `unitarity', the atoms have a
divergent scattering length in the $s$-channel, and the system is
strongly interacting. The shear viscosity to entropy density ratio has
been estimated to be below unity, thus making this system an excellent
fluid (see the review~\cite{Schafer:2009dj}).  On the theory side, the
spectral functions of a Fermi gas have recently been studied in
kinetic theory~\cite{Chao:2010tk}. An important technical
simplification over the relativistic case discussed in this review is
that the shear spectral function goes to zero at high
frequencies~\cite{Taylor:2010ju,Son:2010kq}, which implies that the
Euclidean correlator is far more sensitive to the low-frequency part
of the spectral function relevant to transport properties.

\subsection*{Acknowledgements}
I thank 
Gert Aarts, 
Douglas Beck,
Jorge Casalderrey-Solana, 
Philippe de Forcrand,
Elias Kiritsis, 
Mikko Laine, 
Martin L\"uscher, 
Guy Moore,
John Negele, 
Peter Petreczky,
Krishna Rajagopal, 
Alexi Vuorinen, 
Paul Romatschke,
Thomas Sch\"afer,
Dam Thanh Son,
Misha Stephanov,
Derek Teaney, 
Laurence Yaffe and
Urs Wiedemann
for many stimulating and illuminating discussions on the subject
of this review over the past four years.  Any misstatements or errors
are of course the sole responsibility of the author.

% Completion end of December.

%
% BibTeX users please use
% \bibliographystyle{aipauth4-1}
% \bibliographystyle{apsrev4-1}

% \newpage
\bibliographystyle{JHEP}
\bibliography{/home/meyerh/CTPHOPPER/ctphopper-home/BIBLIO/viscobib}
%
% Non-BibTeX users please use
% \begin{thebibliography}{}
%
% and use \bibitem to create references.
%
% \bibitem{RefJ}
% Format for Journal Reference
% Author, Journal \textbf{Volume}, (year) page numbers.
% Format for books
% \bibitem{RefB}
% Author, \textit{Book title} (Publisher, place year) page numbers
% etc
% \end{thebibliography}

\end{document}